{}

\documentclass[10pt,a4paper]{article}

\newif\ifpublic\publictrue

\ifpublic\else\usepackage{showkeys}\fi
\usepackage[bookmarks=true,hyperfigures=true,hidelinks,bookmarksnumbered,unicode]{hyperref}
\usepackage[nosort]{cite}
\usepackage[parsep]{collref}
\usepackage[english]{babel}
\usepackage[T1]{fontenc}
\usepackage[utf8]{inputenc}
\usepackage{lmodern}
\usepackage{amsfonts,amsbsy,dsfont,euscript,mathrsfs,fixmath}
\usepackage{amsmath,amssymb,mathtools}
\usepackage{scalerel}
\usepackage{enumerate}
\usepackage[x11names]{xcolor}
\usepackage{graphicx}
\usepackage{adjustbox}
\usepackage{multirow}
\usepackage{color, colortbl}

\usepackage{bbold}

\usepackage{cancel}

\def\showkeysrefformat#1{{\normalfont\tiny\ttfamily#1}}
\makeatletter
\def\SK@@ref#1>#2\SK@{{\@inlabelfalse\leavevmode\vbox to\z@{\vss\SK@refcolor\rlap{\vrule\raise .75em \hbox{\showkeysrefformat{#2}}}}}}
\makeatother

\usepackage[a4paper,text={160mm,247mm},centering]{geometry}
\allowdisplaybreaks[3]
\numberwithin{equation}{section}
\def\[{\begin{equation}\begin{aligned}}
\def\]{\end{aligned}\end{equation}}

\usepackage[font=small,labelfont=bf,width=0.85\textwidth]{caption}

\expandafter\def\expandafter\bfseries\expandafter{\bfseries\ifmmode\else\boldmath\fi}
\expandafter\def\expandafter\mdseries\expandafter{\mdseries\ifmmode\else\unboldmath\fi}
\expandafter\def\expandafter\normalfont\expandafter{\normalfont\ifmmode\else\unboldmath\fi}

\RequirePackage{verbatim}

\makeatletter
\newwrite\bibinl@out
\newenvironment{bibtex}[1][\jobname]{%
\immediate\openout\bibinl@out #1.bib%
\immediate\write\bibinl@out{\@percentchar generated from `\jobname' starting line \the\inputlineno^^J}%
\def\verbatim@processline{\immediate\write\bibinl@out{\the\verbatim@line}}%
\@bsphack\let\do\@makeother\dospecials\catcode`\^^M\active\verbatim@start%
}
{\immediate\closeout\bibinl@out\@esphack}
\makeatother

\let\barefrac=\frac
\renewcommand{\frac}[2]{\mathinner{\barefrac{#1}{#2}}}

\let\baresqrt=\sqrt
\makeatletter
\renewcommand{\sqrt}{\@ifnextchar[\@sqrt@space@a\@sqrt@space@b}
\def\@sqrt@space@a[#1]#2{\mathinner{\mathchoice{\mkern-3mu}{\mkern-3mu}{}{}\baresqrt[#1]{#2}}}
\def\@sqrt@space@b#1{\mathinner{\mathchoice{\mkern-3mu}{\mkern-3mu}{}{}\baresqrt{#1}}}
\makeatother

\makeatletter
\let\per@dot@old=\.
\def\.{\ifmmode\def\per@dot@sel{\mkern3mu}\else\def\per@dot@sel{\per@dot@old}\fi\per@dot@sel}
\makeatother

\let\barefootnote=\footnote
\renewcommand{\footnote}[1]{\barefootnote{#1\vspace{3pt}}}

\newcommand{\vfrac}[2]{\ifmmode\mathinner{\textstyle^{#1}\!/\!_{#2}}\else$^{#1}\!/\!_{#2}$\fi}

\DeclareMathOperator{\STr}{STr}

\DeclareMathOperator{\rk}{rk}

\newcommand{\set}[1]{\{#1\}}
\newcommand{\Real}{\mathds{R}}
\newcommand{\Complex}{\mathds{C}}

\makeatletter
\newcommand*\bigcdot{\mathpalette\bigcdot@{.5}}
\newcommand*\bigcdot@[2]{\mathbin{\vcenter{\hbox{\scalebox{#2}{$\m@th#1\bullet$}}}}}
\makeatother

\newcommand{\alg}[1]{\mathfrak{#1}}

\DeclareMathOperator{\ad}{ad}

\newcommand{\com}[2]{[#1,#2]}

\def\<{\big\langle}
\def\>{\big\rangle}

\newcommand{\geom}[1]{#1}
\newcommand{\Man}{\mathcal{M}}
\newcommand{\AdS}{\geom{AdS}}

\newcommand{\Sp}{\geom{S}}

\newcommand{\Texp}{\operatorname{Texp}}

\newcommand{\Act}{\mathcal{S}}

\DeclareFontEncoding{LS1}{}{}
\DeclareFontSubstitution{LS1}{stix}{m}{n}
\DeclareSymbolFont{stixsymbols}{LS1}{stixscr}{m}{n}
\SetSymbolFont{stixsymbols}{bold}{LS1}{stixscr}{b}{n}
\DeclareMathSymbol{\kay}{\mathalpha}{stixsymbols}{"6B}
\DeclareMathSymbol{\hay}{\mathalpha}{stixsymbols}{"68}
\DeclareMathAlphabet{\mathdsl}{U}{bbm}{m}{sl}

\newcommand{\ket}[1]{\big| #1 \big \rangle }

\newcommand{\gf}{\xrightarrow{g.f.}}
\newcommand{\sm}{sigma-model}
\newcommand{\sms}{sigma-models}

\providecommand{\href}[2]{#2}

\makeatletter
\def\mr@ignsp#1 {\ifx\:#1\@empty\else #1\expandafter\mr@ignsp\fi}
\newcommand{\multiref}[1]{\begingroup%
\xdef\mr@no@sparg{\expandafter\mr@ignsp#1 \: }%
\def\mr@comma{}\def\mr@dash{-}%
\@for\mr@refs:=\mr@no@sparg\do{%
\ifx\mr@refs\mr@dash\def\mr@comma{}--\else%
\mr@comma\def\mr@comma{,}\ref{\mr@refs}\fi}%
\endgroup}
\renewcommand{\eqref}[1]{(\multiref{#1})}
\makeatother

\makeatletter
\newcommand{\namedref}[2]{\hyperref[#2]{#1~\ref*{#2}}}
\newcommand{\secref}{\@ifstar{\namedref{Section}}{\namedref{section}}}
\newcommand{\appref}{\@ifstar{\namedref{Appendix}}{\namedref{appendix}}}
\newcommand{\tabref}{\@ifstar{\namedref{Table}}{\namedref{table}}}
\newcommand{\figref}{\@ifstar{\namedref{Figure}}{\namedref{figure}}}
\newcommand{\fref}{\@ifstar{\namedref{Footnote}}{\namedref{footnote}}}
\makeatother

\let\oldbib=\thebibliography
\def\thebibliography{\phantomsection\addcontentsline{toc}{section}{\refname}\oldbib}

\providecommand{\hypersetup}[1]{}
\providecommand{\texorpdfstring}[2]{#1}
\hypersetup{plainpages=false}
\hypersetup{pdfpagemode=UseNone}
\hypersetup{bookmarksnumbered=true}
\hypersetup{pdfstartview=FitH}
\hypersetup{colorlinks=false}
\hypersetup{citebordercolor={1 1 1}}
\hypersetup{urlbordercolor={1 1 1}}
\hypersetup{linkbordercolor={1 1 1}}

\makeatletter
\let\@keywords\@empty
\let\@subject\@empty
\providecommand{\keywords}[1]{\gdef\@keywords{#1}}
\providecommand{\subject}[1]{\gdef\@subject{#1}}
\def\thetitle{\@title}
\def\theauthor{\@author}
\def\thesubject{\@subject}
\def\thedate{\@date}
\def\thekeywords{\@keywords}
\makeatother
\AtBeginDocument{
\hypersetup{pdftitle={\thetitle}}
\hypersetup{pdfauthor={\theauthor}}
\hypersetup{pdfsubject={\thesubject}}
\hypersetup{pdfkeywords={\thekeywords}}}

\definecolor{Gray}{gray}{0.9}

\newif\ifshownote
\shownotetrue

\ifpublic\shownotefalse\fi

\ifshownote

\ifpdf\else\RequirePackage[active]{srcltx}\fi
\RequirePackage{xcolor}
\RequirePackage{ulem}

\newcommand{\remark}[2][]{{\normalfont\sffamily\hspace{1ex}
\def\emph{\textsl}
\def\textbullet{$\bullet$}
\def\tmparga{#1}
\def\tmpargb{BH}\ifx\tmparga\tmpargb\color[rgb]{0.7,0,0}\fi
\def\tmpargb{}\ifx\tmparga\tmpargb\normalfont\color{red}\fi
\def\tmpargb{}\ifx\tmparga\tmpargb\else \textbf{#1:}\fi
#2\hspace{1ex}}}

\else

\newcommand{\remark}[2][]{\ignorespaces}

\fi

\title{Inequivalent light-cone gauge-fixings of strings
\texorpdfstring{\\}{} on \texorpdfstring{$\AdS_n \times \Sp^n$}{AdSn × Sn} backgrounds}
\author{Riccardo Borsato\texorpdfstring{$^{a,}$\footnote{\texttt{riccardo.borsato@usc.es}}}{}, Sibylle Driezen\texorpdfstring{$^{b,}$\footnote{\texttt{sdriezen@phys.ethz.ch}}}{}, Ben Hoare\texorpdfstring{$^{c,}$\footnote{\texttt{ben.hoare@durham.ac.uk}}}{}, Ana L.~Retore\texorpdfstring{$^{c,}$\footnote{\texttt{ana.retore@durham.ac.uk}}}{}, Fiona K.~Seibold\texorpdfstring{$^{d,}$\footnote{\texttt{f.seibold21@imperial.ac.uk}}}{}}

\begin{document}

\pdfbookmark[1]{Title Page}{title}
\thispagestyle{empty}


\vspace*{2cm}
\begin{center}
\begingroup\Large\bfseries\thetitle\par\endgroup
\vspace{1cm}

\begingroup\theauthor\par\endgroup
\vspace{1cm}

\textit{$^a $Instituto Galego de F\'isica de Altas Enerx\'ias (IGFAE) and
Departamento de F\'\i sica de Part\'\i culas,\\
Universidade de Santiago de Compostela}\\
\vspace{1mm}
\textit{$^b $Institut f\"ur Theoretische Physik, ETH Z\"urich,
Wolfgang-Pauli-Stra\ss e 27, Z\"urich 8093, Switzerland}\\
\vspace{1mm}
\textit{$^c $Department of Mathematical Sciences, Durham University, Durham DH1 3LE, UK}\\
\vspace{1mm}
\textit{$^d $Blackett Laboratory, Imperial College, London SW7 2AZ, UK}
\vspace{5mm}


\vspace{2cm}

\textbf{Abstract}\vspace{5mm}

\begin{minipage}{12.5cm}
Light-cone gauge-fixed \sms{} on $AdS_n\times S^n$ backgrounds play an important role in the integrability formulation of the AdS/CFT correspondence.
The string spectrum of the \sm{} is gauge-independent, however the Hamiltonian and scattering matrix of the transverse worldsheet fields are not.
We study how these change for a large family of inequivalent light-cone gauges, which are interpreted as $T\bar T$, $\tilde JT_\tau$, $JT_\sigma$ and $J^\tau$ deformations.
We investigate the moduli space of inequivalent light-cone gauges and, specialising to $\AdS_5 \times \Sp^5$, compute the different light-cone gauge symmetry algebras, well known to be $\mathfrak{psu}(2|2)^{\oplus 2} \oplus \mathfrak{u}(1)^{\oplus 2}$ for the standard gauge-fixing.
Many integrable deformations require a non-standard light-cone gauge, hence our classification and analysis of inequivalent gauges will be important for analysing such models.
\end{minipage}

\end{center}

\newpage

\tableofcontents

\section{Introduction}\label{sec:intro}

The worldsheet reparametrisation invariance of string \sms{} may be viewed as a gauge freedom that can be fixed to identify the physical degrees of freedom.
In the context of the AdS/CFT correspondence~\cite{Maldacena:1997re}, an important class of gauges are the uniform light-cone gauges introduced in \cite{Kruczenski:2004kw,Kruczenski:2004cn,Arutyunov:2004yx,Arutyunov:2005hd}, see~\cite{Arutyunov:2009ga} for a review.
These gauges are engineered such that a certain conserved charge is uniformly distributed along the spatial extent of the string.
Our aim in this paper is to map out the moduli space of inequivalent light-cone gauges, focusing in particular on $AdS_n \times S^n$ backgrounds, the product of $n$-dimensional anti de Sitter space and an $n$-dimensional sphere.
This is particularly of interest in the study of integrable string \sms, such as $AdS_5\times S^5$~\cite{Bena:2003wd} and $AdS_3\times S^3\times T^4$~\cite{Babichenko:2009dk}, and their integrable deformations.

Uniform light-cone gauge can be fixed for any background that has two commuting isometries, one timelike and one spacelike.
In this paper we will take these isometries to be realised by shifts in $t$, a timelike coordinate, and $\varphi$, a spacelike coordinate, such that $t = \varphi = \tau$ is a classical solution of the \sm{} where $\tau$ is the worldsheet time.
Introducing light-cone coordinates $x^+=(t+\varphi)/2$ and $x^-=\varphi-t$, we expand the worldsheet action around the classical solution and gauge-fix the fluctuations of the fields $x^+$ and $p_-$, the momentum conjugate to $x^-$, to zero.
Therefore, the light-cone gauge-fixing essentially demands that these two fields are equal to their classical configuration, $x^+=\tau$ and $p_-=1$.

The light-cone gauge-fixing procedure results in a ``reduced model'' for the fields $x^\mu$ transverse to the longitudinal fields $x^+$ and $x^-$.
The Hamiltonian $H$ of the reduced model is identified with the target-space charge $E-J$, where the energy $E$ and angular momentum $J$ are the Noether charges for shifts in $t$ and $\varphi$ respectively.
The reduced model is invariant under a subalgebra of the full superisometry algebra of the original background, identified as the subalgebra that commutes with the $x^+$ shift isometry.
Relaxing the level-matching condition for worldsheet excitations, this subalgebra is centrally extended by charges depending on the worldsheet momentum.
In the case of $AdS_5\times S^5$, the $\mathfrak{psu}(2,2|4)$ superisometry leads to a centrally-extended $\mathfrak{psu}(2|2)^{\oplus 2}$ residual superalgebra of the reduced model~\cite{Arutyunov:2006ak}, while in the case of $AdS_3\times S^3\times T^4$ (ignoring the torus directions and their superpartners), the $\mathfrak{psu}(1,1|2)^{\oplus2}$ superisometry leads to a central extension of $[\mathfrak{u}(1)\ltimes \mathfrak{psu}(1|1)^{\oplus 2}]^{\oplus 2}$~\cite{Borsato:2014hja,Lloyd:2014bsa}.
These residual superalgebras play a fundamental role in the construction of the exact worldsheet S-matrices, which underpins the integrability description of these models.

While the spectrum, i.e.~the set of eigenvalues of the spacetime energy $E$, does not depend on the choice of gauge, the gauge-fixed Hamiltonian and the worldsheet S-matrix are \textit{gauge-dependent}.
To analyse the moduli space of inequivalent light-cone gauges we fix the classical solution, but change how we identify the longitudinal $x^\pm$ and transverse $x^\mu$ fields.
In particular, we consider target-space coordinate transformations $(x^+,x^-,x^\mu)\to (\tilde x^+,\tilde x^-,\tilde x^\mu)$ and study when they lead to an inequivalent Hamiltonian and S-matrix after gauge-fixing.

The relation between the standard uniform light-cone gauge and $T\bar{T}$ deformations due to a coordinate transformation within the longitudinal sector was first elucidated in~\cite{Baggio:2018gct}.
The Hamiltonian analysis, and the interpretation in terms of current-current deformations, was then later extended to include more general light-cone gauge-fixings in~\cite{Frolov:2019nrr,Frolov:2019xzi}.
In this paper we will build on these results, presenting general derivations for the variations of the Hamiltonian and S-matrix, and the invariance of the spectrum.
Moreover, focusing on the case of symmetric spaces, including $AdS_n \times S^n$, we will also investigate the moduli space of inequivalent gauges and provide perturbative evidence for the general derivations.

\medskip

The motivation for this work comes from the recent construction of large families of integrable deformations, see~\cite{Hoare:2021dix} for a review.
These include the Yang-Baxter deformations~\cite{Klimcik:2002zj,Delduc:2013qra,Kawaguchi:2014qwa,vanTongeren:2015soa}, constructed from solutions to the classical Yang-Baxter equation on the Lie (super)algebra of isometries.
Another class are elliptic deformations, which have only recently been started to be incorporated at the level of string \sms~\cite{Cherednik:1981df,Lacroix:2023qlz,Hoare:2023zti}.
In general, such deformations will break the original group of (super)isometries to a smaller subgroup.
Crucially, in some cases the deformations may break the light-cone isometries that are normally used to gauge-fix the undeformed model.
As a result, gauge-fixing in the presence of the deformation forces us to choose a different set of light-cone isometries, see~\cite{Hoare:2023zti, inpreparation-RiccardoSibylle} for applications to particular models.
Since in the absence of the deformation this can be understood as an alternative light-cone gauge-fixing, the systematic study presented here provides key insights into the quantum integrability description of deformed models.

\medskip

This paper is organised as follows.
In~\secref{sec:ineq-lcgf} we present our strategy for generating inequivalent gauges on generic backgrounds.
In particular, we analyse when target-space coordinate transformations lead to inequivalent gauges and present a general derivation for the effect on the Hamiltonian of the reduced model.
In~\secref{sec:in-sym-cosets} we discuss the classification of inequivalent gauges for symmetric spaces, with particular attention to $AdS_n\times S^n$, and the symmetries of the reduced model.
In~\secref{sec:Smat} we study how the S-matrix changes under the different gauges, both at tree-level and non-perturbatively.
In~\secref{sec:spectrum} we describe how to check the gauge invariance of the spectrum.
Finally, in~\secref{sec:concl} we finish with concluding comments and an outlook.

\section{Inequivalent light-cone gauge-fixings}\label{sec:ineq-lcgf}

In this section we use the procedure of light-cone gauge-fixing reviewed in \appref{sec:appendix}, see also~\cite{Arutyunov:2009ga} and references therein.
In particular, we work with the \sm{} action~\eqref{eq:S-Poly}, which we schematically write as
$ \Act=\int_{-L/2}^{L/2}d\tau d\sigma\, \mathscr L$, where $\mathscr L$ is the Lagrangian density.
The target space is parametrised by $D$ coordinates $x^M$, which we split as $(x^+,x^-,x^\mu)$, and $\tau$ and $\sigma$ are worldsheet time and space coordinates respectively with $\sigma \sim \sigma + L$.
A dot denotes the time derivative $\dot x^M = \partial_\tau x^M$ and a prime, the spatial derivative $x'{}^M = \partial_\sigma x^M$.

We assume that the action is invariant under constant shifts of the two light-cone fields $x^+$ and $x^-$, so that there is a classical point-like string solution of the form $x^+=\tau$.
We expand the action around this classical configuration $x^+=\tau$ in the Hamiltonian formalism and thus introduce a conjugate momentum $p_M$ for each field $x^M$.
More details on this procedure and what follows are collected in \appref{sec:appendix}.

The uniform light-cone gauge is fixed by setting the fluctuations of the fields $x^+$ and $p_-$ to zero.
Therefore, on the gauge these fields coincide with their classical configurations, and we may write just $x^+=\tau$ and $p_-=1$.
After light-cone gauge, one obtains a reduced model for the $D-2$ ``transverse'' fields $x^\mu$, whose Lagrangian density we will denote as $\mathcal L$.
The Hamiltonian density of the reduced model will be denoted as $\mathcal H$, and in the uniform light-cone gauge it is identified as $\mathcal H=-p_+$.

The question that we would like to address here is: Is it possible to fix the light-cone gauge in different ways? In particular, are there alternative light-cone gauge-fixings that, despite the expansion being carried out around the \textit{same} classical solution, yield a different Hamiltonian density $\mathcal H$ for the reduced model?

We will answer this question by comparing the light-cone gauge-fixing $x^+=\tau, \ p_-=1$ with an alternative light-cone gauge-fixing $\tilde x^+=\tau, \ \tilde p_-=1$ after performing target-space coordinate transformations (or, equivalently, local field redefinitions on the worldsheet)
\begin{equation}
x^M=(x^+,x^-,x^\mu)\quad\to\quad \tilde x^M= (\tilde x^+,\tilde x^-,\tilde x^\mu) .
\end{equation}
An important point is that we do not allow for the most general coordinate transformation: we demand that after the transformation the background remains invariant under shifts of the coordinates $\tilde x^\pm$, which allows us to fix the light-cone gauge in the usual way as reviewed in \appref{sec:appendix}.
As we will see, this requirement will constrain the relevant classes of coordinate transformations.

After the transformation to the $\tilde x^M$ coordinates and the alternative gauge-fixing $\tilde x^+=\tau$, $\tilde p_-=1$, in principle one ends up with a Hamiltonian density $\tilde{\mathcal H}$.
We will then distinguish ``equivalent'' and ``inequivalent'' light-cone gauges.
Our definition is that two gauge-fixings are equivalent if the two Hamiltonian densities $\tilde{\mathcal H}$ and $\mathcal H$ are related by a canonical transformation.
We will say that they are inequivalent if there is no canonical transformation relating them.

To start, let us give the simplest possible example of a coordinate transformation (or local field redefinition) that leads to an equivalent gauge.
To simplify the notation, we collect all transverse fields $x^\mu$ in the vector $\vec x$ and consider the transformation,
\begin{equation}
\begin{alignedat}{3}
&x^+=\tilde x^+,\qquad
&& x^-=\tilde x^-,\qquad
&& x^\mu=f^\mu(\tilde{\vec x}),\\
&\tilde p_+=p_+,\qquad
&& \tilde p_-=p_-,\qquad
&& \tilde p_\mu=\frac{\partial f^\nu}{\partial\tilde x^\mu}p_\nu,
\end{alignedat}
\end{equation}
where $f^\mu$ is an invertible function and the second line follows from the first.
Note that here we choose to implement a transformation on the transverse fields only, while the light-cone fields transform trivially.
In particular, the relation $\tilde p_+=p_+$ implies
\begin{equation}
\tilde {\mathcal H}(\tilde{\vec x},\tilde{\vec p})=\mathcal H(\vec x(\tilde{\vec x}),\vec p(\tilde{\vec x},\tilde{\vec p})),
\end{equation}
where we write explicitly the dependence of the Hamiltonian densities on the corresponding fields.
In other words, the two Hamiltonians are the same if the transverse fields and momenta are mapped as
\begin{equation}
x^\mu=f^\mu(\tilde{\vec x}),\qquad\tilde p_\mu=\frac{\partial f^\nu}{\partial\tilde x^\mu}p_\nu.
\end{equation}
The reader will recognise this as a class of canonical transformations that are typically called ``point transformations.'' Because the two Hamiltonians are related by a canonical transformation, in this case, according to our definition, the two light-cone gauges are equivalent.

This result was expected even before considering the Hamiltonians.
Taking into account the above relations
\begin{equation}
x^+=\tilde x^+,\qquad
\tilde p_-=p_-,
\end{equation}
it is obvious that the gauge condition $x^+=\tau,\,p_-=1$ is compatible with the gauge $\tilde x^+=\tau,\,\tilde p_-=1$, because the two are in fact the same condition.
In this case, the coordinate transformation does not affect the gauge condition but only redefines the transverse fields.
This means that the procedure of light-cone gauge-fixing and the field redefinition are two commuting operations.
From this observation it should be clear that to generate inequivalent gauges we must allow the longitudinal coordinates to participate non-trivially in the coordinate transformation.
When doing this, however, we will need to be careful not to spoil the invariance of the action under shifts of the $\tilde x^\pm$ fields, as this is one of our requirements specified above.

Before presenting the concrete examples of interest, let us discuss the general strategy that we will use to construct the transformations and the inequivalent gauges.

\subsection{Inequivalent gauges from current-current deformations}\label{sec:gauges-curr-curr}

Our strategy to construct the coordinate transformations and the corresponding inequivalent gauges is to exploit the symmetries of the \sm{} \textit{before} gauge-fixing.
In particular, let us assume that the action before gauge-fixing is invariant under a global continuous transformation, and therefore an isometry transformation in target space.
This symmetry transformation can be understood as the map
\begin{equation}\label{eq:transf-symm}
x^M = F^M(\tilde x,\lambda),
\end{equation}
where $\lambda$ is the continuous parameter.
Saying that the action is invariant under this map for constant $\lambda$ means that after the transformation the action does not depend on $\lambda$, and the new action agrees with the old one upon the trivial replacement $x\to\tilde x$.
From Noether's theorem, then, it follows that there is a conserved current.
In particular, one considers the infinitesimal transformation
\begin{equation}\label{eq:inf-transf}
\delta x^M = f^M\delta\lambda ,\qquad \text{where } f^M = \left.\frac{\partial F^M}{\partial\lambda}\right|_{\lambda=0},
\end{equation}
so that the infinitesimal variation of the Lagrangian $\mathscr L$ is
\begin{equation}\label{eq:variation-L}
\delta \mathscr L = \partial_\alpha\left(\frac{\partial\mathscr L}{\partial(\partial_\alpha x^M)}f^M\delta \lambda\right),
\end{equation}
up to terms that vanish on the equations of motion.
Invariance of the action for constant $\delta\lambda$ implies that the Lagrangian can change at most by a total derivative, so we write $\delta \mathscr L=\partial_\alpha V^\alpha$ for some $V^\alpha$.
We can then identify the conserved Noether current as
\begin{equation}\label{eq:def-noether-current}
J^\alpha = V^\alpha - \frac{\partial\mathscr L}{\partial(\partial_\alpha x^M)}f^M.
\end{equation}
From now on, for simplicity, we assume that $V^\alpha=0$.

To construct coordinate transformations that generate inequivalent gauges we use the global transformation~\eqref{eq:transf-symm}, but promote the parameter $\lambda$ to be a \textit{function of the fields}.
In particular, let us write
\begin{equation}
\delta\lambda = \gamma c(\tilde x),
\end{equation}
for some function $c$ of $\tilde x^M$.
Here we are introducing a continuous parameter $\gamma$ that we will use to keep track of the transformation, so that $\gamma\to 0$ reduces to the identity.
Importantly, when $\delta\lambda$ is not constant the action is not necessarily invariant.
In other words, the map ceases to be a symmetry transformation and it is understood just as a local field redefinition or target space coordinate transformation.
On-shell (i.e.~on the equations of motion), the infinitesimal variation of the Lagrangian is still given by~\eqref{eq:variation-L}, but because $\delta \lambda$ is no longer constant we now have
\begin{equation}
\delta \mathscr L =-\partial_\alpha\left( J^\alpha\delta \lambda\right)
=-\partial_\alpha J^\alpha \delta \lambda-J^\alpha\partial_\alpha\delta \lambda
=-J^\alpha\partial_\alpha\delta \lambda,
\end{equation}
where we used the conservation of the current.
At this point we can define the \textit{topological current}
\begin{equation}
\tilde J_{(c)}^\alpha = \epsilon^{\alpha\beta}\partial_\beta c,
\end{equation}
which is conserved off-shell (i.e.~without the need of the equations of motion).
We may therefore rewrite the variation of the Lagrangian as
\begin{equation}
\delta \mathscr L = \gamma\epsilon_{\alpha\beta}\tilde J_{(c)}^\alpha J^\beta.
\end{equation}
This formally takes the form of an infinitesimal current-current deformation.
Importantly, $\tilde J_{(c)}^\alpha$ and $J^\alpha$ are different objects: the former is a topological current that is identified by the choice of the function $c$, while the latter is a Noether current identified by the symmetry transformation that we selected.

Finally, let us return to the condition that the action is invariant under shifts of $\tilde x^\pm$ after the coordinate transformation.
As already mentioned, we require this in order to follow the usual procedure for the light-cone gauge-fixing as described in \appref{sec:appendix}.
It is clear that at the level of the infinitesimal transformation we must require the function $c$ to be of the form
\begin{equation}
c(\tilde x^+,\tilde x^-,\vec{\tilde x}) = \gamma_+\tilde x^++\gamma_-\tilde x^-+g(\vec{\tilde x}).
\end{equation}
Here $\gamma_\pm$ are constant parameters (which may be rescaled by redefining the overall deformation parameter $\gamma$) and $g$ is a function of transverse fields only.
This ensures that the derivative of $\delta \lambda$, and therefore $\delta \mathscr L$ itself, may depend on \textit{derivatives} of $\tilde x^\pm$ but not on $\tilde x^\pm$ explicitly, and thus the shift invariance will indeed be preserved.

So far, the discussion has been at the level of the Lagrangian density $\mathscr L$ before gauge-fixing.
When gauge-fixing, the Lagrangian density $\mathscr L$ is evaluated on the solutions to the Virasoro constraints obtained after setting $x^+=\tau,\,p_-=1$.
This procedure sends $\mathscr L$ to $\mathcal L$, the Lagrangian of the reduced model.
Schematically, we may write $\left(\mathscr L\right)_{g.f.}=\mathcal L $, where ``$g.f.$'' denotes the light-cone gauge-fixing procedure.
The transformation of $\mathcal L$ is then simply inherited from that of $\mathscr L$, and we can write
\begin{equation}
\delta \mathcal L = \gamma\epsilon_{\alpha\beta}\left(\tilde J_{(c)}^\alpha\right)_{g.f.}\left( J^\beta\right)_{g.f.}.
\end{equation}
Therefore, the evaluation of the topological and Noether currents on the gauge-fixing constraints will tell us how the Lagrangian of the reduced model transforms.
Taking into account that $\mathcal H=p_\mu\dot x^\mu-\mathcal L$, we can also conclude that the transformation of the Hamiltonian density of the reduced model is
\begin{equation} \label{eq:deltaH-currcurr}
\delta \mathcal H = - \gamma\epsilon_{\alpha\beta}\left(\tilde J_{(c)}^\alpha\right)_{g.f.}\left( J^\beta\right)_{g.f.}.
\end{equation}

\subsubsection{Light-cone currents}\label{ss:lcc}

Of all the Noether and topological currents that we may consider, an important role is played by the ``light-cone currents.'' First, invariance of $\mathscr L$ under shifts of $x^\pm$ implies the conservation of the following two Noether currents:
\begin{equation}
J^\alpha_{(+)}=-\frac{\partial \mathscr L}{\partial(\partial_\alpha x^+)},\qquad
J^\alpha_{(-)}=-\frac{\partial \mathscr L}{\partial(\partial_\alpha x^-)}.
\end{equation}
Second, following analysis above, it is natural to consider the following topological currents:
\begin{equation}
\tilde J^\alpha_{(+)}=\epsilon^{\alpha\beta}\partial_\beta x^+,\qquad
\tilde J^\alpha_{(-)}=\epsilon^{\alpha\beta}\partial_\beta x^-.
\end{equation}
We will now show that upon gauge-fixing these currents become
\begin{equation} \label{eq:lc-currents-claim}
\begin{aligned}
& \left(J^\alpha_{(+)} \right)_{g.f.} =T^\alpha{}_\tau,\qquad\qquad
&& \left(J^\alpha_{(-)}\right)_{g.f.} = -\delta^\alpha_\tau,\\
& \left(\tilde J^\alpha_{(+)} \right)_{g.f.} = \delta^\alpha_\sigma,\qquad\qquad
&& \left(\tilde J^\alpha_{(-)} \right)_{g.f.}= T^\alpha{}_\sigma,
\end{aligned}
\end{equation}
where $T^\alpha{}_\beta$ is the stress-energy tensor of the reduced model.
Importantly, if we call $\mathcal T^\alpha{}_\beta$ the stress-energy tensor of the model before gauge-fixing, it is \textit{not} true that $\left(\mathcal T^\alpha{}_\beta\right)_{g.f.} =T^\alpha{}_\beta$.
In fact, $\mathcal T^\alpha{}_\beta$ is zero as a consequence of the Virasoro constraints, while $T^\alpha{}_\beta$ is not.
The latter is calculated from the Lagrangian $\mathcal L$ in the usual way following Noether's theorem,
\begin{equation}
T^\alpha{}_\beta = \frac{\partial \mathcal L}{\partial(\partial_\alpha x^\mu)}\partial_\beta x^\mu-\delta^\alpha{}_\beta \mathcal L .
\end{equation}
In the Hamiltonian formalism, each component is
\begin{equation}
\begin{aligned}
&T^\tau{}_\tau= \mathcal H,\qquad\qquad
&& T^\sigma{}_\tau = -\frac{\partial \mathcal H}{\partial x'{}^\mu}\frac{\partial \mathcal H}{\partial p_\mu},\\
&T^\tau{}_\sigma= p_\mu x'{}^\mu,\qquad\qquad
&& T^\sigma{}_\sigma =\mathcal H -\frac{\partial \mathcal H}{\partial x'{}^\mu}x'{}^\mu-\frac{\partial \mathcal H}{\partial p_\mu}p_\mu.
\end{aligned}
\end{equation}

To prove the claim \eqref{eq:lc-currents-claim}, let us start with the topological currents.
First, we have
\begin{equation}
\tilde J^\tau_{(+)}=-x'{}^+ \ \gf \ 0,\qquad\qquad
\tilde J^\sigma_{(+)}=\dot x{}^+ \ \gf \ 1,
\end{equation}
where after the arrow we indicate the evaluation of the expression upon light-cone gauge-fixing.
Now considering the other topological current, it is easy to identify the time component as
\begin{equation}
\tilde J^\tau_{(-)}=-x'{}^-\ \gf \ p_\mu x'{}^\mu=T^\tau{}_\sigma.
\end{equation}
To identify the remaining spatial component we may reason as follows.
Consider two conserved currents $J_1^\alpha$ and $J_2^\alpha$, so that $\partial_\alpha J^\alpha_i=0$.
They may be Noether or topological currents, and in the example that we are considering we take $J_1^\alpha=\tilde J^\alpha_{(-)}$ and $J_2^\alpha=T^\alpha{}_\sigma$.
If, as above, we are able to prove that $J^\tau_1=J^\tau_2$ then it follows that $\partial_\sigma(J^\sigma_1-J^\sigma_2)=0$.
In other words, the $\sigma$-components may differ at most by a function of $\tau$, and we can write $J^\alpha_1 = J^\alpha_2 + \epsilon^{\alpha\beta}\partial_\beta c(\tau)$.
The difference $\epsilon^{\alpha\beta}\partial_\beta c(\tau)$ is a topological term that can always be added since it does not spoil the conservation equation.
At this point, redefining one of the currents to include this term, we see that it is always possible to arrive at the equality $J^\alpha_1=J^\alpha_2$.
To summarise, after proving that one component of two conserved currents is the same, we can simply take the full currents to agree.

For completeness, let us be more explicit in the example we are considering.
The relation $x'{}^-=-p_\mu x'{}^\mu$ implies that
\begin{equation}
x^-(\tau, \sigma)=\tilde c(\tau)-\int_0^{\sigma}d\xi\ p_\mu(\tau, \xi) x'{}^\mu(\tau, \xi) .
\end{equation}
Now, using that
\begin{equation}
\dot x^\mu = \frac{\partial \mathcal H}{\partial p_\mu},\qquad\qquad
\dot p_\mu =- \frac{\partial \mathcal H}{\partial x^\mu}+\partial_\sigma \frac{\partial \mathcal H}{\partial x'{}^\mu},
\end{equation}
we have
\begin{equation}
\begin{aligned}
\dot x^- &= \dot {\tilde c} -\int_0^{\sigma}d\xi\ (\dot p_\mu x'{}^\mu+p_\mu \dot x'{}^\mu)\\
& = \dot {\tilde c} -\int_0^{\sigma}d\xi\ \left[- \frac{\partial \mathcal H}{\partial x^\mu}x'{}^\mu+x'{}^\mu\partial_\xi \frac{\partial \mathcal H}{\partial x'{}^\mu}+p_\mu \partial_\xi \frac{\partial \mathcal H}{\partial p_\mu}\right]\\
& = \dot {\tilde c} -\int_0^{\sigma}d\xi\ \left[- \frac{\partial \mathcal H}{\partial x^\mu}x'{}^\mu- \frac{\partial \mathcal H}{\partial x'{}^\mu}x''{}^\mu- \frac{\partial \mathcal H}{\partial p_\mu}p'_\mu+\partial_\xi\left(\frac{\partial \mathcal H}{\partial x'{}^\mu}x'{}^\mu+p_\mu \frac{\partial \mathcal H}{\partial p_\mu}\right)\right]\\
& = \dot {\tilde c} -\int_0^{\sigma}d\xi\ \partial_\xi\left(- \mathcal H+\frac{\partial \mathcal H}{\partial x'{}^\mu}x'{}^\mu+p_\mu \frac{\partial \mathcal H}{\partial p_\mu}\right)\\
&=\dot c +\mathcal H -\frac{\partial \mathcal H}{\partial x'{}^\mu}x'{}^\mu-\frac{\partial \mathcal H}{\partial p_\mu}p_\mu\\
&=\dot c+T^\sigma{}_\sigma,
\end{aligned}
\end{equation}
where the boundary term evaluated at $\xi = 0$ is a function of $\tau$ only, whose sum with $\tilde c$ we denote $c$.
In agreement with the discussion above, we find that the two expressions match up to an unconstrained function of $\tau$.
By redefining, for example, the topological current as $\tilde J^\alpha_{(-)}=\epsilon^{\alpha\beta}\partial_\beta (x^--c(\tau))$, we find the expected agreement.

Let us now turn to the Noether currents.
First, we have
\begin{equation}
J^\tau_{(-)} =-\frac{\partial \mathscr L}{\partial(\dot x^-)}=-p_- \ \gf \ -1.
\end{equation}
In general, after gauge-fixing, the $\sigma$ component of this current will be constant, and (by adjusting the topological term as above) we can fix it to be zero, $J^\sigma_{(-)} \ \gf \ 0$.
Finally, we have
\begin{equation}
J^\tau_{(+)} =-\frac{\partial \mathscr L}{\partial(\dot x^+)}=-p_+ \ \gf \ \mathcal H=T^\tau{}_\tau.
\end{equation}
Having identified the time component, we conclude that the spatial component can be fixed to be $J^\sigma_{(+)} \ \gf \ T^\sigma{}_\tau$.

\subsubsection{Alternative gauges from current-current deformations}\label{s:a-lcgf}

We will organise the presentation of possible alternative gauges in terms of the Noether symmetry that is used to construct the transformation.
We first analyse alternative gauges identified by the Noether symmetries shifting the longitudinal fields $x^\pm$, before analysing those identified by the Noether symmetries acting on the transverse fields only.
The symmetry transformations shifting $x^\pm$ have the advantage of being linear in the deformation parameter at finite order.
The same can be achieved for the Noether symmetries acting only on transverse fields if we adapt our parametrisation of the fields to the symmetry transformation (e.g.~using polar coordinates to study a rotation symmetry, so that the transformation is implemented as a shift of an angular coordinate).
In general, one may also have Noether symmetries that act non-trivially on both the longitudinal and transverse fields, but we will not discuss these explicitly here.

After choosing the Noether symmetry, we will also need to specify the topological current that appears in the infinitesimal deformation.
In particular, we will complete the analysis by looking at the three independent cases in which $c$ is a function of transverse fields only, or of $x^+$, or of $x^-$.

Some of the transformations that we present here have appeared in other papers studying the construction of current-current deformations, see for example~\cite{Baggio:2018gct,Frolov:2019nrr,Frolov:2019xzi}.
The first example was in fact the $T\bar T$ deformation, which, as we will repeat below, can be understood as arising from a light-cone gauge-fixing where the longitudinal fields are subject to a $\gamma$-dependent redefinition.
Moreover, thanks to the reasoning explained at the beginning of~\secref{sec:gauges-curr-curr}, it will be straightforward for us to identify the infinitesimal variation of the Hamiltonian density $\mathcal H$, which sometimes is referred to as the ``flow equation.'' For an alternative derivation of the flow equation, see, for example,~\cite{Frolov:2019xzi}.

Let us also stress that we interpret the deformations as generating \textit{gauge transformations} of the reduced model.
That means that in general the deformation of the Hamiltonian will be accompanied by the deformation of other gauge-dependent quantities, such as the length of the string, in such a way that the spectrum is gauge-independent.
We will demonstrate this explicitly in~\secref{sec:spectrum}.
At the same time, one may reinterpret the deformations listed here as genuine deformations by allowing only the Hamiltonian and not the length of the string to be deformed, in the spirit of~\cite{Baggio:2018gct,Frolov:2019nrr,Frolov:2019xzi}.

\paragraph{Light-cone symmetries and $c(\tilde{\vec x})$.}
\begin{enumerate}
\item
Let us start by considering the Noether symmetry shifting $x^-$, with $c$ a function of transverse fields only.
We then write
\begin{equation}
\begin{alignedat}{3}
&x^+=\tilde x^+,\qquad
&& x^-=\tilde x^-+\gamma c(\tilde{\vec x}),\qquad
&& x^\mu=\tilde x^\mu,\\
& \tilde p_+= p_+,\qquad
&& \tilde p_-= p_-,\qquad
&& \tilde p_\mu= p_\mu+\gamma p_-\partial_\mu c.
\end{alignedat}
\end{equation}
Given the invariance of the fields $x^+$ and $p_-$, we expect it to lead to an equivalent gauge.
In fact, the relation $\tilde p_+= p_+$ implies that the two Hamiltonian densities are the same if we relate the transverse fields as
\begin{equation}
x^\mu=\tilde x^\mu,\qquad
\tilde p_\mu= p_\mu+\gamma \partial_\mu c,
\end{equation}
where the gauge condition $p_-=1$ was used.
Because the momenta are shifted by the \textit{derivative} of a function $c(\vec x)$, it is easy to check that this is indeed a canonical transformation.
We can also confirm this using the interpretation as a current-current deformation.
Using the results from \secref{ss:lcc}, evaluating eq.~\eqref{eq:deltaH-currcurr} gives
\begin{equation}
\delta \mathcal H = - \gamma\epsilon_{\alpha\beta}\left(\tilde J_{(c)}^\alpha\right)_{g.f.}\left( J^\beta_{(-)}\right)_{g.f.}= \gamma \partial_\tau c .
\end{equation}
This is indeed a change by a total derivative.
To conclude, in this case we generate an equivalent gauge.

As a brief comment, let us mention that taking $c$ to be linear in the transverse fields is enough to shift the momenta $p_\mu$ by generic constants.
This freedom is the reason why in \appref{sec:appendix} we could set the classical value of the transverse momenta $\bar p_\mu$ to zero.
\item
Let us now consider a similar transformation, but for the Noether symmetry that shifts $x^+$.
We write
\begin{equation}
\begin{aligned}
&x^+=\tilde x^++\gamma c(\tilde{\vec x}),\qquad
&& x^-=\tilde x^-,\qquad
&&& x^\mu=\tilde x^\mu,\\
& \tilde p_+= p_+,\qquad
&& \tilde p_-= p_-,\qquad
&&& \tilde p_\mu= p_\mu+\gamma p_+\partial_\mu c.
\end{aligned}
\end{equation}
Because $x^+\neq \tilde x^+$, we now expect this to lead to an inequivalent gauge.
In fact, the two gauge conditions $x^+=\tau$ and $\tilde{x}^+=\tau$ are not compatible, since demanding that the fluctuations of both $x^+$ and $\tilde x^+$ are set to zero is possible only if the fluctuations of $c(\tilde{\vec x})$ are also set to zero.
This is clearly impossible for a generic function $c$ as transverse fields do fluctuate.

Despite the relation $\tilde p_+= p_+$, it is not correct to conclude that the two Hamiltonians are the same, because we have
\begin{equation}\label{eq:tr-red-Jtilde-Ttau}
x^\mu=\tilde x^\mu,\qquad \tilde p_\mu= p_\mu-\gamma \mathcal H\partial_\mu c,
\end{equation}
which is not a canonical transformation.

To identify the explicit finite form of the deformed Hamiltonian density $\tilde{\mathcal H}$, we solve for the Virasoro constraint $C_2=0$ as reviewed in \appref{sec:appendix}.
This gives $\tilde{\mathcal H}$ as an explicit deformation of the Hamiltonian density $\mathcal H$.
We briefly explain the reasoning for this example.
While we will not repeat this for other gauge transformations, the reasoning is analogous in each case.
First, note that the expression for $C_2$ is invariant under diffeomorphisms in $D$ dimensions, so that we can trivially write
\begin{equation}
\begin{aligned}
C_2&=G^{MN}p_Mp_N+T^2 G_{MN}x'{}^Mx'{}^N-2T p_MG^{MN}B_{NQ}x'{}^Q+T^2G^{MN}B_{MP}B_{NQ}x'{}^Px'{}^Q,\\
&=\tilde G^{MN}\tilde p_M\tilde p_N+T^2 \tilde G_{MN}\tilde x'{}^M\tilde x'{}^N-2T \tilde p_M\tilde G^{MN}\tilde B_{NQ}\tilde x'{}^Q+T^2\tilde G^{MN}\tilde B_{MP}\tilde B_{NQ}\tilde x'{}^P\tilde x'{}^Q.
\end{aligned}
\end{equation}
The new Hamiltonian is therefore
\begin{equation}
\tilde{\mathcal H}=\frac{\tilde B-\sqrt{\tilde B^2-4\tilde A\tilde C}}{2\tilde A},
\end{equation}
where
\begin{equation}
\begin{aligned}
\tilde A&=\tilde G^{++},\\
\tilde B&= 2\tilde G^{+m}\tilde p_m-2T \tilde G^{+M}\tilde B_{Mn}\tilde x'{}^n,\\
\tilde C&= \tilde G^{mn}\tilde p_m\tilde p_n+T^2 \tilde G_{mn}\tilde x'{}^m\tilde x'{}^n-2T \tilde p_m\tilde G^{mN}\tilde B_{Nq}\tilde x'{}^q+T^2\tilde G^{MN}\tilde B_{Mp}\tilde B_{Nq}\tilde x'{}^p\tilde x'{}^q.
\end{aligned}
\end{equation}
Thus far, these are just the formulae of \appref{sec:appendix} with tildes.
At this point, to see the explicit $\gamma$-dependence, we can use the fact that all these objects transform as covariant tensors, so that
\begin{equation}
\begin{aligned}
\tilde G^{++}&= G^{++}+2\gamma\partial_\mu c\ G^{\mu +}+\gamma^2\partial_\mu c\ \partial_\nu c \ G^{\mu\nu},\\
\tilde G^{+m}&=G^{+m}+\gamma\partial_\mu c\ G^{\mu m},\\
\tilde G^{mn}&=G^{mn},
\end{aligned}
\end{equation}
and similar formulae for the $B$-field.
Having gauge-fixed the fields with tildes, we work with the transverse fields $\tilde x^\mu$, $\tilde p_\mu$.
For ease of notation, and to interpret the Hamiltonian $\tilde{\mathcal H}$ as a \textit{deformation} of $\mathcal H$ where the fields do not change, we will drop the tildes.
In other words we implement the substitution $\tilde x^\mu\to x^\mu,\,\tilde p_\mu\to p_\mu$.
Finally, we arrive at
\begin{equation}
\begin{aligned}
\tilde A&=G^{++}+2\gamma\partial_\mu c\ G^{\mu +}+\gamma^2\partial_\mu c\ \partial_\nu c\ G^{\mu\nu},\\
\tilde B&= 2(G^{+m}+\gamma\partial_\mu c\ G^{\mu m}) p_m-2T ( G^{+M}+\gamma\partial_\mu c\ G^{\mu M}) B_{Mn} x'{}^n,\\
\tilde C&= G^{mn} p_m p_n+T^2 G_{mn} x'{}^m x'{}^n-2T p_m G^{mN} B_{Nq} x'{}^q+T^2 G^{MN} B_{Mp} B_{Nq} x'{}^p x'{}^q,
\end{aligned}
\end{equation}
where we explicitly see the complicated $\gamma$-dependence of the Hamiltonian $\tilde{\mathcal H}$ through $\tilde A$ and $\tilde B$.

To conclude, let us note that according to the reinterpretation as a current-current deformation we find that the variation of the Hamiltonian corresponds to
\begin{equation} \label{eq:dH-Jt-0}
\delta \mathcal H = - \gamma\epsilon_{\alpha\beta}\left(\tilde J_{(c)}^\alpha \right)_{g.f.}\left(J^\beta_{(+)}\right)_{g.f.}=
- \gamma\epsilon_{\alpha\beta}\tilde J_{(c)}^\alpha T^\alpha{}_\tau
= \gamma \partial_\alpha c \ T^\alpha{}_\tau.
\end{equation}
In~\cite{Frolov:2019xzi} this deformation was called a $\tilde JT_0$ deformation; here we will call it $\tilde JT_\tau$.
While, according to our definition, it leads to an inequivalent gauge transformation, we will show later that it has no effect on the S-matrix.
Indeed, notice that on-shell (in particular when using the conservation of the stress-energy tensor) the above infinitesimal transformation is just a total derivative.
\end{enumerate}

\paragraph{Light-cone symmetries and $c(\tilde{x}^\pm)$.}
Taking into account that the Noether symmetry may shift either $x^+$ or $x^-$, and that we may choose the function $c$ to be linear in either $x^+$ or $x^-$, there are a total of four cases to consider.
\begin{enumerate}
\item Let us start with the symmetry shifting $x^+$ and take $c=\gamma\tilde x^+$.
Then
\begin{equation}
\begin{alignedat}{3}
&x^+=(1+\gamma)\tilde x^+,\qquad
&& x^-=\tilde x^-,\qquad
&& x^\mu=\tilde x^\mu,\\
& \tilde p_+=(1+\gamma) p_+,\qquad
&& \tilde p_-= p_-,\qquad
&& \tilde p_\mu= p_\mu.
\end{alignedat}
\end{equation}
Strictly speaking this yields an inequivalent gauge, but it is clear from the above formulae that it corresponds simply to rescaling $\tau$, and consequently the overall Hamiltonian.
Therefore, we may say that this gauge is ``almost equivalent''.
According to the reinterpretation as a current-current deformation, we have
\begin{equation}
\delta \mathcal H = - \gamma\epsilon_{\alpha\beta}\left(\tilde J_{(+)}^\alpha \right)_{g.f.}\left(J^\beta_{(+)}\right)_{g.f.}= \gamma\mathcal H.
\end{equation}

\item Consider now the symmetry shifting $x^-$ and take $c=\gamma\tilde x^+$.
Then
\begin{equation}
\begin{alignedat}{3}
&x^+=\tilde x^+,\qquad
&& x^-=\tilde x^-+\gamma\tilde x^+,\qquad
&& x^\mu=\tilde x^\mu,\\
& \tilde p_+=p_++\gamma p_-,\qquad
&& \tilde p_-= p_-,\qquad
&& \tilde p_\mu= p_\mu.
\end{alignedat}
\end{equation}
This leads to an equivalent gauge since $x^+$ and $p_-$ do not transform.
In fact, it corresponds simply to a shift of the Hamiltonian by a constant $\tilde{\mathcal H}=\mathcal H-\gamma$.
According to the interpretation as a current-current deformation, we indeed have
\begin{equation}
\delta \mathcal H = - \gamma\epsilon_{\alpha\beta}\left(\tilde J_{(+)}^\alpha \right)_{g.f.}\left(J^\beta_{(-)}\right)_{g.f.}= -\gamma.
\end{equation}

\item Consider the symmetry shifting $x^-$ and take $c=\gamma\tilde x^-$.
Then
\begin{equation}
\begin{alignedat}{3}
&x^+=\tilde x^+,\qquad
&& x^-=(1+\gamma)\tilde x^-,\qquad
&& x^\mu=\tilde x^\mu,\\
& \tilde p_+= p_+,\qquad
&& \tilde p_-=(1+\gamma) p_-,\qquad
&& \tilde p_\mu= p_\mu.
\end{alignedat}
\end{equation}
Strictly speaking this again yields an inequivalent gauge, but it corresponds to just rescaling $p_-$ and $x^-$.
In the reduced model, this can be compensated by rescaling $\sigma$ and the tension $T$.
Also in this case, we may say that this is an ``almost equivalent'' gauge.
According to the interpretation as current-current deformations, we have
\begin{equation}
\delta \mathcal H = - \gamma\epsilon_{\alpha\beta}\left(\tilde J_{(-)}^\alpha\right)_{g.f.} \left(J^\beta_{(-)}\right)_{g.f.}= -\gamma T^\sigma{}_\sigma.
\end{equation}
This is indeed the variation of the Hamiltonian when rescaling the worldsheet coordinate $\sigma$.
In fact, taking $\delta\sigma=\gamma\sigma$ and formally calculating the infinitesimal variation of the Lagrangian, one finds
\begin{equation}
\delta\mathcal L =\partial_\alpha(T^\alpha{}_\sigma\ \delta\sigma)=T^\alpha{}_\sigma\partial_\alpha\delta\sigma=\gamma T^\sigma{}_\sigma.
\end{equation}

\item Finally, consider the symmetry shifting $x^+$ and take $c=\gamma\tilde x^-$.
Then
\begin{equation} \label{eq:prom-for-TT}
\begin{alignedat}{3}
&x^+=\tilde x^++\gamma\tilde x^-,\qquad
&& x^-=\tilde x^-,\qquad
&& x^\mu=\tilde x^\mu,\\
& \tilde p_+= p_+,\qquad
&& \tilde p_-= p_-+\gamma p_+,\qquad
&& \tilde p_\mu= p_\mu.
\end{alignedat}
\end{equation}
Both $x^+$ and $p_-$ transform non-trivially, and this leads to an inequivalent gauge.
Recalling how we fix $x^\pm$ in terms of $t$ and $\varphi$ in \appref{sec:appendix}, this corresponds to the so-called $a$-gauge of~\cite{Arutyunov:2006gs,Arutyunov:2009ga}
\begin{equation}
\tilde x^+=(1-a)t+a\varphi,\qquad
\tilde x^-=\varphi- t,
\end{equation}
if we identify $a=1/2-\gamma$.
As a current-current deformation, we have
\begin{equation}
\delta \mathcal H = - \gamma\epsilon_{\alpha\beta}\left(\tilde J_{(-)}^\alpha\right)_{g.f.}\left( J^\beta_{(+)}\right)_{g.f.}= -\gamma\epsilon_{\alpha\beta}T^\alpha{}_\sigma T^\beta{}_\tau,
\end{equation}
which corresponds to the well-known interpretation as a $T\bar T$ deformation that was given in~\cite{Baggio:2018gct,Frolov:2019nrr}.
We will not write the explicit finite form of the deformed Hamiltonian density $\tilde{\mathcal H}$, which may be found for example in~\cite{Frolov:2019nrr}.

\end{enumerate}

\paragraph{Transverse symmetries and $c(\tilde x^\pm)$.}
Let us now consider the case of a symmetry transformation that acts non-trivially only on transverse fields.
If the function $c$ entering the definition of the topological current $\tilde J^\alpha_{(c)}$ depends on transverse fields only, then we would end up with a ``point-canonical'' transformation as in the discussion at the beginning of~\secref{sec:ineq-lcgf}.
Hence, the only way to generate inequivalent gauges is to take $c$ either linear in $\tilde x^+$ or in $\tilde x^-$.

\begin{enumerate}
\item We first consider the case $c=\gamma \tilde x^-$, so that
\begin{equation} \label{eq:CT-JT-1}
\begin{alignedat}{3}
&x^+=\tilde x^+,\qquad
&& x^-=\tilde x^-,\qquad
&& x^\mu=F^\mu(\tilde{\vec{ x}},\lambda(\tilde x^-)) ,\\
&\tilde p_+= p_+ ,\qquad
&& \tilde p_-= p_-+\frac{d\lambda}{d\tilde x^-}\frac{\partial F^\mu}{\partial\lambda}p_\mu,\qquad
&& \tilde p_\mu= \frac{\partial F^\nu}{\partial\tilde x^\mu}p_\nu,
\end{alignedat}
\end{equation}
where
\begin{equation}\label{eq:prom-xm-jt1}
\lambda(\tilde x^-)=c(\tilde x^-)+\ldots=\gamma\tilde x^-+\ldots\,.
\end{equation}
That is we identify the leading order of $\lambda$ with the function $c$, as in~\secref{sec:gauges-curr-curr}.
If the symmetry transformation is non-linear, the parameter $\lambda$ of the finite transformation may also depend on higher-order terms in $x^-$.
These terms are identified by demanding that shifts of $\tilde x^-$ remain symmetries.
We have not needed to consider this subtlety up to now since shifts of $x^\pm$ are linear transformations, hence the infinitesimal and the finite transformations coincide.

If we also assume (as done in~\cite{Frolov:2019xzi}) that we work in adapted target-space coordinates, so that the symmetry transformation simply acts as the shift of a transverse field that we call $\theta$,
\unskip\footnote{If $\theta$ is compact it has the interpretation of an angle, but it may also be non-compact.}
we can then write the finite transformation as
\begin{equation} \label{eq:CT-JT-2}
\begin{alignedat}{4}
&x^+=\tilde x^+,\qquad
&& x^-=\tilde x^-,\qquad
&& x^\mu=\tilde x^\mu,\qquad
&& \theta = \tilde\theta+\gamma \tilde x^-,\\
&\tilde p_+= p_+ ,\qquad
&& \tilde p_-= p_-+\gamma p_\theta,\qquad
&& \tilde p_\mu= p_\mu,\qquad
&& \tilde p_\theta = p_\theta.
\end{alignedat}
\end{equation}
Even without this assumption, it is obvious that the two conditions $p_-=1$ and $\tilde p_-=1$ are not compatible, so we expect an inequivalent gauge.
In fact, according to the interpretation as a current-current deformation, we have
\begin{equation}
\delta \mathcal H = - \gamma\epsilon_{\alpha\beta}\left(\tilde J_{(-)}^\alpha \right)_{g.f.}\left(J^\beta\right)_{g.f.}= -\gamma \epsilon_{\alpha\beta} T^\alpha{}_\sigma J^\beta .
\end{equation}
Here $J^\alpha$ is the Noether current of the transverse symmetry that we are using to generate the transformation.
In~\cite{Frolov:2019xzi} this deformation was called a $JT_1$ deformation; we will call it $JT_\sigma$.

\item
Consider now the choice $c=\gamma\tilde x^+$, so that
\begin{equation}
\begin{alignedat}{3}
&x^+=\tilde x^+,\qquad
&& x^-=\tilde x^-,\qquad
&& x^\mu=F^\mu(\tilde{\vec{ x}},\lambda(\tilde x^+)) ,\\
& \tilde p_+= p_++\frac{d\lambda}{d\tilde x^+}\frac{\partial F^\mu}{\partial\lambda}p_\mu,\qquad
&& \tilde p_-= p_-,\qquad
&& \tilde p_\mu= \frac{\partial F^\nu}{\partial\tilde x^\mu}p_\nu,
\end{alignedat}
\end{equation}
where, as in the previous discussion, we identify
\begin{equation} \label{eq:prom-xp-j}
\lambda(\tilde x^+)=c(\tilde x^+)+\ldots=\gamma\tilde x^++\ldots\,.
\end{equation}
In this case the gauge-fixing conditions in the two coordinate systems are compatible, because neither $x^+$ nor $p_-$ transform.
Although the Hamiltonian densities $\tilde{\mathcal H}$ and $\mathcal H$ are related by a canonical transformation, this is \textit{time-dependent} since $F^\mu$ depends on $x^+=\tilde x^+=\tau$.
Therefore, $\tilde{\mathcal H}$ is related to $\mathcal H$ by an extra shift as indicated in the relation between $\tilde p_+$ and $p_+$.
The fact that a time-dependent canonical transformation generates an extra shift of the Hamiltonian density also follows from the definition $\mathcal H=p_\mu\dot x^\mu-\mathcal L$, where the shift comes from the explicit time derivative of $x^\mu$.
According to the interpretation as a current-current deformation, we have
\begin{equation}
\delta \mathcal H = - \gamma\epsilon_{\alpha\beta}\left(\tilde J_{(+)}^\alpha \right)_{g.f.}\left(J^\beta\right)_{g.f.}= \gamma J^\tau ,
\end{equation}
where $J^\alpha$ is the Noether current of the transverse symmetry.
We will call this transformation a $J^\tau$ deformation.

Choosing adapted coordinates in target space so that the symmetry acts simply as the shift of a coordinate $\theta$, we have
\begin{equation}
\begin{alignedat}{4}
&x^+=\tilde x^+,\qquad
&& x^-=\tilde x^-,\qquad
&& x^\mu=\tilde x^\mu,\qquad
&& \theta = \tilde\theta+\gamma \tilde x^+,\\
&\tilde p_+= p_+ +\gamma p_\theta,\qquad
&& \tilde p_-= p_-,\qquad
&& \tilde p_\mu= p_\mu,\qquad
&& \tilde p_\theta = p_\theta.
\end{alignedat}
\end{equation}
Therefore, the finite deformation of the Hamiltonian density is
\begin{equation}
\tilde {\mathcal H} =\mathcal H+\gamma J^\tau,
\end{equation}
where we use that the time component of the Noether current and the momentum conjugate to $\theta$ are related as $J^\tau=-p_\theta$.
Note that the deformed Hamiltonian (defined as the spatial integration of the Hamiltonian density) is given by
\begin{equation}
\tilde H=H+\gamma Q,
\end{equation}
where $Q=\int d\sigma\, J^\tau$ is the Noether charge.

A similar discussion holds if we instead assume that the symmetry transformation is an $SO(2)$ rotation of coordinates $x^2,x^3$
\begin{equation}
x^2=\cos\lambda\, \tilde x^2+\sin\lambda\, \tilde x^3,\qquad
x^3=\cos\lambda\, \tilde x^3-\sin\lambda\, \tilde x^2.
\end{equation}
Introducing the vector $x=(x^2,x^3)$ we can write
\begin{equation}
x=R(\lambda)\tilde x,\qquad
p=R(\lambda)\tilde p,\qquad
R(\lambda)=\left(\begin{array}{cc}
\cos\lambda& \sin\lambda\\
-\sin\lambda& \cos \lambda
\end{array}\right).
\end{equation}
We also have
\begin{equation}
\frac{dR(\lambda)}{d\lambda}=\left(\begin{array}{cc}
-\sin\lambda& \cos \lambda\\
-\cos\lambda& -\sin\lambda
\end{array}\right)
=R(\lambda)
\left(\begin{array}{cc}
0& 1\\
-1&0
\end{array}\right).
\end{equation}
Now taking $x^M=\tilde x^M$ for $M\neq 2,3$, we promote the above redefinition to $\lambda=\gamma\tilde x^+$, which implies
\begin{equation}
\begin{aligned}
\tilde p_+&=p_++\frac{\partial x^i}{\partial\tilde x^+}p_i = p_++\frac{d \lambda}{d\tilde x^+}\frac{\partial x^i}{\partial\lambda}p_i
=p_++\gamma \tilde x^T\left(\frac{dR(\lambda)}{d\lambda}\right)^Tp \\
&=p_++\gamma(\tilde x^3\tilde p_2-\tilde x^2\tilde p_3).
\end{aligned}
\end{equation}
The deformed Hamiltonian density is then
\begin{equation}
\tilde{\mathcal H} = \mathcal H +\gamma(\tilde x^2\tilde p_3-\tilde x^3\tilde p_2),
\end{equation}
which is again of the form $\tilde{ \mathcal H}=\mathcal H+\gamma J^\tau$.
\end{enumerate}
Having analysed all the relevant coordinate transformations outlined at the beginning of this section, this concludes our discussion of inequivalent gauges.

\subsection{Recap of inequivalent gauges} \label{s:recap-ilcgf}

For the reader's convenience, let us recap the inequivalent gauges that we have identified:
\begin{enumerate}
\item
The $\tilde JT_\tau$ deformation obtained by the shift $x^+=\tilde x^++\gamma c(\tilde{\vec x})$.
\item
The $T\bar T$ deformation obtained by the shift $x^+=\tilde x^++\gamma \tilde x^-$.
\item
The $JT_\sigma$ deformation obtained by promoting the parameter of a transverse symmetry to a function of $\tilde x^-$ (for example, $\theta=\tilde \theta+\gamma\tilde x^-$).
\item
The $J^\tau$ deformation obtained by promoting the parameter of a transverse symmetry to a function of $\tilde x^+$ (for example, $\theta=\tilde \theta+\gamma\tilde x^+$).

\end{enumerate}

\section{Inequivalent light-cone gauges for symmetric spaces}\label{sec:in-sym-cosets}

Symmetric spaces, which include anti de Sitter space and the spheres, play an important role in integrable worldsheet theories of strings.
Therefore, we now investigate the moduli space of inequivalent light-cone gauges for the symmetric space \sm.

\medskip

Symmetric spaces $\Man = G/H$ are isomorphic to homogeneous spaces for which the Lie algebra $\mathfrak{g}$ of the Lie group $G$ admits a $\mathbb{Z}_2$ grading $\mathfrak{g} = \mathfrak{g}^{(0)}\oplus \mathfrak{g}^{(2)}$, i.e.~such that
\unskip\footnote{We employ a notation that is natural for semisymmetric spaces, which admit a $\mathbb Z_4$ decomposition.
We do so because of our motivation to eventually describe superstrings on spacetimes such as $AdS_5\times S^5$.}
\begin{equation}
[\mathfrak{g}^{(i)}, \mathfrak{g}^{(j)}] \subset \mathfrak{g}^{(i+j \ \text{mod}\ 4)} , \qquad i,j=0,2 ,
\end{equation}
where $\mathfrak{g}^{(0)} = \mathrm{Lie}(H)$.
Introducing the projectors $P^{(i)}$ on the subspaces $\mathfrak{g}^{(i)}$ and the notation $M^{(i)} \equiv P^{(i)} M$ for generic Lie algebra elements $M\in \mathfrak{g}$, the symmetric space \sm{} action can be written as
\begin{equation} \label{eq:action-symmcoset}
\Act = -\frac{T}{4} \int_\Sigma d \tau d \sigma\ \gamma^{\alpha\beta} \mathrm{STr} \big( A_\alpha P^{(2)} A_\beta \big) ,
\end{equation}
with $A_\alpha = g^{-1}\partial_\alpha g$, $g : \Sigma \rightarrow G/H$ a coset parametrisation, and $\mathrm{STr}$ an ad-invariant non-degenerate bilinear form on $\mathfrak{g}$.
Furthermore we have $\gamma^{\alpha\beta}=\sqrt{|h|}h^{\alpha\beta}$ where $h_{\alpha\beta}$ is the worldsheet metric.
The action is invariant under global left-acting transformations by $G$ and local right-acting transformations by $H$, whose combination we denote as $G_L \times H_R$.
The equations of motion are
\begin{equation} \label{eq:eom-symmcoset}
\partial_\alpha \big( \gamma^{\alpha\beta} A_\beta^{(2)} \big) + \gamma^{\alpha\beta} [A_\alpha^{(0)} ,A_\beta^{(2)} ] = 0 ,
\end{equation}
subject to the Virasoro constraints
\begin{equation}\label{eq:virasoro-symmcoset}
\mathcal T_{\alpha\beta} = \mathrm{STr} \big( A_\alpha^{(2)} A_\beta^{(2)} \big) -\frac{1}{2} \gamma_{\alpha\beta} \gamma^{\gamma\delta} \mathrm{STr} \big( A_\gamma^{(2)} A_\delta^{(2)} \big) = 0 .
\end{equation}

Our starting assumption in the light-cone gauge-fixing procedure relies on having a parametrisation of $G/H$ that realises at least two manifest abelian isometries corresponding to shifts of a timelike coordinate $t$ and a spacelike coordinate $\varphi$.
The most general coset parametrisation satisfying these criteria is
\begin{equation} \label{eq:g-par}
g= \exp (\Lambda_t t + \Lambda_\varphi \varphi ) g_X , \qquad
[\Lambda_t,\Lambda_\varphi] = 0 ,
\end{equation}
where the field $g_X$ is a generic parametrisation of the transverse fields $x^\mu$.
Recalling that $x^+ = (t+\varphi)/2$ and $x^- = \varphi - t$, this parametrisation can be equivalently written as
\begin{equation}\label{eq:Lpm}
g=\exp(\Lambda_+ x^+ + \Lambda_- x^-) g_X , \qquad \Lambda_+ = \Lambda_t + \Lambda_\varphi ~, \quad \Lambda_- = \frac12(\Lambda_\varphi - \Lambda_t) ~, \qquad [\Lambda_+,\Lambda_-] = 0 ~.
\end{equation}
Shifts in the longitudinal coordinates $t$, $\varphi$, or $x^\pm$, are realised by left-acting transformations generated by $\Lambda_{t}$, $\Lambda_{\varphi}$, or $\Lambda_\pm$, respectively.

In this section we make the assumption that the background is a Cartesian product of a Lorentzian (non-compact) symmetric space $\Man_\mathfrak{a} = G_{\mathfrak{a}}/H_{\mathfrak{a}}$ and a (compact) Euclidean symmetric space $\Man_\mathfrak{s} = G_{\mathfrak{s}}/H_{\mathfrak{s}}$, such as $AdS_n \times S^n$.
We define the projectors $P_{\alg{a}}$ and $P_{\alg{s}}$ onto the Lie algebras $\alg{g}_{\alg{a}}$ and $\alg{g}_{\alg{s}}$, which, due to the Cartesian product structure, commute with $P^{(0)}$ and $P^{(2)}$.
We also make the assumption that the symmetric spaces are of rank-1, i.e. the maximal abelian subalgebra of $\alg{g}_{\alg{a}}^{(2)}$ and $\alg{g}_{\alg{s}}^{(2)}$ is 1-dimensional.
We will not assume that $t\in \mathcal M_{\mathfrak{a}}$ and $\varphi\in \mathcal M_{\mathfrak{s}}$, i.e.~$t$ and $\varphi$ may mix coordinates of $\mathcal{M}_{\mathfrak{a}}$ and $\mathcal{M}_{\mathfrak{s}}$.

The classical point-like string that we use for light-cone gauge-fixing takes the form
\begin{equation}\label{eq:sol}
t = \varphi = \tau , \qquad g_X = 1 , \qquad \gamma^{\alpha\beta} = T^{-1}\eta^{\alpha\beta} .
\end{equation}
In general, we may consider arbitrary constant $g_X = g_0$, however we can always use the global $G_L$ symmetry to choose $g_X = 1$ at the expense of a compensating rotation of $\Lambda_+$, i.e.~$\Lambda_+\to g_0\Lambda_+g_0^{-1}$.
Since we have not specified $\Lambda_+$, other than that it commutes with $\Lambda_-$, which is also unspecified, we take $g_X = 1$ on the classical solution without loss of generality.

Defining
\unskip\footnote{Note that only if we take $t\in \mathcal M_{\mathfrak{a}}$ and $\varphi\in \mathcal M_{\mathfrak{s}}$ we have $\Lambda_{\alg{a}}=\Lambda_t$ and $\Lambda_{\alg{s}}=\Lambda_\varphi$.}
\begin{equation}
\Lambda_{\alg{a},\alg{s}} = P_{\alg{a},\alg{s}} \Lambda_+ ,
\end{equation}
and substituting~\eqref{eq:sol} into the equations of motion~\eqref{eq:eom-symmcoset} and Virasoro constraints~\eqref{eq:virasoro-symmcoset}, we find the conditions
\begin{equation}\label{eq:eom}
[\Lambda_{\mathfrak{a}}^{(0)},\Lambda_{\mathfrak{a}}^{(2)}] = 0 , \qquad
[\Lambda_{\mathfrak{s}}^{(0)},\Lambda_{\mathfrak{s}}^{(2)}] = 0 , \qquad
\STr\big(\Lambda_{\mathfrak{a}}^{(2)}\Lambda_{\mathfrak{a}}^{(2)}\big) +
\STr\big(\Lambda_{\mathfrak{s}}^{(2)}\Lambda_{\mathfrak{s}}^{(2)}\big) = 0 .
\end{equation}
Therefore, $\Lambda_{\mathfrak{a}}^{(0)}$ and $\Lambda_{\mathfrak{s}}^{(0)}$ are valued in the centralisers of $\Lambda_{\mathfrak{a}}^{(2)}$ and $\Lambda_{\mathfrak{s}}^{(2)}$ respectively.
Since we assume $\alg{g}_{\alg{s}}$ is compact, it follows that $\STr\big(\Lambda_{\mathfrak{s}}^{(2)}\Lambda_{\mathfrak{s}}^{(2)}\big) \geq 0$, hence we must have $\STr\big(\Lambda_{\mathfrak{a}}^{(2)}\Lambda_{\mathfrak{a}}^{(2)}\big) \leq 0$ for the final equation in eq.~\eqref{eq:eom} to admit a solution.
For simplicity we assume that these quantities are non-vanishing,
\unskip\footnote{Note that this restriction excludes the $AdS$ light-cone gauge~\cite{Giombi:2009gd} for which $\STr(\Lambda_{\mathfrak{a}}^{(2)}\Lambda_{\mathfrak{a}}^{(2)}) = 0$.}
hence by rescaling $x^+$ and $\tau$ we are free to fix the normalisation of $\Lambda_{\mathfrak{a}}^{(2)}$ and $\Lambda_{\mathfrak{s}}^{(2)}$.
In the following we will take
\begin{equation}
\STr\big(\Lambda_{\mathfrak{a}}^{(2)}\Lambda_{\mathfrak{a}}^{(2)}\big) = -2 ,
\qquad
\STr\big(\Lambda_{\mathfrak{s}}^{(2)}\Lambda_{\mathfrak{s}}^{(2)}\big) = 2 .
\end{equation}

At this point we note that we could use the local $H_R$ symmetry to remove the $\Lambda_\alg{a}^{(0)}$ and $\Lambda_\alg{s}^{(0)}$ dependence of the classical point-like string solution
$g= \exp (\Lambda_+\tau) $.
We can further conjugate $\Lambda_\alg{a}^{(2)}$ and $\Lambda_\alg{s}^{(2)}$ by a constant element of $H$ to specified elements of $\alg{g}_{\alg{a}}^{(2)}$ and $\alg{g}_{\alg{s}}^{(2)}$ with the same norm.
This reflects the fact that there is a unique point-like string solution with non-vanishing momentum in both $\Man_\alg{a}$ and $\Man_\alg{s}$ up to global symmetry transformations.
However, the first of these transformations in particular does not preserve the parametrisation~\eqref{eq:g-par} with $g_X$ transverse only, therefore we instead take $\Lambda_\alg{a}$ and $\Lambda_\alg{s}$ to satisfy~\eqref{eq:eom}, but otherwise leave them unfixed.

Only a subset of the original $G_L \times H_R$ symmetry preserves our choice of parametrisation.
Included in the residual symmetries we have global $H$ transformations acting vectorially as
\begin{equation}\label{eq:resglob}
H_V : \quad \Lambda_{\mathfrak{a},\mathfrak{s}}\to h_0\Lambda_{\mathfrak{a},\mathfrak{s}}h_0^{-1} , \quad g_X \to h_0 g_X h_0^{-1} , \quad h_0 \in H ,
\end{equation}
and local right-acting transformations that only depend on the transverse fields and reduce to the identity on the classical solution.
We fix the latter symmetry by setting $g_X = \exp X$, with $X\equiv X^{(2)}\in \mathfrak{g}^{(2)}$.
We have now parametrised the group-valued field in terms of $\dim \alg{g}^{(2)} + 2$ scalar fields.
This is two more than if we had fully fixed the gauge symmetry, and indeed our parametrisation includes a redundancy $x^\pm \to x^\pm + c^\pm(X)$, $X \to X - c^\pm(X) \Lambda_\pm^{(2)} + \dots$, together with a compensating gauge transformation to restore the original form.
The two functions $c^\pm (X)$ can be used to fix the two components of $X$ in the $\Lambda_\pm^{(2)}$ directions, giving a minimal set of transverse fields that we denote by $x$:
\begin{equation}\label{transparam}
g_X = \exp(x + f^+(x) \Lambda_+^{(2)} + f^-(x) \Lambda_-^{(2)}).
\end{equation}
Since the functions $f^\pm$ originate from shifts of the longitudinal coordinates $x^\pm$ by functions of the transverse coordinates, they can lead to different gauge-fixings.
Therefore, for now we leave them unspecified.

In order to understand the freedom that remains in our choice of $\Lambda_{\mathfrak{a}}$ and $\Lambda_{\mathfrak{s}}$ after imposing~\eqref{eq:eom}, we observe that the $H_V$ symmetry~\eqref{eq:resglob} preserves our gauge choice $g_X = \exp X$.
As we have restricted to rank-1 cosets, this means that we can take $\Lambda_{\mathfrak{a}}$ and $\Lambda_{\mathfrak{s}}$ to lie in given Cartan subalgebras $\alg{t}_{\mathfrak{a}} \subset \alg{g}_{\mathfrak{a}}$ and $\alg{t}_{\mathfrak{s}} \subset \alg{g}_{\mathfrak{s}}$ with the properties
\unskip\footnote{We first use the conjugation to fix $\Lambda_{\mathfrak{a}}^{(2)}$ and $\Lambda_{\mathfrak{s}}^{(2)}$.
Since the cosets are rank-1, we can conjugate between any two elements of $\mathfrak{g}_{\mathfrak{a}}^{(2)}$ or $\mathfrak{g}_{\mathfrak{s}}^{(2)}$ that have the same norm.
The remaining freedom is then conjugation by elements of the centraliser group of $\Lambda_{\mathfrak{a}}^{(2)}$ and $\Lambda_{\mathfrak{s}}^{(2)}$, which we can use to rotate $\Lambda_{\mathfrak{a}}^{(0)}$ and $\Lambda_{\mathfrak{s}}^{(0)}$ to be valued in a Cartan subalgebra of the centraliser algebra.
If the centraliser is non-compact, there may be inequivalent choices for its Cartan subalgebra.
However, since this is not the case for $AdS_n \times S^n$, as we will discuss in~\secref{s:AdSn-Sn}, we will not address this potential subtlety here.}
\begin{equation}\begin{gathered}
\STr\big(\alg{t}_{\mathfrak{a}}^{(2)}\alg{t}_{\mathfrak{a}}^{(2)}\big) < 0 , \qquad
[\alg{t}_{\mathfrak{a}}^{(0)},\alg{t}_{\mathfrak{a}}^{(2)}] \subset \{0\} ,
\\
\STr\big(\alg{t}_{\mathfrak{s}}^{(2)}\alg{t}_{\mathfrak{s}}^{(2)}\big) > 0 , \qquad
[\alg{t}_{\mathfrak{s}}^{(0)},\alg{t}_{\mathfrak{s}}^{(2)}] \subset \{0\} .
\end{gathered}\end{equation}
Given that the normalisations of $\Lambda_{\mathfrak{a}}^{(2)}$ and $\Lambda_{\mathfrak{s}}^{(2)}$ are fixed, the remaining freedom is thus $\rk {\alg{g}_{\alg{a}}} -1$ parameters in $\Lambda_{\mathfrak{a}}^{(0)}$ and $\rk {\alg{g}_{\alg{s}}} -1$ parameters in $\Lambda_{\mathfrak{s}}^{(0)}$.
The origin of these parameters can be understood as the rotation $g_X \to \exp(\Lambda_+^{(0)}x^+) g_X \exp(-\Lambda_+^{(0)}x^+)$, hence by the summary in \secref{s:recap-ilcgf}, they are expected to correspond to $J^\tau$ deformations.

Finally, we would like to understand the freedom that we have in choosing $\Lambda_-$, which is thus far unspecified other than that it should commute with $\Lambda_+$ and is such that $t$ and $\varphi$ are timelike and spacelike respectively.
We will leave a full analysis of the possible choices of $\Lambda_-$, which depends on $\Lambda_+$ and any residual $H_V$ symmetry that preserves $\Lambda_+$ for the future.
Here we investigate one possible solution, which is to take $\Lambda_-$ to be valued in the same Cartan subalgebra as $\Lambda_+$.
This is the general solution when $\Lambda_+$ is a generic element of the Cartan subalgebra.
Then, of the $\rk \alg{g}$ parameters in $\Lambda_-$ one can be fixed by rescaling $x^-$, another one, the part proportional to $\Lambda_+$, can be understood as a shift of $x^+$ by $x^-$, hence corresponds to the $T\bar T$ deformation, and the remaining $\rk \alg{g} -2$ can be taken to parametrise $\alg{t}_{\mathfrak{a}}^{(0)}$ and $\alg{t}_{\mathfrak{s}}^{(0)}$.
\unskip\footnote{Note that there may be bounds on these parameters that depend on the form of $\Lambda_+$ to ensure that $t$ and $\varphi$ are timelike and spacelike respectively.}
Therefore, the origin of these parameters can be understood as the rotation $g_X \to \exp(\Lambda_-^{(0)}x^-) g_X \exp(-\Lambda_-^{(0)}x^-)$, and by the summary in \secref{s:recap-ilcgf}, they are expected to correspond to $J T_\sigma$ deformations.

In total, through this analysis, we have found five sets of freedom in our parametrisation, four leading to inequivalent gauge-fixings, mirroring the summary in \secref{s:recap-ilcgf}, and one to a total derivative.
In particular, the two functions $f^+(x)$ and $f^-(x)$ correspond to a $\tilde J T_\tau$ deformation and a total derivative respectively, while the $\rk \alg{g}-2$ parameters in each of $\Lambda_+^{(0)}$ and $\Lambda_-^{(0)}$ correspond to $J^\tau$ and $J T_\sigma$ deformations respectively.
Finally, the component of $\Lambda_-$ proportional to $\Lambda_+$ corresponds to the $T\bar T$ deformation.

\medskip

In the perturbative analyses in sections \ref{s:pp-wave}, \ref{s:AdSn-Sn} and \ref{s:treelevel-checks}, we will make the following simplifying assumption.
We assume that $t$ and $\varphi$ are coordinates on $\mathcal{M}_{\alg{a}}$ and $\mathcal{M}_{\alg{s}}$ respectively, i.e.~$\Lambda_t \in \alg{g}_{\alg{a}}$ and $\Lambda_\varphi \in \alg{g}_{\alg{s}}$.
In particular, this means that $P_{\alg{a}} \Lambda_+ = - 2 P_{\alg{a}} \Lambda_- = \Lambda_t$ and $P_{\alg{s}} \Lambda_+ = 2 P_{\alg{s}} \Lambda_- = \Lambda_\varphi$, hence the projections of $\Lambda_+$ and $\Lambda_-$ are not independent.
We also assume that the transverse coordinates are split into a set of coordinates on $\mathcal{M}_{\mathfrak{a}}$ and a set on $\mathcal{M}_{\mathfrak{s}}$.
As a result, an alternative gauge-fixing that leads to a $\tilde J T_\tau$ will always come with a total derivative: shifts of $x^\pm$ are now restricted such that $t$ and $\varphi$ remain in $\mathcal{M}_{\mathfrak{a}}$ and $\mathcal{M}_{\mathfrak{s}}$ respectively, and the transverse coordinates are still split.
Similarly, $J^\tau$ and $JT_\sigma$ deformations will be tied together, with a single parameter controlling both.
Strictly speaking, this would also remove the $T\bar T$ deformation, however, we can reintroduce this by hand and will do so when studying the tree-level S-matrix in \secref{s:treelevel-checks}.

We now carry out a more detailed analysis of the light-cone gauge moduli space for the simplified case of $\Real \times \Man_{\mathfrak{s}}$, with the generalisation to $\Man_{\mathfrak{a}} \times \Man_{\mathfrak{s}}$ straightforward up to the identification of $\Lambda_{\mathfrak{a}}$.
We will then discuss explicitly how to appropriately identify $\Lambda_{\mathfrak{a}}$ and $\Lambda_{\mathfrak{s}}$ for $AdS_n \times S^n$ and the residual symmetry algebras of different light-cone gauge-fixed theories in~\secref{s:AdSn-Sn}.

\subsection{Inequivalent light-cone gauges for strings on \texorpdfstring{$ \Real\times \Man_{\mathfrak{s}} $}{R × Ms}}\label{s:pp-wave}

In order to probe the moduli space of inequivalent light-cone gauge-fixings around the point-like string solution~\eqref{eq:sol}, it is useful to study the pp-wave limit.
For simplicity, we consider the space $\Real \times \Man_\mathfrak{s}$ such that the metric reads
\begin{equation}\label{eq:metms}
ds^2 = -dt^2 + \frac{1}{2} \STr\big((g_X^{-1} d\varphi \Lambda_{\mathfrak{s}} g_X + g_X^{-1}dg_X)P^{(2)}(g_X^{-1} d\varphi \Lambda_{\mathfrak{s}} g_X + g_X^{-1}dg_X)\big) .
\end{equation}
We set $t = x^+ - \frac12 \epsilon^2 x^-$ and $\varphi = x^++ \frac12 \epsilon^2 x^-$, with $\epsilon$ a small constant parameter, $g_X = \exp(X)$ and expand $X$ according to eq.~\eqref{transparam} as
\begin{equation}\label{eq:Xparam}
X = \epsilon x + \epsilon f(\epsilon x) \Lambda_{\alg{s}}^{(2)}, \qquad \STr(x \Lambda_{\alg{s}}^{(2)}) = 0 .
\end{equation}
Finally, we recall that we normalise $\Lambda_{\mathfrak{s}}$ such that $\STr(\Lambda_{\mathfrak{s}}^{(2)}\Lambda_{\mathfrak{s}}^{(2)}) = 2$.

Using that $[\Lambda_{\mathfrak{s}}^{(0)}, \Lambda_{\mathfrak{s}}^{(2)}] = 0$, and expanding to quadratic order in $\epsilon$ we find
\begin{equation}
\begin{split}
ds^2 & = - 2\epsilon \, dx^+ df(\epsilon x)
\\ & \quad + \epsilon^2 \Big(2 dx^+ dx^- +\frac 12 \STr\big(dx_1^2\big)
- \frac12 (dx^+)^2 \Big(\STr\big([x_1,\Lambda_{\mathfrak{s}}^{(2)}]^2\big) - \STr\big([x_1,\Lambda_{\mathfrak{s}}^{(0)}]^2\big) \Big)
\\ & \hspace{150pt}
- dx^+ \STr\big(\Lambda_{\mathfrak{s}}^{(0)} [x_1,dx_1]\big) \Big) + {\cal O}(\epsilon^3) .
\end{split}
\end{equation}
For the pp-wave limit to be finite and non-degenerate, we rescale the string tension $T \to T \epsilon^{-2}$ and require that $\epsilon f(\epsilon x) = \epsilon^2 f_0(x) + \mathcal{O}(\epsilon^3)$.
\unskip\footnote{While for a finite pp-wave limit in the \sm{} we cannot have an $\mathcal{O}(\epsilon)$ term in this expansion, if we light-cone gauge-fix around $x^+ = \tau$, the divergent piece will be a total derivative $\sim \partial_\tau f$ that we can drop.}
The metric now simplifies to
\begin{equation}\label{eq:ds2-ppwave}
\begin{split}
ds^2 & = 2 dx^+ dx^- +\frac 12 \STr\big(dx^2\big) - \frac12 (dx^+)^2 \Big(\STr\big([x,\Lambda_{\mathfrak{s}}^{(2)}]^2\big) - \STr\big([x,\Lambda_{\mathfrak{s}}^{(0)}]^2\big)\Big)
\\ & \qquad \qquad - dx^+ \STr\big(\Lambda_{\mathfrak{s}}^{(0)} [x,dx]\big) +2 dx^+ df_0(x) + {\cal O}(\epsilon^3) .
\end{split}
\end{equation}
The freedom in this limit is thus captured by $\Lambda_{\mathfrak{s}}^{(0)}$ for the longitudinal sector, which contains $\rk \alg{g}_\mathfrak{s} -1$ parameters, and the function $f(x)$ for the transverse sector.

To interpret these freedoms let us note that before taking the pp-wave limit we can remove $\Lambda_{\mathfrak{s}}^{(0)}$ from our parametrisation~\eqref{eq:g-par} with $g_X = \exp X$ and $X$ given in eq.~\eqref{eq:Xparam}, by redefining
\begin{equation}\label{eq:redef}
x \to \exp(-\varphi \Lambda_{\mathfrak{s}}^{(0)}) x \exp(\varphi\Lambda_{\mathfrak{s}}^{(0)}) ,
\end{equation}
where we assume the function $f$ is invariant.
From the summary in~\secref{s:recap-ilcgf} we see that this can be understood as a combination of a $JT_\sigma$ and a $J^\tau$ deformation.
After taking the pp-wave limit, $\Lambda_{\mathfrak{s}}^{(0)}$ can similarly be removed from \eqref{eq:ds2-ppwave} by the redefinition $x \to \exp(-x^+ \Lambda_{\mathfrak{s}}^{(0)}) x \exp(x^+\Lambda_{\mathfrak{s}}^{(0)})$, therefore only the $J^\tau$ deformation survives, while, as we will see, the $JT_\sigma$ deformation contributes at higher orders in the transverse fields.

Similarly, we can in principle remove $f(x)$ from the pp-wave metric by shifting $x^- \to x^- + f_0(x)$.
However, if we demand that this does not transform $t$, then at higher orders we will also need to shift
$x^+ \to x^+ - \frac14 f_0(x)$ and we have an inequivalent gauge-fixing corresponding to a $\tilde{J}T_\tau$ deformations as follows from the summary in~\secref{s:recap-ilcgf}.

\medskip

Based on the pp-wave analysis above, we now fix light-cone gauge in the \sm{} on $\Real \times \Man_{\mathfrak{s}}$ with the goal of understanding the effect of inequivalent gauge-fixings.
Here we work in the Lagrangian formalism, while analogous results for $AdS_5 \times S^5$ for the Hamiltonian and tree-level S-matrix $ \mathbf{T} $ will be derived in~\secref{s:treelevel-checks-Jtilde}.

We start from the metric~\eqref{eq:metms} and expand in powers of the transverse field $X = P^{(2)} X$.
Introducing the operators $D_0 = d + d\varphi \ad_{\Lambda^{(0)}_\alg{s}}$ and expanding to quartic order in $X$ we find the metric is given by
\begin{equation}
\begin{aligned}
ds^2=-dt^2 & +d\varphi^2\left(1+\frac{1}{2}\STr\left[\Lambda_{\mathfrak{s}}^{(2)}\ad_X^2\Lambda_{\mathfrak{s}}^{(2)}\right]+\frac{1}{6}\STr\left[\Lambda_{\mathfrak{s}}^{(2)}\ad_X^4\Lambda_{\mathfrak{s}}^{(2)}\right]\right)\\
&+d\varphi\left(\STr\left[\Lambda_{\mathfrak{s}}^{(2)}D_0X\right]+\frac{2}{3}\STr\left[\Lambda_{\mathfrak{s}}^{(2)}\ad_X^2D_0X\right]\right)\\
&+\frac{1}{2}\STr\left[D_0XD_0X\right]+\frac{1}{6}\STr\left[D_0X\ad_X^2D_0X\right]+\mathcal{O}(X^5).
\end{aligned}
\label{eq:metricorder4}
\end{equation}
In order to light-cone gauge-fix in the Lagrangian formalism we exploit the results of \cite{Arutyunov:2014jfa}, which follows the method of~\cite{Kruczenski:2004cn,Zarembo:2009au}.
In terms of the light-cone coordinates
\begin{equation}
\varphi=x^++(1-a)x^- ,\qquad t=x^+-ax^- , \qquad a\in [0,1] ,
\label{eq:lcgfxpxm}
\end{equation}
the metric can be written as
\begin{equation}\label{eq:metparam}
ds^2=G_{++}dx^{+2}+2G_{+-}dx^{+}dx^{-}+G_{--}dx^{-2}+2G_+dx^++2G_-dx^-+G_t ,
\end{equation}
where $G_\pm$ has terms linear in $dX$ and $G_t$ quadratic terms in $dX$.
Using~\eqref{eq:lcgfxpxm}, we can straightforwardly read off the elements of $ ds^2 $ from \eqref{eq:metricorder4}.

The light-cone gauge-fixed action is given by
\begin{equation}
\mathcal{S}_{g.f.}=-T\int_{\Sigma} d\tau d\sigma \, (\sqrt{-M}+\frac{1}{2}E) = \int d\tau d\sigma \, \mathcal{L},
\label{eq:action}
\end{equation}
with
\begin{align}
M & =\frac{1}{G_{--}^2}\left((\mathring{G}_{++}+2\mathring{G}_{+,\tau}+\mathring{G}_{t,\tau\tau})(1+\mathring{G}_{t,\sigma\sigma})-(\mathring{G}_{+,\sigma}+\mathring{G}_{t,\tau\sigma})^2\right),
\label{eq:lcgfM}
\\
E & =-\frac{2}{G_{--}}\left(G_{+-}+G_{-,\tau}\right) ,
\label{eq:lcgfE}
\end{align}
and we recall that $ \mathcal{L} $ is the gauge-fixed Lagrangian.
The notation here is as follows: $ G_{\pm,\alpha} $ denotes $ G_{\pm} $ with $d$ replaced by $\partial_\alpha$, while $G_{t,\alpha\beta}$ denotes $G_t$ with one $d$ replaced by $\partial_\alpha$ and the other by $\partial_\beta$.
This latter step is unambiguous by the symmetry of $G_t$.
Additionally, the components of $ \mathring{G}$ are defined as
\begin{equation}\begin{split}
\mathring{G}_{++}&=G_{--}G_{++}-G_{+-}^2 ,
\\	\mathring{G}_{+,\alpha}&=G_{--}G_{+,\alpha}-G_{+-}G_{-,\alpha} ,
\\	\mathring{G}_{t,\alpha\beta}&=G_{--}G_{t,\alpha\beta}-G_{-,\alpha}G_{-,\beta} .
\label{eq:mathringGAB}
\end{split}\end{equation}

\subsubsection{\texorpdfstring{$ \tilde{J}T_\tau $}{JtildeTτ} deformation for \texorpdfstring{$ \Real\times \Man_{\mathfrak{s}} $}{R × Ms}}\label{s:JTildeT0RxSn}

We start by focusing on the $ \tilde{J}T_\tau $ deformation and therefore for simplicity assume $ \Lambda_{\mathfrak{s}}^{(0)}=0 $.
In particular, this implies that $ D_0X=dX $.
Therefore, to quartic order in $X$ we find
\begin{equation}
\begin{aligned}
G_{++}&=\frac{1}{2}\STr\left[\Lambda_{\mathfrak{s}}^{(2)}\ad_X^2\Lambda_{\mathfrak{s}}^{(2)}\right]+\frac{1}{6}\STr\left[\Lambda_{\mathfrak{s}}^{(2)}\ad_X^4\Lambda_{\mathfrak{s}}^{(2)}\right]+\mathcal{O}(X^5)
\coloneqq V_2(X)+V_4(X)+\mathcal{O}(X^5),\\
G_{--} & =1-2a+(1-a)^2G_{++}, \qquad G_{+-}=1+(1-a)G_{++},\\
G_+&=\frac{1}{2}\STr\left[\Lambda_{\mathfrak{s}}^{(2)}dX\right]+\frac{1}{3}\STr\left[\Lambda_{\mathfrak{s}}^{(2)}\ad_X^2dX\right]+\mathcal{O}(X^5)
\coloneqq L_1(X)+L_3(X)+\mathcal{O}(X^5),\\
G_-&=(1-a)G_{+},\\
G_t&=\frac{1}{2}\STr\left[dXdX\right]+\frac{1}{6}\STr\left[dX\ad_X^2dX\right]+\mathcal{O}(X^5)
\coloneqq K_2(X)+K_4(X)+\mathcal{O}(X^5).
\end{aligned}
\label{eq:metricJtildeT0}
\end{equation}
The indices on $ V_i $, $ L_i $ and $ K_i $ denote the power of $ X $.

We can now compute the light-cone gauge-fixed Lagrangian as defined in eq.~\eqref{eq:action} up to quartic order.
We rescale $ X\rightarrow T^{-\frac12} X $, substitute the metric \eqref{eq:metricJtildeT0} in the action \eqref{eq:action} using the expressions \eqref{eq:lcgfM,-,eq:mathringGAB} and expand to obtain
\begin{equation}
\mathcal{L}(X)=T^{\frac12}\mathcal{L}_1(X)+\mathcal{L}_2(X)+T^{-\frac12}\mathcal{L}_3(X)+T^{-1}\mathcal{L}_4(X)+\mathcal{O}(T^{-\frac{3}{2}}X^5),
\end{equation}
with
\begin{align}
&\mathcal{L}_1=L_{1;\tau},\label{eq:Lag1}\\
&\mathcal{L}_2=\frac{1}{2}\left(K_{2;\tau,\tau}-K_{2;\sigma,\sigma}-L_{1;\tau}^2+L_{1;\sigma}^2+V_2\right),\label{eq:Lag2}\\
&\mathcal{L}_3=-L_{1,\tau}V_2+L_{3,\tau}\nonumber\\
&\hspace{1cm}+a\left(\frac{1}{2}L_{1;\tau}(V_2-K_{2;\tau,\tau}-K_{2;\sigma,\sigma}+L_{1,\tau}^2-L_{1;\sigma}^2)+L_{1;\sigma}K_{2;\tau,\sigma}\right),\label{eq:Lag3}\\
&\mathcal{L}_4=\frac{1}{8}\left(K_{2;\tau,\tau}+K_{2;\sigma,\sigma}-2K_{2;\tau,\sigma}-\left(L_{1;\tau}-L_{1;\sigma}\right)^2\right)\left(K_{2;\tau,\tau}+K_{2;\sigma,\sigma}-2K_{2;\tau,\sigma}-\left(L_{1;\tau}+L_{1;\sigma}\right)^2\right)\nonumber\\
&\hspace{1cm}-L_{1,\tau}L_{3,\tau}+L_{1,\sigma}L_{3,\sigma}-\frac{1}{4}\left(K_{2;\tau,\tau}+K_{2;\sigma,\sigma}-3L_{1,\tau}^2+L_{1,\sigma}^2+\frac{3}{2}V_2\right)V_2\nonumber\\
&\hspace{1cm}+\frac{1}{2}\left(K_{4;\tau,\tau}-K_{4;\sigma,\sigma}+V_4\right)+\frac{a^2}{2}\left(K_{2;\tau,\tau}-L_{1,\tau}^2\right)\left(L_{1,\tau}^2-L_{1,\sigma}^2\right)\nonumber\\
&\hspace{1cm}+a\left(\left(K_{2;\tau,\sigma}-L_{1;\tau}L_{1;\sigma}\right)^2-\frac{1}{4}\left(K_{2;\tau,\tau}+K_{2;\sigma,\sigma}-L_{1,\tau}^2-L_{1,\sigma}^2\right)^2+\frac{1}{4}V_2^2\right).\label{eq:Lag4}
\end{align}
Note that, as for $G_\pm$ and $G_t$ above, the labels $ \tau $ and $\sigma$ on $L_i$ and $K_i$ indicate that $d$ should be replaced by $\partial_\tau $ and $\partial_\sigma$ respectively, where the symmetry of $K_i$ again means that this procedure is unambiguous.

From the analysis in \secref{s:pp-wave}, inequivalent gauge-fixings corresponding to $ \tilde{J}T_\tau $ deformations are parametrised by a function $f(x)$, which can be introduced as
\unskip\footnote{Here $f(x)$ can be related to $c(x)$ in the shift $\varphi \to \varphi + c(x)$ (at leading order they are equal).
The shift in $\varphi$ can be split into a shift in $x^-$, which corresponds to a total derivative after light-cone gauge-fixing and was visible in the pp-wave analysis, and a shift in $x^+$ corresponding to a $\tilde{J}T_\tau$ deformation.}
\begin{equation}
X=x+f(x)\Lambda_{\mathfrak{s}}^{(2)},
\label{eq:X}
\end{equation}
where we take $\STr[x\Lambda_{\mathfrak{s}}^{(2)}] = 0$.
We will now show that, up to total derivatives and redefinitions of the transverse fields, $ \mathcal{L}_j $ for $ j=1,...,4 $ does not depend on $ f(x) $.

Substituting~\eqref{eq:X} into the expansion of the light-cone gauge Lagrangian~\eqref{eq:Lag1,-,eq:Lag4} we obtain
\begin{align}
&\mathcal{L}_1=\partial_\tau f,\label{eq:L1}\\
&\mathcal{L}_2=\frac{1}{2}\left(K_{2;\tau,\tau}(x)-K_{2;\sigma,\sigma}(x)+V_2(x)
\right),\label{eq:L2}\\
&\mathcal{L}_3=L_3(x)-\frac{1}{3}\partial_\tau\left(V_2(x)f\right)+
a\left(K_{2;\tau,\sigma}(x)\partial_{\sigma}f+\frac{1}{2}\partial_{\tau}f\left(V_2(x)-K_{2;\tau,\tau}(x)-K_{2;\sigma,\sigma}(x)\right)\right),\label{eq:L3}\\
&\mathcal{L}_4=\frac{1}{2}\left(K_{4;\tau,\tau}(x)-K_{4;\sigma,\sigma}(x)+V_4(x)\right)
+\frac{1}{8}\left(K_{2;\tau,\tau}(x)+K_{2;\sigma,\sigma}(x) -V_2(x)\right)^2
-\frac{1}{2}\left(K_{2;\tau,\sigma}(x)^2+V_2(x)^2\right)\nonumber\\
&\hspace{1cm}-\frac{a}{4}\left(\left(K_{2;\tau,\tau}(x)+K_{2;\sigma,\sigma}(x)\right)^2-4K_{2;\tau,\sigma}(x)^2-V_2(x)^2\right)-\frac12(\partial_{\tau}-\partial_{\sigma})\left(L_3(x)f-\frac{1}{6}(\partial_{\tau}V_2(x))f^2\right)\nonumber\\
&\hspace{1cm}+\frac{a^2}{2}K_{2;\tau,\tau}(x)\left((\partial_\tau{f})^2-(\partial_\sigma{f})^2\right)
-\frac{1}{6}f^2\STr[(\ad_{\Lambda_{\mathfrak{s}}^{(2)}}^2 x) \mathcal{D}x]-\frac{1}{3}f\STr[(\ad_x\ad_{\Lambda_{\mathfrak{s}}^{(2)}}x)\mathcal{D}x],\label{eq:L4}
\end{align}
where
\begin{equation}
\mathcal{D}=\frac{1}{2}\left(\frac{\partial^2}{\partial\tau^2}-\frac{\partial^2}{\partial\sigma^2}-\ad_{\Lambda_{\mathfrak{s}}^{(2)}}^2\right) .
\end{equation}
From this point on we drop total derivatives.
Doing so, we can rewrite the above expansion as
\begin{align}
&\mathcal{L}_1=0,\label{eq:L1tt}\\
&\mathcal{L}_2= - \frac12\STr[x\mathcal{D}x] ,\label{eq:L2tt}\\
&\mathcal{L}_3=L_3(x)+ a \STr[f\dot x \mathcal{D} x] ,\label{eq:L3tt}\\
&\mathcal{L}_4=\frac{1}{2}\left(K_{4;\tau,\tau}(x)-K_{4;\sigma,\sigma}(x)+V_4(x)\right)
+\frac{1}{8}\left(K_{2;\tau,\tau}(x)+K_{2;\sigma,\sigma}(x) -V_2(x)\right)^2
-\frac{1}{2}\left(K_{2;\tau,\sigma}(x)^2+V_2(x)^2\right)\nonumber\\
&\hspace{1cm}-\frac{a}{4}\left(\left(K_{2;\tau,\tau}(x)+K_{2;\sigma,\sigma}(x)\right)^2-4K_{2;\tau,\sigma}(x)^2-V_2(x)^2\right)\nonumber\\
&\hspace{1cm}+\frac{a^2}{2}\left(\STr[f\dot{x} \mathcal{D} f\dot{x}]-\STr[f^2\dot{x} \mathcal{D}\dot{x}]\right)-\frac{1}{6}f^2\STr[(\ad_{\Lambda_{\mathfrak{s}}^{(2)}}^2 x) \mathcal{D}x]-\frac{1}{3}f\STr[(\ad_x\ad_{\Lambda_{\mathfrak{s}}^{(2)}}x)\mathcal{D}x].\label{eq:L4tt}
\end{align}
Finally, we can implement the following field redefinition:
\begin{align}
&x\rightarrow x+T^{-\frac12}a f\dot{x}+T^{-1}\big(\frac{1}{2}a^2 (f^2\dot{x})^{\dot{}}
-\frac16 f^2 \ad_{\Lambda_{\mathfrak{s}}^{(2)}}^2 x - \frac13 f\ad_x\ad_{\Lambda_{\mathfrak{s}}^{(2)}}x\big) +\cdots,\\
&f\rightarrow f+ T^{-\frac12} a f \dot{f}+\cdots ,
\label{eq:fieldredefJtildeT0}
\end{align}
which completely eliminates the dependence of the quartic light-cone gauge Lagrangian on $f$.

Therefore, we find that the effect of inequivalent gauge-fixings corresponding to $ \tilde{J}T_{\tau} $ deformations in the light-cone gauge-fixed Lagrangian up to quartic order can be removed by a field redefinition if we drop total derivatives.
It follows that the light-cone gauge S-matrix at tree-level will not depend on $f$, and we will see an explicit example of this in~\secref{s:treelevel-checks-Jtilde} for $AdS_5\times S^5$.

\subsubsection{\texorpdfstring{$J^\tau$}{Jτ} and \texorpdfstring{$ JT_\sigma $}{JTσ} deformation for \texorpdfstring{$ \Real\times \Man_{\mathfrak{s}} $}{R × Ms}}\label{s:JTsigmaRxSn}

Again based on the pp-wave analysis in~\secref{s:pp-wave}, we now study the effect of inequivalent gauge-fixings corresponding to $J^\tau$ and $JT_\sigma$ deformations in the gauge-fixed theory.
We will do this in the Lagrangian formalism, fixing light-cone gauge in the \sm{} on $\Real \times \Man_{\mathfrak{s}}$ as in~\secref{s:JTildeT0RxSn}, but now setting $f =0$ and keeping $ \Lambda_{\mathfrak{s}}^{(0)} $ non-zero.
Recall that in this analysis the $J^\tau$ and $JT_\sigma$ deformations are tied together since $\Lambda^{(0)}$ originates from the redefinition~\eqref{eq:redef}.

Since we now consider $\Lambda_{\mathfrak{s}}^{(0)}\neq0 $, the metric~\eqref{eq:metparam}
has extra terms compared to eq.~\eqref{eq:metricJtildeT0} and can be written as
\begin{align}
G_{++}&=\frac{1}{2}\STr\left[\Lambda_{\mathfrak{s}}^{(2)}\ad_X^2\Lambda_{\mathfrak{s}}^{(2)}\right]+\frac{1}{6}\STr\left[\Lambda_{\mathfrak{s}}^{(2)}\ad_X^4\Lambda_{\mathfrak{s}}^{(2)}\right]-\frac{1}{2}\STr\left[\Lambda_{\mathfrak{s}}^{(0)}\ad_X^2\Lambda_{\mathfrak{s}}^{(0)}\right]-\frac{1}{6}\STr\left[\Lambda_{\mathfrak{s}}^{(0)}\ad_X^4\Lambda_{\mathfrak{s}}^{(0)}\right] \nonumber\\
&\quad-\frac{2}{3}\STr\left[\Lambda_{\mathfrak{s}}^{(2)}\ad_X^3\Lambda^{(0)}_{\mathfrak{s}}\right]+\mathcal{O}(X^5)
\coloneqq V_2(X)+V_4(X)+\bar{V}_2(X)+\bar{V}_4(X)+\bar{V}_3(X)+\mathcal{O}(X^5),\nonumber\\
G_{--}&=1-2a+(1-a)^2G_{++},\qquad G_{+-}=1+(1-a)G_{++},\nonumber\\
G_+&=\frac{1}{2}\STr\left[\Lambda_{\mathfrak{s}}^{(2)}dX\right]+\frac{1}{3}\STr\left[\Lambda_{\mathfrak{s}}^{(2)}\ad_X^2dX\right]-\frac{1}{2}\STr\left[dX\ad_X\Lambda^{(0)}_{\mathfrak{s}}\right]-\frac{1}{6}\STr\left[dX\ad_X^3\Lambda^{(0)}_{\mathfrak{s}}\right]\\
&\coloneqq L_1(X)+L_3(X)+\bar{L}_2(X)+\bar{L}_4(X)+\mathcal{O}(X^5),\nonumber\\
G_-&=(1-a)G_{+},\nonumber\\
G_t&=\frac{1}{2}\STr\left[dXdX\right]+\frac{1}{6}\STr\left[dX\ad_X^2dX\right]+\mathcal{O}(X^5)
\coloneqq K_2(X)+K_4(X)+\mathcal{O}(X^5),\nonumber
\end{align}
where we have introduced new functions $ \bar{V}_i $ and $ \bar{L}_i $, which depend on $\Lambda_{\mathfrak{s}}^{(0)}$ and whose index again indicates the power of $X$.

Setting $ X=x $ where $\STr[x\Lambda^{(2)}_\mathfrak{s}] = 0$, rescaling $ x\rightarrow T^{-\frac{1}{2}}x $, and computing the light-cone gauge-fixed Lagrangian as defined in \eqref{eq:action} to quartic order, we find
\begin{equation}
\mathcal{L}(x)=T^{\frac12}\mathcal{L}_1(x)+\mathcal{L}_2(x)+T^{-\frac12}\mathcal{L}_3(x)+T^{-1}\mathcal{L}_4(x)+\mathcal{O}(T^{-\frac{3}{2}}x^5),
\end{equation}
with
\begin{align}\label{eq:l1jt}
\mathcal{L}_1=&\,0,\\
\mathcal{L}_2=&\,\frac{1}{2}\left(V_2+K_{2;\tau,\tau}-K_{2;\sigma,\sigma}\right)+\frac{1}{2}\bar{V}_2+\bar{L}_{2;\tau},\\
\mathcal{L}_3=&\,L_{3;\tau}+\frac{1}{2}\bar{V}_3,\\
\mathcal{L}_4=&\,\frac{1}{8}(1-2a)\left((K_{2;\tau,\tau}+K_{2;\sigma,\sigma})^2-4K_{2;\tau,\sigma}^2-(V_2+\bar{V}_2)^2\right)-\frac{1}{2}a\left(K_{2;\tau,\tau}+K_{2;\sigma,\sigma}-V_2-\bar{V}_2\right)\bar{L}_{2;\tau}\nonumber\\
&+aK_{2;\tau,\sigma}\bar{L}_{2;\sigma}+\frac{1}{4}\left(V_2+\bar{V}_2\right)\left(K_{2;\tau,\tau}+K_{2;\sigma,\sigma}+V_2+\bar{V}_2+4\bar{L}_{2;\tau}\right)-\frac{1}{2}\left(\bar{L}_{2;\tau}^2-\bar{L}_{2;\sigma}^2\right)\nonumber\\
&+\frac{1}{2}\left(K_{4;\tau,\tau}-K_{4;\sigma,\sigma}+V_4+\bar{V}_4+2\bar{L}_{4;\tau}\right),\label{eq:l4jt}
\end{align}
where we have used that $ \STr[x\Lambda_{\mathfrak{s}}^{(2)}]=0 $ implies $L_1(x) = 0$.
For clarity, we have also suppressed the dependence of the functions $V_i$, $L_i$ and $K_i$ on $x$.

To see that the effect of $\Lambda_{\mathfrak{s}}^{(0)}$ is a combination of $J^\tau$ and $JT_\sigma$ deformations as claimed, we start by noting that eqs.~\eqref{eq:l1jt,-,eq:l4jt} are invariant under the transformation
\begin{equation}\label{eq:symtransrot}
x\rightarrow e^{-\alpha \Lambda_{\mathfrak{s}}^{(0)}}x\,e^{\alpha \Lambda_{\mathfrak{s}}^{(0)}},
\end{equation}
for constant $ \alpha $ since $[\Lambda_{\mathfrak{s}}^{(0)},\Lambda_{\mathfrak{s}}^{(2)}] = 0$.
We can therefore remove the $J^\tau$ deformation by promoting $\alpha$ to be time dependent and rotating
\begin{equation}
x\rightarrow e^{-\tau \Lambda_{\mathfrak{s}}^{(0)}}x\,e^{\tau \Lambda_{\mathfrak{s}}^{(0)}},
\end{equation}
under which ($i=1,2$)
\begin{equation}
\begin{aligned}
& \bar{L}_{2i,\tau}\rightarrow\bar{L}_{2i,\tau}-\bar{V}_{2i}, &&\qquad
\bar{L}_{2i-1,\tau}\rightarrow\bar{L}_{2i-1,\tau}-\frac12\bar{V}_{2i-1},
\\ & K_{2i;\tau,\tau}\rightarrow K_{2i;\tau,\tau}-2\bar{L}_{2i;\tau}+\bar{V}_{2i},&&\qquad K_{2i;\tau,\sigma}\rightarrow K_{2i;\tau,\sigma}-\bar{L}_{2i;\sigma},
\end{aligned}
\end{equation}
where $\bar V_{1} = 0$.
The remaining functions do not transform.
The transformed Lagrangian is then given by
\begin{equation}
\mathcal{L}^{\tau}(x)=T^{\frac12}\mathcal{L}_1^{\tau}(x)+\mathcal{L}_2^{\tau}(x)+T^{-\frac12}\mathcal{L}_3^{\tau}(x)+T^{-1}\mathcal{L}_4^{\tau}(x)+\mathcal{O}(T^{-\frac{3}{2}}x^5),
\end{equation}
with
\begin{align}
\mathcal{L}_1^{\tau}=&\,0,\\
\mathcal{L}_2^{\tau}=&\,\frac{1}{2}\left(V_2+K_{2;\tau,\tau}-K_{2;\sigma,\sigma}\right),\\
\mathcal{L}_3^{\tau}=&\,L_{3;\tau},\\
\mathcal{L}_4^{\tau}=&\,\frac{1}{8}\left(1-2a\right)\left((K_{2;\tau,\tau}+K_{2;\sigma,\sigma})^2-4K_{2;\tau,\sigma}^2-V_2^2\right)\nonumber\\
&-(1-a)\left(K_{2;\tau,\sigma}\bar{L}_{2;\sigma}+\frac{1}{2}\left(K_{2;\tau,\tau}+K_{2;\sigma,\sigma}+V_2\right)\bar{L}_{2;\tau}\right)\nonumber\\
&-\frac{1}{4}V_2\left(K_{2;\tau,\tau}+K_{2;\sigma,\sigma}+V_2\right)+\frac{1}{2}\left(K_{4;\tau,\tau}-K_{4;\sigma,\sigma}+V_4\right).
\end{align}

Computing the conserved current associated to the symmetry~\eqref{eq:symtransrot} we find
\begin{equation}
J_{\tau}=\bar{L}_{2;\tau}, \qquad J_{\sigma}=-\bar{L}_{2;\sigma},
\label{eq:conservedcurrents}
\end{equation}
while the components of the stress-energy tensor are given by
\begin{equation}
T_{\tau\tau}=-\frac{1}{2}\left(K_{2;\tau,\tau}+K_{2;\sigma,\sigma}-V_2\right),\quad T_{\sigma\sigma}=-\frac{1}{2}\left(K_{2;\tau,\tau}+K_{2;\sigma,\sigma}+V_2\right), \quad T_{\tau\sigma}=T_{\sigma\tau}=K_{2;\tau,\sigma}.
\end{equation}
Constructing the $ T\bar{T} $ and $ JT_{\sigma} $ operators as
\begin{equation}
\hat{O}_{T\bar{T}}=\epsilon_{\alpha\beta}T^{\alpha}_{\,\.\sigma}T^{\beta}_{\,\.\tau} \quad \text{and} \quad \hat{O}_{JT_{\sigma}}=\epsilon_{\alpha\beta}T^{\alpha}_{\,\.\sigma}J^{\beta},
\end{equation}
we see that we can rewrite $ \mathcal{L}_4^{\tau} $ as
\begin{equation}
\mathcal{L}_4^{\tau}=\mathcal{L}_4\Big|_{\Lambda^{(0)}_{\mathfrak{s}}\rightarrow 0,\,a\rightarrow \frac{1}{2}}+\left(\frac{1}{2}-a\right)\hat{O}_{T\bar{T}}+(1-a)\hat{O}_{JT_{\sigma}},
\end{equation}
demonstrating the form of the $ JT_{\sigma} $ deformation explicitly.

\subsection{Strings on \texorpdfstring{$AdS_n\times S^n$}{AdSn × Sn}} \label{s:AdSn-Sn}

In the remainder of this section, we will focus on string \sms{} on $AdS_n\times S^n$ backgrounds and explore the light-cone gauge freedom in their longitudinal sector in more detail, including analysing the symmetries of the resulting light-cone gauge-fixed theories.

The target spacetimes of these \sms{} can be realised as the symmetric cosets
\begin{equation}
\Man_\mathfrak{a} \times \Man_\mathfrak{s}= \frac{SO(n-1,2)}{SO(n-1,1)} \times \frac{SO(n+1)}{SO(n)} .
\end{equation}
Hence the Lie group $G$ is the product of a non-compact and a compact group.
Their Lie algebras $\mathfrak{g}_\mathfrak{a} = \mathfrak{so}(n-1,2)$ and $\mathfrak{g}_\mathfrak{s} = \mathfrak{so}(n+1)$ can be spanned respectively by antihermitian matrices $J_{IJ}$, $I,J=0,\ldots, n$ and $R_{AB}$, $A,B=1,\ldots,n+1$, satisfying
\begin{equation} \label{eq:commrel}
\begin{alignedat}{3}
[J_{IJ} , J_{KL}] &= \eta_{IK}J_{JL} - \eta_{JK} J_{IL} +\eta_{JL} J_{IK} - \eta_{IL}J_{JK}, \qquad &&J_{IJ}&&=-J_{JI}, \\
[R_{AB} , R_{CD}] &= \delta_{AC} R_{BD} - \delta_{BC} R_{AD} + \delta_{BD} R_{AC}-\delta_{AD} R_{BC}, \qquad &&R_{AB}&&=-R_{BA} ,
\end{alignedat}
\end{equation}
with $\eta_{IJ} = \mathrm{diag} ( -1 , 1, \ldots ,1 , -1 )$.
This realises a symmetric space with $\mathfrak{g}^{(2)}=\mathrm{span} (J_{in}, R_{a n+1})$, for $i=0,\ldots, n-1 $, and $a = 1,\ldots,n$, and the invariant subalgebra $\mathfrak{g}^{(0)}$ spanned by the remaining orthogonal generators.
To explore the light-cone gauge freedom in the longitudinal sector, we need to identify the Cartan subalgebras $\mathfrak{t}_\mathfrak{a}$ and $\mathfrak{t}_\mathfrak{s}$.

\subsubsection{Identifying the Cartan subalgebra }

For compact groups, there is a unique Cartan subalgebra up to inner automorphisms by Cartan's torus theorem.
The rank of $\mathfrak{g}_\mathfrak{s} =\mathfrak{so}(n+1)$ is $\lfloor \frac{n+1}{2} \rfloor$ and we can take the Cartan subalgebra to be spanned, e.g.,~by
\begin{equation}
\mathfrak{t}_{\mathfrak{s}} = \mathrm{span} \left\{R_{n(n+1)} , \bigcup_{i=1}^{\lfloor \frac{n-1}{2} \rfloor} R_{(2i-1)(2i)} \right\} ,
\end{equation}
where we have introduced brackets on indices for readability.
For example, in the case of $\mathfrak{so}(6)$ we take $\mathfrak{t}_{\mathfrak{s}} = \mathrm{span} \left\{R_{56},R_{12},R_{34}\right\}$.
Because of its definite signature, a generic element $\Lambda_\mathfrak{s} \in \mathfrak{t}_\mathfrak{s}$ is spacelike under $\mathrm{STr}$.
In contrast, for non-compact groups, there can be distinct Cartan subalgebras not related by inner automorphisms.
To identify the space of inequivalent gauge-fixings, we should therefore take into account all these possibilities.
However, the Virasoro constraint \eqref{eq:virasoro-symmcoset} of the $AdS_n\times S^n$ string singles out one Cartan subalgebra \cite{Miramontes:2008wt} up to inner automorphisms.
To elaborate, let us consider for simplicity $\gamma^{\alpha\beta}=\eta^{\alpha\beta}$, such that in the light-cone coordinates $\sigma^\pm = \frac{1}{2}(\tau \pm \sigma)$ we have $\mathcal T_{+-}=0$ identically.
Because of the Cartesian product structure of the spacetime, we can write $A_\alpha = A_{\alpha {\mathfrak{a}}}+ A_{\alpha {\mathfrak{s}}}$, with $A_{\alpha {\mathfrak{a}}}$ ($A_{\alpha {\mathfrak{s}}}$) the projections of $A_\alpha$ on the subalgebra $\mathfrak{g}_\mathfrak{a} $ ($\mathfrak{g}_\mathfrak{s} $) for the $AdS_n$ ($S^n$) space.
The other components $\mathcal T_{\pm \pm}$ of the energy-momentum tensor can similarly be split into a contribution from $AdS_n$ and $S^n$, i.e.~
\begin{equation}
\mathcal T_{\pm\pm} = \mathcal T^{\mathfrak{a}}_{\pm\pm} + \mathcal T^{\mathfrak{s}}_{\pm\pm} \overset{!}{=} 0 , \qquad \mathcal T^{\mathfrak{a}(\mathfrak{s})}_{\pm\pm} = \mathrm{STr} \left( A_{\pm \mathfrak{a}(\mathfrak{s})}^{(2)} A_{\pm \mathfrak{a}(\mathfrak{s})}^{(2)} \right) .
\end{equation}
The conformal symmetry of the worldsheet means that it is always possible to choose coordinates such that $\mathcal T^{\mathfrak{a}({\mathfrak{s}})}_{\pm\pm}=\mu_{\mathfrak{a}({\mathfrak{s}})}$ are real constants.
As before, for $S^n$, which is a space of definite signature, $\mu_{\mathfrak{s}}$ is positive-definite under $\mathrm{STr}$.
For $AdS_n$, on the other hand, a space of indefinite signature, $\mu_{\mathfrak{a}}$ can be negative, null or positive.
These cases lead to three inequivalent 1-dimensional Cartan subspaces of $P^{(2)}(\mathfrak{g}_\mathfrak{a} )$ \cite{Miramontes:2008wt}.
\unskip\footnote{For $n=2$ there are actually two possibilities with $\mu_{\mathfrak{a}}=0$, see \cite{Miramontes:2008wt}.}
The Virasoro constraints however require that $\mu_{\mathfrak{s}} = - \mu_{\mathfrak{a}}>0$.
\unskip\footnote{The choices of Cartan resulting in $\mu_{\mathfrak{a}}=0$ allow to consider bosonic string configurations on $AdS_n$ only.
As mentioned above, we will not consider such examples in this paper.}
For all $n$, this singles out the 1-dimensional Cartan subspace generated by $J_{0n}$ up to inner automorphisms.
The centraliser algebra of $J_{0n}$ is the \textit{compact} subalgebra $\mathfrak{so}(n-1)$ of $\mathfrak{g}^{(0)}_{\mathfrak{a}}=\mathfrak{so}(n-1,1)$.
This means that, up to conjugations by the compact subgroup $SO(n-1)\subset SO(n-1,1)$, the requirement that $J_{0n}$ is an element of the Cartan subalgebra polarises the full Cartan subalgebra of $\mathfrak{g}_\mathfrak{a} =\mathfrak{so}(n-1,2)$ to be the $\lfloor \frac{n+1}{2} \rfloor$-dimensional subspace spanned by
\begin{equation}
\mathfrak{t}_{\mathfrak{a}} = \mathrm{span} \left\{J_{0n} ,\bigcup_{i=1}^{\lfloor \frac{n-1}{2} \rfloor} J_{(2i-1)(2i)} \right\} .
\end{equation}
For example, in the case of $\mathfrak{so}(4,2)$ we take $\mathfrak{t}_{\mathfrak{a}} = \mathrm{span} \left\{J_{05},J_{12},J_{34}\right\}$.
Generic elements $P^{(2)}(\Lambda_\mathfrak{a}) \in \mathfrak{t}_\mathfrak{a}$ are now guaranteed to be timelike under $\mathrm{STr}$.

As a side remark, let us note that the above discussion holds more generally for $AdS_p\times S^q$ spaces with $p\neq q$.

\subsubsection{Relation to \texorpdfstring{$JT_\sigma$}{JTσ} and \texorpdfstring{$J^\tau$}{Jτ} deformations} \label{s:AdSn-Sn-JT}

Let us now consider $n=5$ and explore the $\rk \alg{g}_\mathfrak{a}+\rk \alg{g}_\mathfrak{s} -2 = 4$ parameter freedom in the longitudinal sector of $AdS_5\times S^5$.
The following can be readily extended to different values of $n$.
Based on the discussion above, we parametrise \eqref{eq:g-par} with
\begin{equation} \label{eq:lambda-a-s-gen}
\Lambda_\mathfrak{a} = \alpha_0 J_{05} + \alpha_1 J_{12} + \alpha_2 J_{34} , \qquad \Lambda_\mathfrak{s} = \beta_0 R_{56} + \beta_1 R_{12} + \beta_2 R_{34} ,
\end{equation}
where $\alpha_{1,2}$ and $\beta_{1,2}$ are free real parameters, and the transverse fields as in \cite{Arutyunov:2009ga}, i.e.
\begin{equation}
g_X = g_X(z,y)= \left( \frac{1+ \sum_{i=1}^4 z_i J_{i5} }{\sqrt{1-\frac{z^2}{4}}} \right) \oplus \left( \frac{1+ \sum_{i=1}^4 y_i R_{i6} }{\sqrt{1+\frac{y^2}{4}}}\right) ,
\end{equation}
where $z_i$ and $y_i$ are the transverse coordinates of $AdS_5$ and $S^5$ respectively, and $z^2 = z_iz_i$, $y^2 = y_iy_i$.
Because of our assumptions outlined at the beginning of this section, $\Lambda_\mathfrak{a}^{(2)}$ and $ \Lambda_\mathfrak{s}^{(2)}$ must be non-vanishing and thus we must require $\alpha_0\neq0 \neq \beta_0$.
Furthermore, $\alpha_0$ and $\beta_0$ will not be true parameters, as they can be rescaled to fix a definite normalisation of $\Lambda_\mathfrak{a}$ and $ \Lambda_\mathfrak{s}$.
The metric reads
\begin{equation} \label{eq:ads5s5-metric-jtpars}
\begin{aligned}
ds^2 ={}& -\alpha_0^2 \left(\frac{1+\frac{z^2}{4}}{1-\frac{z^2}{4}}\right)^2 dt^2 + \beta_0^2 \left(\frac{1-\frac{y^2}{4}}{1+\frac{y^2}{4}}\right)^2 d\varphi^2 \\
&+ \frac{(dz_1 - \alpha_1 z_2 dt)^2+(dz_2 + \alpha_1 z_1 dt)^2}{(1-\frac{z^2}{4})^2} + \frac{(dz_3 - \alpha_2 z_4 dt)^2+(dz_4 + \alpha_2 z_3 dt)^2}{(1-\frac{z^2}{4})^2} \\
& + \frac{(dy_1 - \beta_1 y_2 d\varphi)^2+(dy_2 + \beta_1 y_1 d\varphi)^2}{(1+\frac{y^2}{4})^2} + \frac{(dy_3 - \beta_2 y_4 d\varphi)^2+(dy_4 + \beta_2 y_3 d\varphi)^2}{(1+\frac{y^2}{4})^2} ,
\end{aligned}
\end{equation}
and thus indeed the parameters $\alpha_0$ and $\beta_0$ can be reabsorbed by a rescaling of $t$ and $\varphi$ (and $\alpha_{1,2}$ and $\beta_{1,2}$).
From now on we will set $\alpha_0 = \beta_0=1$.

At this stage, we indeed have a 4-dimensional moduli space in the longitudinal sector parametrised by $(\alpha_1 , \alpha_{2} , \beta_{1}, \beta_2 )$.
This freedom can be understood as coming from the action of the generators $J_{12}$, $J_{34}$, $R_{12}$ and $R_{34}$, where the would-be symmetry parameters are promoted to linear functions of the coordinates $t$ and $\varphi$ (or equivalently $x^\pm$).
The parameters $(\alpha_1 , \alpha_{2} , \beta_{1}, \beta_2 )$ thus correspond to $JT_\sigma$ and $J^\tau$ deformations.
Let us see this explicitly.
Starting from the standard $AdS_5\times S^5$ light-cone gauge-fixed theory with $\alpha_{1,2}=\beta_{1,2}=0$, there is an $\mathfrak{so}(4)\oplus \mathfrak{so}(4)\cong \mathfrak{su}(2)^{\oplus 2}\oplus \mathfrak{su}(2)^{\oplus 2}$ algebra in the centraliser of $\Lambda_\mathfrak{a}+ \Lambda_\mathfrak{s}$, which acts as $SO(4) \times SO(4)$ rotations of the $z_i$ and $y_i$ fields (see e.g.~\cite{Arutyunov:2009ga}).
Of these, there are $2+2$ abelian isometries that can maximally be realised.
In the above coordinate system these can be chosen to correspond to rotations in the planes
\unskip\footnote{Of course, it is possible to go to a coordinate system in which these rotations are realised as shifts of angles.}
\begin{equation} \label{eq:so4so4cartanisometries}
\begin{alignedat}{4}
&(z_1, z_2) \quad &&: \quad &&\text{generated by} \quad J_{12} \quad &&\text{as} \quad G_L \ : \ e^{\zeta_{12}J_{12}}, \\
&(z_3, z_4) \quad &&: \quad &&\text{generated by} \quad J_{34} \quad &&\text{as} \quad G_L \ : \ e^{\zeta_{34}J_{34}}, \\
&(y_1, y_2) \quad &&: \quad &&\text{generated by} \quad R_{12}\quad &&\text{as} \quad G_L \ : \ e^{\psi_{12}R_{12}}, \\
&(y_3, y_4) \quad &&: \quad &&\text{generated by} \quad R_{34} \quad &&\text{as} \quad G_L \ : \ e^{\psi_{34}R_{34}},
\end{alignedat}
\end{equation}
with $\zeta_{12}$, $\zeta_{34}$, $\psi_{12}$, $\psi_{34}$ constant isometry parameters.
These are actually global $G_L$ transformations by $h \in H_L \subset G_L$.
That they amount to rotations in the corresponding planes can be seen by noticing that, for example, $R_{i6}$ transforms as an $SO(4)$ vector under the rotations generated by $R_{ij}$, i.e.~$hR_{i6}h^{-1}=M_i{}^jR_{j6}$, with $M_i{}^j$ an orthogonal matrix, and that multiplications of $g_X$ from the right by $h^{-1}$ are in $H_R$.

We can now promote the parameters of the $AdS$ isometries to be linear in $t$ and the parameters of the sphere isometries to be linear in $\varphi$,
\begin{equation} \label{eq:ads5s5-promisom}
\begin{aligned}
\zeta_{12} = \alpha_1 t , \qquad \zeta_{34} = \alpha_2 t , \qquad \psi_{12} = \beta_1 \varphi, \qquad \psi_{34} = \beta_2 \varphi ,
\end{aligned}
\end{equation}
resulting in the following coordinate transformation $x^M \rightarrow \tilde{x}^M$:
\begin{equation} \label{eq:ads5s5-ct}
\begin{alignedat}{3}
z_1 &= \cos (\alpha_1 \tilde t) \tilde z_1 - \sin (\alpha_1 \tilde t) \tilde z_2 , \qquad &&z_2 &&= \cos (\alpha_1 \tilde t)\tilde z_2 + \sin (\alpha_1 \tilde t) \tilde z_1 , \\
z_3 &= \cos (\alpha_2 \tilde t)\tilde z_3 - \sin (\alpha_2 \tilde t)\tilde z_4 , \qquad
&&z_4 &&= \cos (\alpha_2 \tilde t)\tilde z_4 + \sin (\alpha_2 \tilde t)\tilde z_3 , \\
y_1 &= \cos (\beta_1 \tilde \varphi) \tilde y_1 - \sin (\beta_1 \tilde \varphi) \tilde y_2 , \qquad
&&y_2 &&= \cos (\beta_1 \tilde \varphi)\tilde y_2 + \sin (\beta_1 \tilde \varphi)\tilde y_1 , \\
y_3 &= \cos (\beta_2 \tilde \varphi)\tilde y_3 - \sin (\beta_2 \tilde \varphi)\tilde y_4 , \qquad
&&y_4 &&= \cos (\beta_2 \tilde \varphi)\tilde y_4 + \sin (\beta_2 \tilde \varphi)\tilde y_3 ,
\end{alignedat}
\end{equation}
and $t=\tilde{t}$ and $\varphi = \tilde{\varphi}$.
Up to local $H_R$ transformations, one can show that this corresponds to the field redefinition $g\rightarrow\tilde g$ with
\begin{equation}
\begin{aligned}
g&= \exp (J_{05} t + R_{56}\varphi) \ g_X (z,y) , \\
\tilde g &= \exp ((J_{05} + \alpha_1 J_{12} + \alpha_2 J_{34}) \tilde t + (R_{56}+\beta_1 R_{12}+\beta_{2} R_{34}) \tilde \varphi) \ g_X (\tilde z, \tilde y ) ,
\end{aligned}
\end{equation}
thus giving the parametrisation \eqref{eq:lambda-a-s-gen} after dropping the tildes.
Let us note that we do not mix $\tilde{t}$ and $\tilde{\varphi}$ in \eqref{eq:ads5s5-promisom} since we are assuming that $\Lambda_\mathfrak{a}$ and $\Lambda_\mathfrak{s}$ should generically remain elements of $\mathfrak{g}_{\mathfrak{a}}$ and $\mathfrak{g}_{\mathfrak{s}}$ respectively.
With
\begin{equation} \label{eq:s33-xpxm-with-a}
\tilde t = \tilde x^+ - a \tilde{x}^- , \qquad \tilde\varphi = \tilde{x}^+ + (1-a) \tilde{x}^- ,
\end{equation}
for some real parameter $a\in[0,1]$, this means that the resulting light-cone gauge-fixed theory will indeed be a combination of $JT_\sigma$ deformations (due to the promotions linear in $\tilde x^-$) and $J^\tau$ deformations (due to the promotions linear in $\tilde x^+$), as follows from the summary in \secref{s:recap-ilcgf}.
Furthermore, the introduction of the parameter $a$ will correspond to a $T\bar{T}$-deformation.

In ~\secref{s:JT1-S-matrix}, we will verify this at the level of the gauge-fixed Hamiltonian and tree-level S-matrix.
For this, it will be useful to give the explicit expressions of the time components of the currents for the rotational isometries \eqref{eq:so4so4cartanisometries}.
With the definitions \eqref{eq:inf-transf} and \eqref{eq:def-noether-current} and $\lambda \in \{ \zeta_{12},\zeta_{34},\psi_{12},\psi_{34}\}$ we find
\begin{equation} \label{eq:t-currents-cancoords}
\begin{aligned}
J^\tau_{(\mathfrak{a}, 12)} &= z_2 p_{z_1} - z_1 p_{z_2}, \qquad
J^\tau_{(\mathfrak{a},34)} = z_4 p_{z_3} - z_3 p_{z_4} , \\
J^\tau_{(\mathfrak{s},12)} &= y_2 p_{y_1} - y_1 p_{y_2} , \qquad
J^\tau_{(\mathfrak{s},34)} = y_4 p_{y_3} - y_3 p_{y_4} .
\end{aligned}
\end{equation}

\subsubsection{Residual light-cone symmetries} \label{s:res-symms}

Let us now continue with the background \eqref{eq:ads5s5-metric-jtpars} with $\alpha_0 = \beta_0 = 1$ and discuss the residual symmetries of the resulting inequivalent gauge-fixings.
For this, we consider the point-like solution
\begin{equation} \label{eq:point-like-sol}
{x}^+ = \tau , \qquad {x}^- = 0 , \qquad {z}_i = a_i , \qquad {y}_i = b_i , \qquad {\gamma}^{\alpha\beta} = T^{-1} \eta^{\alpha\beta} ,
\end{equation}
with $x^+ = (1-a)t + a \varphi$, $x^- = \varphi - t$ as usual.
Demanding this ansatz solves the equations of motion and the Virasoro constraints, as well as giving vanishing transverse canonical momenta
\unskip\footnote{Recall that this can be achieved by the shift $x^-\to x^-+c_\mu x^\mu$ with constant $c_\mu$, which results in an equivalent gauge-fixing, as explained in~\secref{s:a-lcgf}.}
$\bar{p}_\mu = 0$, we find that we must set $a_i=b_i=0$.
The classical solution then takes precisely the form \eqref{eq:can-momenta-class}.

After fixing the uniform light-cone gauge $x^+=\tau$, $p_-=1$, the residual bosonic time-independent charges of the gauge-fixed theory will come from those $G_L$ transformations that are generated by the centraliser $\mathfrak{c}$ of the abelian algebra generated by $\Lambda_\mathfrak{a}$ and $\Lambda_\mathfrak{s}$.
Depending on the values of the parameters $\alpha_1,\,\alpha_2$ in the $AdS$ sector the centraliser $\mathfrak{c}_\mathfrak{a}$ is given in \tabref{tab:ca}.
For the sphere sector the centraliser $\mathfrak{c}_\mathfrak{s}$ depends on the values of the parameters $\beta_1,\,\beta_2$ and is given in \tabref{tab:cs}.
We have identified these algebras by their dimension, dual Coxeter number and signature.
Furthermore, we have used automorphisms of the centralisers of $\Lambda^{(2)}_\mathfrak{a}$ and $\Lambda^{(2)}_\mathfrak{s}$ in order to reduce their possible embeddings within $\mathfrak{so}(4,2) \cong \mathfrak{su}(2,2)$ or $\mathfrak{so}(6) \cong \mathfrak{su}(4)$.
\unskip\footnote{For example, one can also consider $\Lambda'_\mathfrak{a} = J_{05} + J_{12} + \alpha J_{34}$ or $\Lambda''_\mathfrak{a} = J_{05} - J_{12} + \alpha J_{34}$ with $\alpha$ generic, which will have the same centraliser algebra as that of $\Lambda_\mathfrak{a} = J_{05} + \alpha J_{12} + J_{34}$, though embedded differently in $\mathfrak{so}(4,2) \cong \mathfrak{su}(2,2)$.
The different embeddings can be related by means of automorphisms in the centraliser of $\Lambda^{(2)}_\mathfrak{a}$ and $\Lambda^{(2)}_\mathfrak{s}$, which here is $\mathfrak{so}(4)\oplus \mathfrak{so}(4)$.
For these examples, the cases of $\Lambda_\mathfrak{a}$ and $ \Lambda'_{\mathfrak{a}}$ are related by the automorphism replacing the indices as $(1\leftrightarrow 3, 2 \leftrightarrow 4)$, while the cases of $\Lambda'_\mathfrak{a} $ and $ \Lambda''_{\mathfrak{a}}$ are related by replacing $(1\leftrightarrow 2)$.}
Since we have fixed our choice of $\Lambda^{(2)}_\mathfrak{a}$ and $\Lambda^{(2)}_\mathfrak{s}$ from the beginning we do not allow for more generic automorphisms of $\mathfrak{so}(4,2) \cong \mathfrak{su}(2,2)$ or $\mathfrak{so}(6) \cong \mathfrak{su}(4)$.
This means that, for example, the second and third lines of \tabref{tab:cs} cannot be mapped to each other.

\begin{table}[t]
\centering
\begin{adjustbox}{center}
\begin{tabular}{|c|c|c|c|}
\hline

\rowcolor{Gray}
{ $\alpha_1$} & {$\alpha_2$} & {$\mathfrak{c}_\mathfrak{a}$} & {Basis} \\
\hline

$\alpha_1$ & $\alpha_2$ & $\mathfrak{u}(1)^{\oplus 3}$ & $\{ J_{05} , J_{12} , J_{34} \} $ \\
\hline
$\alpha$ & $ \alpha$
& $\mathfrak{u}(1)^{\oplus 2} \oplus \mathfrak{su}(2)$ & $\{ J_{05} , J_{12} + J_{34} \} \oplus \{ J_{13} + J_{24} , J_{14}- J_{23} , J_{12} - J_{34} \}$ \\
\hline
$\alpha$ & $1$
& $\mathfrak{u}(1)^{\oplus 2} \oplus \mathfrak{su}(1,1)$ & $\{ J_{12} , J_{05} + J_{34} \} \oplus \{ J_{03} + J_{45} , J_{04}- J_{35} , J_{05} - J_{34} \}$ \\
\hline
\multirow{2}{*}{$1$} & \multirow{2}{*}{$1$}
& \multirow{2}{*}{$\mathfrak{u}(1) \oplus \mathfrak{su}(2,1)$ } & $\{ J_{05} + J_{12} + J_{34} \} \oplus \{J_{05}- J_{12} ,J_{34}- J_{12} , J_{14}- J_{23}, \.$ \\
& & & $\.J_{13}+ J_{24} , J_{04}- J_{35}, J_{03}+ J_{45} , J_{02}- J_{15} , J_{01}+ J_{25} \} $ \\
\hline
$0$ & $0$
& $\mathfrak{u}(1) \oplus \mathfrak{su}(2) \oplus \mathfrak{su}(2)$ & $\{ J_{05} \} \oplus \{ J_{ij} \ | \ i,j=1,\ldots,4 \}$ \\
\hline
\end{tabular}
\end{adjustbox}
\caption{The centraliser of $\Lambda_\mathfrak{a} = J_{05} + \alpha_1 J_{12} + \alpha_2 J_{34}$ in $\mathfrak{so}(4,2) \cong \mathfrak{su}(2,2)$.
The first line corresponds to generic $\alpha_1,\alpha_2$.
The $\mathfrak{u}(1)$ elements are all in the centre of $\mathfrak{c}_\mathfrak{a}$.\label{tab:ca}}
\end{table}
\begin{table}[t]
\centering
\begin{adjustbox}{center}
\begin{tabular}{|c|c|c|c|}
\hline

\rowcolor{Gray}
{ $\beta_1$} & $\beta_2$ & {$\mathfrak{c}_\mathfrak{s}$} & {Basis} \\
\hline

$\beta_1$ & $\beta_2$ & $\mathfrak{u}(1)^{\oplus 3}$ & $\{ R_{56} , R_{12} , R_{34} \} $ \\
\hline
$\beta$ & $ \beta$
& $\mathfrak{u}(1)^{\oplus 2} \oplus \mathfrak{su}(2)$ & $\{ R_{56} , R_{12} + R_{34} \} \oplus \{ R_{13} + R_{24} , R_{14}- R_{23} , R_{12} - R_{34} \}$ \\
\hline
$\beta$ & $ 1$
& $\mathfrak{u}(1)^{\oplus 2} \oplus \mathfrak{su}(2)$ & $\{ R_{12} , R_{34} + R_{56} \} \oplus \{ R_{35} + R_{46} , R_{36}- R_{45} , R_{34} - R_{56} \}$ \\
\hline
\multirow{2}{*}{$ 1$} & \multirow{2}{*}{$ 1$}
& \multirow{2}{*}{$\mathfrak{u}(1) \oplus \mathfrak{su}(3)$ } & $\{ R_{56} + R_{12}+ R_{34} \} \oplus \{R_{56}- R_{12} ,R_{36}- R_{45} , R_{35}- R_{46}, \.$ \\
& & & $\.R_{34}- R_{12} , R_{16}- R_{25}, R_{15}+ R_{26} , R_{14}- R_{23} , R_{13}+ R_{24} \} $ \\
\hline
$0$ & $0$
& $\mathfrak{u}(1) \oplus \mathfrak{su}(2) \oplus \mathfrak{su}(2)$ & $\{ R_{56} \} \oplus \{ R_{ij} \ | \ i,j=1,\ldots,4 \}$ \\
\hline
\end{tabular}
\end{adjustbox}
\caption{The centraliser of $\Lambda_\mathfrak{s} = R_{56} + \beta_1 R_{12} + \beta_2 R_{34}$ in $\mathfrak{so}(6) \cong \mathfrak{su}(4)$.
The first line corresponds to $\beta_1,\beta_2$ generic.
The $\mathfrak{u}(1)$ elements are all in the centre of $\mathfrak{c}_\mathfrak{s}$.
}
\label{tab:cs}
\end{table}

Interestingly, there is an enhancement of the residual symmetries for specific points in the moduli space of gauge-fixings.
For generic parameters the symmetry algebra is the smallest possible.
For $\alpha_{1,2}=\beta_{1,2}=0$ we recover the bosonic $ \mathfrak{u}(1)^{\oplus 2} \oplus \mathfrak{su}(2)^{\oplus 2} \oplus \mathfrak{su}(2)^{\oplus 2}$ symmetry algebra (with $\Lambda_{\mathfrak{a}}=J_{05}$ and $\Lambda_{\mathfrak{s}}=R_{56}$ corresponding to the central $\mathfrak{u}(1)^{\oplus 2}$) of the standard light-cone gauge-fixed theory \cite{Arutyunov:2006ak,Arutyunov:2009ga}, which is 14-dimensional.
An intriguing case is $\alpha_{1,2}=\beta_{1,2}=1$ leading to the largest number of bosonic symmetries, namely the 18-dimensional $\mathfrak{u}(1)^{\oplus 2} \oplus \mathfrak{su}(2,1)\oplus \mathfrak{su}(3) $ algebra, where the $\mathfrak{u}(1)^{\oplus 2}$ elements are again in the centre and given by $\Lambda_{\mathfrak{a}}=J_{05}+J_{12}+J_{34}$ and $\Lambda_{\mathfrak{s}}=R_{56}+R_{12}+R_{34}$.

In the $AdS_5\times S^5$ superstring setting, where $\mathfrak{g} = \mathfrak{psu}(2,2|4)$, the bosonic residual symmetry will be further enhanced with supercharges.
In the light-cone gauge with $a=\frac{1}{2}$, the bosonic \textit{and} fermionic generators that give rise to charges independent of $x^+ = \tau$ have to commute with $\Lambda_+ = \Lambda_{\mathfrak{a}}+\Lambda_{\mathfrak{s}}$.
We will call this superalgebra $\mathfrak{c}_+$.
Further specifying the gauge with $\alpha_{1,2}=\beta_{1,2}=0$ leads to the usual $\mathfrak{c}_+=\mathfrak{u}(1)^{\oplus 2} \oplus \mathfrak{psu}(2|2)^{\oplus 2}$ superalgebra of light-cone symmetries, which has in total 8 complex supercharges \cite{Arutyunov:2006ak,Arutyunov:2009ga}.
For the case with $\alpha_{1,2}=\beta_{1,2}=1$ we find the following algebra embedded in $\mathfrak{psu}(2,2|4)$
\begin{equation} \label{eq:largest-lcsymms}
\mathfrak{c}_+=\mathfrak{p}(\mathfrak{u}(1|1) \oplus \mathfrak{su}(2,1|3)) ,
\end{equation}
where we quotient out by the identity $\mathbb{1}_8$.
\unskip\footnote{It would be interesting to explore connections with non-relativistic string theories and spin matrix theories in zero-temperature critical limits of ${\cal N}=4$ super-Yang-Mills where similar symmetry subgroups appear, see, e.g.~\cite{Harmark:2014mpa}.}
This algebra has 10 complex supercharges, of which one is in $\mathfrak{u}(1|1)$.
Its structure is most easily obtained working in a representation of $\mathfrak{su}(2,2|4)$ in which $\Lambda_{\mathfrak{a}}$ and $\Lambda_{\mathfrak{s}}$ are diagonal, and the reality condition reads
\begin{equation}
M^\dag {\cal I} + {\cal I} M = 0 , \qquad {\cal I} = \text{diag}(-1,1,1,-1 \ |\ 1,1,1,1) ,
\end{equation}
for all $M\in \mathfrak{psu}(2,2|4)$.
The matrix realisation of the $\mathfrak{c}_+$ superalgebra \eqref{eq:largest-lcsymms} then schematically is
\begin{equation}
\begin{aligned}
\mathfrak{u}(1|1) &= \text{span}\left\{ \left(\begin{array}{c|c}
\Lambda_{\mathfrak{a}}& \\\hline
& ~~~~
\end{array} \right), \left(\begin{array}{c|c}
~~~~ & \\\hline
& \Lambda_{\mathfrak{s}}
\end{array} \right) , \left(\begin{array}{cc|cc}
& & \theta & ~~ \\
& & & \mathbf{0}_3 \\ \hline
\theta^\dag& ~~ & & \\
& \mathbf{0}_3& &
\end{array} \right) \right\} , \\[.1cm]
\mathfrak{su}(2,1|3) &= \text{span} \left\{ \left(\begin{array}{cc|cc}
0& & & \\
& L & & \\ \hline
& & ~ & \\
& & &~~~~
\end{array} \right), \left(\begin{array}{cc|cc}
~& & & \\
& ~~~~ & & \\ \hline
& & 0& \\
& & & R
\end{array} \right) ,
\left(\begin{array}{cc|cc}
& & 0 & ~~ \\
& & & - {\tt I} Q^\dag \\ \hline
0& ~~ & & \\
& ~~Q~~ & &
\end{array} \right) \right\} ,
\end{aligned}
\end{equation}
where $L\in \mathfrak{su}(2,1)$, $R\in \mathfrak{su}(3)$, $Q\in \Complex^{3\times 3}$, $\theta\in \Complex$ and $\mathtt{I}=\text{diag}(1,1,-1)$.
To work with explicit matrix realisations (before diagonalisation of $\Lambda_{\mathfrak{a}}, \Lambda_{\mathfrak{s}}$ and ${\cal I}$) we refer, e.g.,~to appendix B of~\cite{Borsato:2022ubq} (see also~\cite{Minahan:2010js,Arutyunov:2009ga}).

One can repeat a similar exercise for the other cases in \tabref{tab:ca} and \tabref{tab:cs}.
Already for $a=\frac{1}{2}$ there are many possible combinations of $\Lambda_+ = \Lambda_{\mathfrak{a}}+\Lambda_{\mathfrak{s}}$ to consider, but many of these choices lead to a centraliser $\mathfrak c_+$ with no supercharges.

\section{Effect of inequivalent light-cone gauges on the S-matrix}\label{sec:Smat}

In this section we analyse the effect of the different light-cone gauge-fixings on the perturbative and exact worldsheet S-matrix.
Taking into account our motivations, we will focus on the case of factorised scattering.
Therefore, we only need to consider the $2\to 2$ S-matrix.
The arguments are generalisable beyond this case, however we will not consider this here avoiding subtleties that do not arise in our setup.

After decompactifying the worldsheet, the S-matrix relates incoming states at time $\tau=-\infty$ with outgoing states at $\tau=+\infty$.
These asymptotic states are thought of as collections of wave-packets that have a well-defined momentum and are well-separated.
On the spatial line they can be ordered, and for incoming states we take
\begin{equation}
\ket{p_1,p_2,\ldots,p_N}^{in}_{\mu_1,\mu_2,\ldots,\mu_N},
\end{equation}
where $p_1>p_2>\ldots>p_N$.
In this way, each incoming particle has a right-moving momentum greater than the momenta of the particles to its right, hence all the particles will scatter with each other.
Here $\mu_1,\mu_2,\ldots,\mu_N$ are labels that identify the possible different flavours of the $N$ particles.
Because of the restriction to the case of factorised scattering, the outgoing particles will have the same set of momenta as the incoming particles but, because scattering has occurred, their ordering will be reversed
\begin{equation}
\ket{p_N,p_{N-1},\ldots,p_1}^{out}_{\nu_N,\nu_{N-1},\ldots,\nu_1}.
\end{equation}
Here $\nu_N,\nu_{N-1},\ldots,\nu_1$ label the flavours of the outgoing particles.
We will therefore work in the convention that the S-matrix reduces to the permutation operator when turning off interactions.
In the case of the two-body S-matrix, for example, we write a generic entry as $S_{\mu_1\mu_2}^{\nu_2\nu_1}$, where $\mu_1,\mu_2$ are the flavours of the incoming particles with momenta $p_1,p_2$ respectively, and $\nu_2,\nu_1$ are the flavours of the outgoing particles with momenta $p_2,p_1$ respectively.
The non-trivial part of the S-matrix is given by the T-matrix $\mathbf T$ defined as $S=\Pi+\frac{i}{T}\mathbf T$, where $\Pi$ is the (graded) permutation.

We will now focus on the four inequivalent gauge-fixings summarised in~\secref{s:recap-ilcgf}.
We will carry out an analysis at tree level for $AdS_5 \times S^5$, before giving arguments for the non-perturbative S-matrix.

\subsection{Tree-level} \label{s:treelevel-checks}

\subsubsection{\texorpdfstring{$T\bar{T}$}{TTbar}, \texorpdfstring{$JT_\sigma$}{JTσ} and \texorpdfstring{$J^\tau$}{Jτ} deformations: \texorpdfstring{$\AdS_5 \times \Sp^5$}{AdS5 × S5} tree-level S-matrix} \label{s:JT1-S-matrix}

We first illustrate the effect of the inequivalent gauge-fixings discussed in \secref{sec:gauges-curr-curr} on the perturbative S-matrix for bosonic strings propagating in $\AdS_5 \times \Sp^5$, focusing on the $T\bar{T}$, $JT_\sigma$ and $J^\tau$ deformations.
The analysis can be straightforwardly generalised to $AdS_n\times S^n$ with different $n$.
Our starting point is the metric \eqref{eq:ads5s5-metric-jtpars}, with $\alpha_0=\beta_0=1$ and free deformation parameters $\alpha_1,\alpha_2$ and $\beta_1,\beta_2$.
We slightly generalise the light-cone gauge-fixing discussed in \appref{sec:appendix} by including the gauge parameter $a \in[0,1]$ as in \eqref{eq:s33-xpxm-with-a}, so that
\begin{equation} \begin{aligned}
x^+ &= (1-a) t + a \varphi \ \gf \ \tau, &\qquad p_- &= -a p_t + (a-1) p_\varphi \ \gf \ 1, \\
x^- &= - t +\varphi, &\qquad p_+ &= p_t + p_\varphi ,
\end{aligned}
\end{equation}
as is compatible with the classical solution \eqref{eq:point-like-sol} (or equivalently \eqref{eq:can-momenta-class}).

\paragraph{Complex coordinates.} The effect of the deformation is best seen in a basis of eigenstates of the charges associated with the currents \eqref{eq:t-currents-cancoords}.
As discussed in \secref{s:AdSn-Sn-JT}, on the real transverse coordinates $(z_j,y_j)$, $j=1,2,3,4$, the symmetries act as rotations.
This motivates the introduction of the complex fields,
\begin{equation} \begin{aligned}
u_1 &= \frac{1}{\sqrt{2}} (y_1 + i y_2), &\, \bar{u}_1 &= \frac{1}{\sqrt{2}} (y_1 - i y_2), &\, u_2 &= \frac{1}{\sqrt{2}} (y_3 + i y_4), &\, \bar{u}_2 &= \frac{1}{\sqrt{2}} (y_3 - i y_4), \\
u_3 &= \frac{1}{\sqrt{2}} (z_1 + i z_2), &\, \bar{u}_3 &= \frac{1}{\sqrt{2}} (z_1 - i z_2), &\, u_4 &= \frac{1}{\sqrt{2}} (z_3 + i z_4), &\, \bar{u}_4 &= \frac{1}{\sqrt{2}} (z_3 - i z_4),
\end{aligned}
\end{equation}
with canonically conjugate momenta
\begin{equation} \begin{aligned}
P_{u_1}&= \frac{1}{\sqrt{2}} (p_{y_1} - i p_{y_2}), \qquad P_{\bar{u}_1} = \frac{1}{\sqrt{2}} (p_{y_1} + i p_{y_2}) = \bar{P}_{u_1}, \\
P_{u_2}&= \frac{1}{\sqrt{2}} (p_{y_3} - i p_{y_4}), \qquad P_{\bar{u}_2} = \frac{1}{\sqrt{2}} (p_{y_3} + i p_{y_4}) = \bar{P}_{u_2}, \\
P_{u_3}&= \frac{1}{\sqrt{2}} (p_{z_1} - i p_{z_2}), \qquad P_{\bar{u}_3} = \frac{1}{\sqrt{2}} (p_{z_1} + i p_{z_2}) = \bar{P}_{u_3}, \\
P_{u_4}&= \frac{1}{\sqrt{2}} (p_{z_3} - i p_{z_4}), \qquad P_{\bar{u}_4} = \frac{1}{\sqrt{2}} (p_{z_3} + i p_{z_4}) = \bar{P}_{u_4},
\end{aligned}
\end{equation}
and we refer to the transverse fields and conjugate momenta collectively by $u_j, \bar{u}_j$ and $P_{u_j}, \bar{P}_{u_j}$ with $j=1,2,3,4$.
It will also be convenient to rename the currents of \eqref{eq:t-currents-cancoords} as
\begin{equation} \label{eq:t-currents-new}
J_1 = J_{(\alg{s},12)}, \qquad J_2 = J_{(\alg{s},34)}, \qquad J_3 = J_{(\alg{a},12)}, \qquad J_4 = J_{(\alg{a},34)},
\end{equation}
and identify
\begin{equation}
\beta_3 \equiv \alpha_1, \qquad \beta_4 \equiv \alpha_2.
\end{equation}

\paragraph{Light-cone gauge-fixed Hamiltonian.}
The light-cone gauge-fixed Hamiltonian density $\mathcal H$ can be computed as explained in \appref{sec:appendix}.
It admits an expansion in powers of the transverse fields.
For the case at hand the expansion starts at quadratic order and only includes terms with an even number of transverse fields, $\mathcal H = \mathcal H_2 +\mathcal H_4 + \dots$.
The quadratic Hamiltonian density is given by
\begin{equation}
\begin{aligned} \label{eq:H2-def-AdS5}
\mathcal H_2 &= \mathcal H_2^0 + \sum_{j=1}^4 \beta_j J^\tau_j ,
\end{aligned}
\end{equation}
where the undeformed quadratic Hamiltonian density describes a collection of four free complex fields,
\begin{equation}
\mathcal H_2^0 = \sum_{j=1}^4 \left( |P_{u_j}|^2 + |u'_j|^2 + |u_j|^2\right),
\end{equation}
and the currents \eqref{eq:t-currents-cancoords} (with the notation \eqref{eq:t-currents-new}) read
\begin{equation} \label{eq:Jtau-def-AdS5}
\begin{aligned}
J_j^\tau &= - i (P_{u_j} u_j - \bar{P}_{u_j} \bar{u}_j) .
\end{aligned}
\end{equation}
As expected, these coincide with the $\tau$-component of the currents associated to the four $\mathfrak{u}(1)$ symmetries of $\mathcal H_2^0$, realised as
\begin{equation}
u_j \rightarrow e^{i \beta_j} u_j, \qquad j=1,2,3,4.
\end{equation}
The $\sigma$-components of these currents are given by
\begin{equation} \label{eq:Jsigma-def-AdS5}
\begin{aligned}
J^\sigma_{j} &= i (u_j \bar{u}'_j - \bar{u}_j u'_j).
\end{aligned}
\end{equation}
The undeformed quadratic Hamiltonian (or rather its associated quadratic Lagrangian), is also invariant under shifts of $\tau$ and $\sigma$.
The conserved current associated to these symmetries is simply the energy-momentum tensor, whose explicit form (to quadratic order) is
\begin{gather}
T^\tau{}_\tau = \sum_{j=1}^4 \left(|P_{u_j}|^2+ |u'_j|^2+ |u_j|^2 \right), \qquad
T^\sigma{}_\sigma = \sum_{j=1}^4 \left(- |P_{u_j}|^2- |u'_j|^2+ |u_j|^2\right), \\
T^\tau{}_\sigma = -T^{\sigma}{}_{\tau} = \sum_{j=1}^4 \left(P_{u_j} u'_j + \bar{P}_{u_j} \bar{u}'_j \right).
\end{gather}
From these conserved currents we then construct the $JT_\sigma$ and $T\bar{T}$ operators
\begin{equation}
O_{J T_\sigma}^j = -\epsilon_{\alpha\beta} T^\alpha{}_\sigma J^\beta_j , \qquad O_{T \bar{T}} = - \epsilon_{\alpha\beta} T^\alpha{}_\sigma T^\beta{}_\tau ,
\end{equation}
where we recall our convention for the antisymmetric tensor $\epsilon^{\tau \sigma} = - \epsilon_{\tau \sigma}=-1$.
The four $J T_\sigma$ operators correspond to the four currents~\eqref{eq:t-currents-new}.
The quartic Hamiltonian density can then be written
\begin{equation} \label{eq:H4-def-AdS5}
\mathcal H_4 = \mathcal H_4^0 + (1-a) \beta_1 O_{J T_\sigma}^{1} + (1-a) \beta_2 O_{J T_\sigma}^{2}- a \beta_3 O_{J T_\sigma}^{3} - a \beta_4 O_{J T_\sigma}^{4} - \left(a-\frac{1}{2} \right) O_{T \bar{T}},
\end{equation}
where the undeformed quartic Hamiltonian is
\begin{equation}
\begin{aligned}
\mathcal H_4^0 &= \left(|u_3|^2 + |u_4|^2 \right)\left(2|u'_3|^2+ 2|u'_4|^2+|P_{u_1}|^2 + |P_{u_2}|^2 + |u'_1|^2 + |u'_2|^2 \right)\\
&\qquad -\left(|u_1|^2 + |u_2|^2\right) \left(2 |u'_1|^2+ 2 |u'_2|^2 +|P_{u_3}|^2 + |P_{u_4}|^2 + |u'_3|^2 + |u'_4|^2 \right).
\end{aligned}
\end{equation}
We therefore see that the way the quadratic and quartic Hamiltonians are deformed by the parameters $\beta_j$ precisely matches with the discussion in \secref{sec:gauges-curr-curr}.

\paragraph{Oscillator expansion.}
To solve the Hamilton equations of motion associated to $\mathcal H_2$ and quantise the fields we introduce the oscillator expansion
\begin{align}
u_{j} &= \frac{1}{\sqrt{2 \pi}} \int d p \frac{1}{\sqrt{ 2\omega}} \left( e^{-i \omega_{j,+} \tau+ i p \sigma} a_{j,+}(p) + e^{i \omega_{j,-}\tau- i p \sigma} a_{j,-}^\dagger(p) \right), \\
\bar{u}_j &= \frac{1}{\sqrt{2 \pi}} \int d p \frac{1}{\sqrt{2 \omega}} \left( e^{-i \omega_{j,-} \tau+ i p \sigma} a_{j,-}(p) + e^{i \omega_{j,+}\tau- i p \sigma} a_{j,+}^\dagger(p) \right),
\end{align}
with the relativistic and shifted dispersion relation
\begin{equation}
\omega = \sqrt{p^2+1}, \qquad \omega_{j,\pm} = \omega \pm \beta_j, \qquad j=1,2,3,4.
\end{equation}
The annihilation and creation operators satisfy the canonical commutation relations (with all the other commutation relations vanishing)
\begin{equation}
\com{a_{j,\pm}(p)}{a^\dagger_{k,\pm}(q)} = \delta_{j k} \delta(p-q), \qquad j,k=1,2,3,4.
\end{equation}
The oscillator representation of the canonically conjugate momenta directly follows from the equations of motion, giving
\begin{align}
\bar{P}_{u_j} = \partial_\tau u_j + i \beta_j u_j = \frac{1}{\sqrt{2 \pi}} \int dp \frac{1}{\sqrt{2 \omega}} (-i \omega) \left( e^{-i \omega_{j,+} \tau+ i p \sigma} a_{j,+}(p) - e^{i \omega_{j,-}\tau- i p \sigma} a_{j,-}^\dagger(p) \right), \\
P_{u_j} = \partial_\tau \bar{u}_j - i \beta_j \bar{u}_j=\frac{1}{\sqrt{2 \pi}} \int dp \frac{1}{\sqrt{2 \omega}} (-i \omega) \left( e^{-i \omega_{j,-} \tau+ i p \sigma} a_{j,-}(p) - e^{i \omega_{j,+}\tau- i p \sigma} a_{j,+}^\dagger(p) \right).
\end{align}
Note that, while the exponents in the plane-wave ansatz depend on the shifted energies $\omega_{j,\pm}$, since the momentum is not just given by the $\tau$-derivative of the corresponding field, but also includes a contribution from the $J^\tau_j$ deformation in the quadratic Hamiltonian \eqref{eq:H2-def-AdS5}, the shift is precisely cancelled.
This explains why the normalisation of the fields and momenta depends on the relativistic dispersion $\omega$.
With these expressions for the fields and momenta in terms of oscillators, the quadratic Hamiltonian takes the canonical form,
\begin{equation}
H_2 = \int d \sigma \mathcal H_2 = \int dp \sum_{j=1}^4 \sum_{s = \pm} \left(\omega_{j,s} a^\dagger_{j,s} a_{j,s} \right),
\end{equation}
while the charges are
\begin{equation}
Q_k = \int d \sigma J_{k}^\tau = \int dp \sum_{j=1}^4 \sum_{s = \pm} \left( s \, \delta_{jk} \, a^\dagger_{j,s} a_{j,s} \right).
\end{equation}
These results are summarised in \tabref{table:particles-def-AdS5}.
\begin{table} \centering
\begin{tabular}{|c|c|c|c|c|c|} \hline \rowcolor{Gray}
State & $H_2$ & $Q_1$ & $Q_2$ & $Q_3$ & $Q_4$ \\ \hline
$\ket{p}_{1,\pm} = a_{1, \pm}^\dagger(p) \ket{0}$ & $\omega_{1,\pm} = \omega \pm \beta_1$ & $\pm 1$ & 0 & 0 & 0 \\
$\ket{p}_{2,\pm} = a_{2, \pm}^\dagger(p) \ket{0}$ & $\omega_{2,\pm} = \omega \pm \beta_2$ & 0 & $\pm 1$ & 0 & 0 \\
$\ket{p}_{3,\pm} = a_{3, \pm}^\dagger(p) \ket{0}$ & $\omega_{3,\pm} = \omega \pm \beta_3$ & 0 & 0 & $\pm 1$ & 0 \\
$\ket{p}_{4,\pm} = a_{4, \pm}^\dagger(p) \ket{0}$ & $\omega_{4,\pm} = \omega \pm \beta_4$ & 0 & 0 & 0 & $\pm 1$ \\[.1cm] \hline
\end{tabular}
\caption{This table summarises the particle content in the light-cone gauge-fixed $\AdS_5 \times \Sp^5$ theory.
Eight different states can be created from the vacuum using the eight different creation operators.
These states are eigenstates of the quadratic Hamiltonian $H_2$ and the four charges $Q_j$, with eigenvalues as given in the table.}
\label{table:particles-def-AdS5}
\end{table}

\paragraph{Tree-level S-matrix.} Plugging the oscillator expansion into the quartic Hamiltonian \eqref{eq:H4-def-AdS5} gives terms involving four oscillators of the form
\begin{equation} \begin{aligned}
H_4 = \int d \sigma \, \mathcal H_4 &= \int dp_1 dp_2 dp_3 dp_4 \, \mathbf T_{i s_1 , j s_2}^{l s_4 , k s_3} a^\dagger_{l,s_4}(p_4) a^\dagger_{k,s_3}(p_3)a_{j,s_2}(p_2) a_{i,s_1}(p_1) \\
&\qquad \times \delta(p_1+p_2-p_3-p_4) \delta(\omega_1 + \omega_2 - \omega_3 - \omega_4),
\end{aligned}
\end{equation}
where we use the shorthand
\begin{equation}
\omega_1 = \omega_{i,s_1}(p_1), \qquad \omega_2 = \omega_{j,s_2}(p_2), \qquad \omega_3 = \omega_{k,s_3}(p_3),\qquad \omega_4 = \omega_{l,s_4}(p_4).
\end{equation}
Only terms with equal number of creation and annihilation operators contribute, which is a consequence of the integrability of the model.
From this we can read off the tree-level S-matrix $S=\Pi+\frac{i}{T}{\mathbf T}$
with the non-trivial elements given by
\begin{equation} \label{eq:T-def-AdS5}
\begin{alignedat}{3}
\mathbf T_{i s_1, j s_2}^{l s_4 , k s_3} &= (+2 \mathcal A+\mathcal O_{is_1 js_2}) \delta_{i}^k \delta_{j}^l \delta_{s_1}^{s_3} \delta_{s_2}^{s_4} + \mathcal B \left(\delta_i^k \delta_j^l \delta_{s_1}^{s_3} \delta_{s_2}^{s_4} + \delta_i^l \delta_j^k \delta_{s_1}^{s_4} \delta_{s_2}^{s_3} \right), \, &&i,j,k,l=1,2, \quad &&s_1=s_2, \\
\mathbf T_{i s_1, j s_2}^{l s_4 , k s_3} &= (+2 \mathcal A+\mathcal O_{is_1 js_2}) \delta_{i}^k \delta_{j}^l \delta_{s_1}^{s_3} \delta_{s_2}^{s_4} + \mathcal B \left(\xi_{il} \xi_{jk} \delta_{s_1}^{s_3} \delta_{s_2}^{s_4} + \xi_{ik} \xi_{jl} \delta_{s_1}^{s_4} \delta_{s_2}^{s_3} \right), ~~~ &&i,j,k,l=1,2, \quad &&s_1=-s_2, \\
\mathbf T_{i s_1, j s_2}^{l s_4 , k s_3} &= (-2 \mathcal A+\mathcal O_{is_1 js_2}) \delta_{i}^k \delta_{j}^l \delta_{s_1}^{s_3} \delta_{s_2}^{s_4} - \mathcal B \left(\delta_i^k \delta_j^l \delta_{s_1}^{s_3} \delta_{s_2}^{s_4} + \delta_i^l \delta_j^k \delta_{s_1}^{s_4} \delta_{s_2}^{s_3} \right), \, &&i,j,k,l=3,4, \quad &&s_1=s_2, \\
\mathbf T_{i s_1, j s_2}^{l s_4 , k s_3} &= (-2 \mathcal A+\mathcal O_{is_1 js_2}) \delta_{i}^k \delta_{j}^l \delta_{s_1}^{s_3} \delta_{s_2}^{s_4} - \mathcal B \left(\xi_{il} \xi_{jk} \delta_{s_1}^{s_3} \delta_{s_2}^{s_4} + \xi_{ik} \xi_{jl} \delta_{s_1}^{s_4} \delta_{s_2}^{s_3} \right), \, &&i,j,k,l=3,4, \quad &&s_1=-s_2, \\
\mathbf T_{i s_1, j s_2}^{l s_4 , k s_3} &= (+2 \mathcal G+\mathcal O_{is_1 js_2}) \delta_{i}^k \delta_{j}^l \delta_{s_1}^{s_3} \delta_{s_2}^{s_4}, \, &&i,k=1,2, \qquad &&j,l =3,4, \\
\mathbf T_{i s_1, j s_2}^{l s_4 , k s_3} &= (-2 \mathcal G+\mathcal O_{is_1 js_2}) \delta_{i}^k \delta_{j}^l \delta_{s_1}^{s_3} \delta_{s_2}^{s_4}, \, &&i,k=3,4, \qquad &&j,l =1,2,
\end{alignedat}
\end{equation}
where
\begin{equation}
\begin{gathered}
\mathcal A = \frac{1}{4} \frac{(p_1-p_2)^2}{p_1 \omega_2 - p_2 \omega_1}, \qquad \mathcal B = \frac{p_1 p_2}{p_1 \omega_2 - p_2 \omega_1}, \qquad \mathcal G = -\frac{1}{4}(p_1 \omega_2 + p_2 \omega_1), \\
\mathcal O_{is_1 js_2}= - a \sum_{n=3}^4 \beta_n (\delta_j^n s_2 p_1 - \delta_i^n s_1 p_2)+(1-a) \sum_{n=1}^2 \beta_n (\delta_j^n s_2 p_1 - \delta_i^n s_1 p_2)-\left(a-\frac{1}{2}\right) (\omega_2 p_1 - \omega_1 p_2),
\end{gathered}
\end{equation}
and quantity $\xi$ is defined such that its only non-vanishing components are
\begin{equation} \label{eq:eps-convs}
\xi_{12} = \xi_{21} = \xi_{34} = \xi_{43}=1.
\end{equation}
The terms involving $\mathcal A$, $\mathcal B$ and $\mathcal G$ reproduce the standard tree-level S-matrix of the bosonic $\AdS_5 \times \Sp^5$ string in the $a=1/2$ gauge.
The effect of the free parameters characterising different gauge choices is gathered in the contribution $\mathcal O$.
This contribution only modifies the term proportional to the permutation operator (which in our conventions corresponds to free propagation).
One can check explicitly that the tree-level S-matrix \eqref{eq:T-def-AdS5} satisfies charge conservation for $Q_j$ using that the only non-vanishing scattering processes obey $s_1 + s_2 = s_3 + s_4$.
Finally, let us conclude by mentioning that the tree-level S-matrix still satisfies the classical Yang-Baxter equation, indicating that the model is classically integrable for all choices of light-cone gauge-fixing as expected.
This will be made more rigorous when we consider the exact S-matrix in \secref{sec:np}, of which the tree-level S-matrix calculated here is the first order in the large tension expansion.

\subsubsection{\texorpdfstring{$\tilde{J}T_\tau$}{JtildeTτ} deformation}\label{s:treelevel-checks-Jtilde}

To illustrate the effect of the $\tilde{J}T_\tau$ deformation on the tree-level S-matrix we start with the metric of (undeformed) $\AdS_5 \times \Sp^5$ and perform a shift
\begin{equation}
t \rightarrow t + c(x^\mu), \qquad \varphi \rightarrow \varphi + c(x^\mu),
\end{equation}
with a function $c(x^\mu)$ of the (real) transverse coordinates $x^\mu=(z_1,z_2,z_3,z_4,y_1,y_2,y_3,y_4)$.
For concreteness we assume that this function can be expanded in powers of the transverse fields, starting at linear order,
\begin{equation} \label{eq:c-def-AdS5}
c = c_1 + c_2 + \dots = \beta_\mu x^\mu + \beta_{\mu \nu} x^{\mu} x^{\nu} + \dots,
\end{equation}
with free parameters $\beta_\mu, \beta_{\mu \nu}, \dots$.
For the purpose of computing the tree-level S-matrix we use the light-cone gauge-fixed Hamiltonian up to quartic order in the fields, hence it is sufficient to consider the expansion of $c$ up to quadratic order.

According to the discussion in \secref{s:a-lcgf} (see also eq.~\eqref{eq:dH-Jt-0}) we expect the light-cone gauge-fixed Hamiltonian to change as
\begin{equation}
\delta \mathcal H = \partial_\alpha c T^\alpha{}_\tau = - c \, \partial_\alpha T^\alpha{}_\tau + \text{total derivatives}.
\end{equation}
Up to total derivatives, the variation of the Hamiltonian therefore vanishes on-shell.
Therefore, in general we expect that $\delta {\cal H}$ can be removed by means of a field redefinition or canonical transformation of the transverse variables.
Let us illustrate this explicitly at leading order in fields for the function $c$ in eq.~\eqref{eq:c-def-AdS5}.
The quadratic Hamiltonian does not depend on the parameters $\beta_\mu$, $\beta_{\mu \nu}$, $\dots$ and simply reads
\begin{equation}
\mathcal H_2 = \mathcal H_2^0 = \frac{1}{2}(p_\mu p_\mu + x'_\mu x'_\mu+x_\mu x_\mu),
\end{equation}
whose associated equations of motion are
\begin{equation}
\dot{x}_\mu = p_\mu, \qquad \dot{p}_\mu = - x_\mu + x_\mu'', \qquad \Rightarrow \qquad \mathcal E_\mu := \ddot{x}_\mu - x''_\mu + x_\mu =0.
\end{equation}
The Hamiltonian now also has a cubic term,
\begin{equation}
\mathcal H_3 = (\left.\partial_\mu c \right|_{x=0}) \partial_\alpha x^\mu T^\alpha{}_\tau = \beta_\mu \left( p_\mu \mathcal H_2 - x'_\mu p_\nu x'_\nu \right),
\end{equation}
where in the energy-momentum tensor is computed from the quadratic Hamiltonian $\mathcal H_2$.
To see that this cubic contribution can be removed by an appropriate field redefinition, we switch to the Lagrangian formalism.
After integrating by parts, the cubic contribution can be written in terms of the equations of motion as
\begin{equation}
\mathcal L_2 = \frac{1}{2} ( \dot{x}_\mu \dot{x}_\mu - x'_\mu x'_\mu - x_\mu x_\mu), \qquad \mathcal L_3 = - \beta_\mu x^\mu \mathcal E_\nu \dot{x}^\nu.
\end{equation}
This can be removed using the field redefinition
\begin{equation} \label{eq:redef-tJT0}
x^\mu \rightarrow x^\mu + \beta_\nu x^\nu \dot{x}^\mu.
\end{equation}
In the Hamiltonian formalism the redefinition becomes
\begin{equation}
x_\mu \rightarrow x_\mu + \beta_\nu x^\nu p_\mu, \qquad p_\mu \rightarrow p_\mu - \beta_\mu \mathcal H_2 - \beta_\nu x^\nu x_\mu,
\end{equation}
which mixes fields $x_\mu$ and momenta $p_\mu$.
One can check that this corresponds to a canonical transformation to first order in the fields, meaning that $\{x_\mu, p_\nu\} = \delta_{\mu \nu} + \dots$ where the ellipses denote terms that are at least quadratic in the fields.
Interestingly, we can understand this canonical transformation as an improved version of~\eqref{eq:tr-red-Jtilde-Ttau}, which is the non-canonical transformation corresponding to the $\tilde JT_\tau$ deformation, which here we want to neutralise.

The canonical transformation is such that $\delta \mathcal H_2 + \mathcal H_3 =0$.
The quartic light-cone gauge-fixed Hamiltonian, from which the tree-level S-matrix is deduced, is then given by $\delta \mathcal H_3 + \mathcal H_4$.
We find that the ${\mathbf T}$-matrix obtained from the resulting quartic Hamiltonian does not depend on the function $c(x^\mu)$, as expected from the general results of \secref{s:JTildeT0RxSn}.
\unskip\footnote{In principle, one can also verify this without using field redefinitions.
${\cal H}_3$ vanishing on-shell ensures that the $1\rightarrow 2$ and $2\rightarrow 1$ processes vanish.
However, to compute the $2\rightarrow 2$ S-matrix, one needs to consider diagrams involving two cubic vertices, i.e.~with four external particles and one internal particle, as well as quartic diagrams.}
This suggests that the $2\rightarrow 2$ S-matrix does not depend on a change of gauge that induces a $\tilde{J} T_\tau$ deformation.
We will argue that this is indeed the case at the level of the exact S-matrix in the next section.
Finally, let us mention that when the function $c$ starts at quadratic order in fields, then ${\cal H}_3 = 0$, while the variation of ${\cal H}_4$ vanishes on-shell and thus the $2\rightarrow 2$ S-matrix is manifestly independent of $c$.

\subsection{Non-perturbative}\label{sec:np}

Following on from the explicit tree level calculations, our aim is to now understand the effect of the gauge transformations on the S-matrix non-perturbatively.
In order to do so, we first note that the inequivalent gauge transformations come in two types: they are either bilinear in the currents (the $\tilde J T_\tau$, the $T\bar T$ and the $JT_\sigma$ deformations) or linear (the $J^\tau$ deformation).
Here we analyse the two cases separately.

Before turning to the details of the arguments, let us summarise the result of the gauge transformations on the S-matrix.
If the Hamiltonian of the reduced model is deformed by a current-current deformation
\begin{equation}
\delta \mathcal H = -\gamma \epsilon_{\alpha\beta} J^\alpha_1 J^\beta_2,
\end{equation}
where $J^\alpha_i$ with $i=1,2$ are conserved currents, then the S-matrix $\tilde S$ of the deformed model is related to the undeformed S-matrix $S$ simply as
\begin{equation} \label{eq:smatrix-currcurr}
\tilde S_{\mu_1\mu_2}^{\nu_2\nu_1}
= e^{-\frac{i\gamma}{2}\epsilon^{ij}(q_i^{\nu_1}q_j^{\nu_2}+q_i^{\mu_1}q_j^{\mu_2})}S_{\mu_1\mu_2}^{\nu_2\nu_1}.
\end{equation}
Here $\epsilon^{12}=1$ and $q_i^{\mu_1}$, for example, denotes the charge $i$ (corresponding to the current $J_i$) of the particle with flavour $\mu_1$.
Our argument will only use the fact that the currents $J_i$ are conserved.
In particular, they may be Noether currents for spacetime or internal symmetries, topological currents, or any other kind of conserved current.
The above formula agrees with known deformations of the S-matrix in the case of the $T\bar T$ deformation~\cite{Dubovsky:2017cnj,Baggio:2018gct,Frolov:2019nrr,Sfondrini:2019smd,Arutyunov:2009ga}, the $JT$ deformation~\cite{Anous:2019osb}, as well as TsT deformations~\cite{Dubovsky:2023lza}.
\unskip\footnote{It is well known that TsT deformations are the integrated version of current-current deformations, where the currents correspond to global internal Noether symmetries of the \sm~\cite{Hassan:1992gi,Henningson:1992rn,Dubovsky:2023lza}, see also the review~\cite{Borsato:2023dis}.
If we consider a \sm{} and perform a TsT deformation along transverse fields only, then the Hamiltonian density of the light-cone gauge-fixed model is indeed deformed as $\delta \mathcal H = -\gamma \epsilon_{\alpha\beta} J^\alpha_1 J^\beta_2$.
See~\cite{vanTongeren:2021jhh} for examples with TsT deformations also involving the light-cone directions $x^\pm$.}
It also agrees with the results of~\cite{Doyon:2021tzy} where generalisations of the $T\bar T$ deformation by extensive charges were discussed.
\unskip\footnote{Note that, while in~\cite{Doyon:2021tzy} it is assumed that the scattering is diagonal in the space of flavours, we will not require this.}
The S-matrix $\tilde S$ is a twisted version of the original $S$ (see~\secref{sec:Smat-symm} for more details).
This means that integrability in the original (gauge-fixed) model is preserved for different (gauge) deformations.

When the Hamiltonian of the reduced model is instead deformed by a $J^\tau$ deformation
\begin{equation}
\delta \mathcal H = \gamma J^\tau,
\end{equation}
then the S-matrix $\tilde S$ of the deformed model is equal to the undeformed S-matrix $S$
\begin{equation}
\tilde S_{\mu_1\mu_2}^{\nu_2\nu_1}
= S_{\mu_1\mu_2}^{\nu_2\nu_1}.
\end{equation}
As we will argue, in this case the deformation of the Hamiltonian can be completely reabsorbed into the ``free part'' $H_2$ of the Hamiltonian that is responsible for the time-evolution of the asymptotic states.
Therefore, although the asymptotic states evolve in time with a deformed dispersion relation, the scattering matrix remains undeformed.

\subsubsection{Current-current deformations}

To prove the formula \eqref{eq:smatrix-currcurr}, let us start with the case of a Hamiltonian deformed by a current-current deformation, which we write explicitly as $\delta \mathcal H = -\gamma(J^\tau_1 J^\sigma_2-J^\sigma_1 J^\tau_2)$.
Let us also define
\begin{equation}
\mathcal Q_i(\sigma)=\int^\sigma_{-\infty}d\sigma'\ J^\tau_i(\sigma').
\end{equation}
This field can be thought of as measuring the charge corresponding to $J_i$ up to the worldsheet point $\sigma$.
The total charge $Q_i=\int^\infty_{-\infty}d\sigma'\ J^\tau_i(\sigma')$ is related to it as $Q_i=\mathcal Q_i(\infty)$.
When it is not ambiguous, we will omit the explicit dependence of $\mathcal Q_i$ on $\sigma$.
First, it is easy to check that
\begin{equation}
J^\tau_1 J^\sigma_2-J^\sigma_1 J^\tau_2 = -\tfrac12 (J^\alpha_1\partial_\alpha \mathcal Q_2-J^\alpha_2\partial_\alpha \mathcal Q_1).
\end{equation}
Indeed, we have
\begin{equation}
\begin{aligned}
J^\alpha_1\partial_\alpha \mathcal Q_2-J^\alpha_2\partial_\alpha \mathcal Q_1&=
J^\tau_1(\sigma)\int^\sigma_{-\infty}d\sigma'\ \partial_\tau J^\tau_2(\sigma')+J^\sigma_1(\sigma)J^\tau_2(\sigma)-J^\tau_2(\sigma)\int^\sigma_{-\infty}d\sigma'\ \partial_\tau J^\tau_1(\sigma')-J^\sigma_2(\sigma)J^\tau_1(\sigma) \\
&=
-J^\tau_1(\sigma)\int^\sigma_{-\infty}d\sigma'\ \partial_\sigma' J^\sigma_2(\sigma')+J^\sigma_1(\sigma)J^\tau_2(\sigma)+J^\tau_2(\sigma)\int^\sigma_{-\infty}d\sigma'\ \partial_\sigma' J^\sigma_1(\sigma')-J^\sigma_2(\sigma)J^\tau_1(\sigma) \\
&=-2(J^\tau_1 J^\sigma_2-J^\sigma_1 J^\tau_2) ,
\end{aligned}
\end{equation}
where we have used current conservation and that in the decompactification limit fields fall off to zero at infinity.

We can now compute the infinitesimal deformation of the Hamiltonian to be
\begin{equation}
\begin{aligned}
\delta H &= \int_{-\infty}^{\infty}d\sigma\ \delta\mathcal H = \frac{\gamma}{2}\int_{-\infty}^{\infty}d\sigma\ (J^\alpha_1\partial_\alpha \mathcal Q_2-J^\alpha_2\partial_\alpha \mathcal Q_1)\\
&=\frac{\gamma}{2}\int_{-\infty}^\infty d\sigma\ [\partial_\alpha(J^\alpha_1\mathcal Q_2-J^\alpha_2\mathcal Q_1)-\cancel{\partial_\alpha J^\alpha_1}\mathcal Q_2+\cancel{\partial_\alpha J^\alpha_2}\mathcal Q_1]\\
&=\frac{\gamma}{2}[\partial_\tau\int_{-\infty}^\infty d\sigma\ (J^\tau_1\mathcal Q_2-J^\tau_2\mathcal Q_1)+\cancel{\int_{-\infty}^\infty d\sigma\ \partial_\sigma(J^\sigma_1\mathcal Q_2-J^\alpha_2\mathcal Q_1)}]\\
&=\frac{\gamma}{2}\partial_\tau \mathcal Q_{12},
\end{aligned}
\end{equation}
where we again use current conservation and that fields fall off at infinity, and we define the non-local quantity
\begin{equation}
\mathcal Q_{12}=\int_{-\infty}^\infty d\sigma\ (J^\tau_1\mathcal Q_2-J^\tau_2\mathcal Q_1)
=\int_{-\infty}^\infty d\sigma\ \int_{-\infty}^\sigma d\sigma'\ (J^\tau_1(\sigma)J^\tau_2(\sigma')-J^\tau_2(\sigma)J^\tau_1(\sigma')).
\end{equation}
Classically, the time-derivative of a field is given by the Poisson bracket with the Hamiltonian, hence we have $\delta H = \frac{\gamma}{2}\{H,\mathcal Q_{12}\}$.
Quantum mechanically, this becomes $\delta H = \frac{i\gamma}{2}[H,\mathcal Q_{12}]$.
We may interpret this as a differential equation for the deformed Hamiltonian $\tilde H$ as a function of the deformation parameter $\gamma$:
\begin{equation}
\frac{d\tilde H}{d\gamma} = \frac{i}{2}[\tilde H,\mathcal Q_{12}].
\end{equation}
In the Heisenberg picture, using that $\tilde H|_{\gamma=0}=H$, this is solved by
\begin{equation}
\tilde H = e^{-\frac{i\gamma}{2}\mathcal Q_{12}} H e^{\frac{i\gamma}{2}\mathcal Q_{12}}.
\end{equation}
Assuming that the scattering matrix for the undeformed theory is known, we would like to determine that of the deformed theory.
Scattering is obtained by first rewriting the Hamiltonian as the sum $H=H_2+V$, where $H_2$ is the free part without interactions (typically quadratic in the fields, for example a Klein-Gordon Hamiltonian) and $V$ is the part with interactions only.
The asymptotic states evolve with $H_2$, and the S-matrix is given by the time-ordered exponential of the interacting part of the Hamiltonian,
\begin{equation}
S= \Texp\left[-i\int_{-\infty}^{\infty}d\tau\ V\right].
\end{equation}
The considerations above suggest that in the deformed theory we should define
\begin{equation}
\tilde H_2 = e^{-\frac{i\gamma}{2}\mathcal Q_{12}} H_2 e^{\frac{i\gamma}{2}\mathcal Q_{12}},
\end{equation}
so that
\begin{equation}\label{eq:Sgamma}
\tilde V = e^{-\frac{i\gamma}{2}\mathcal Q_{12}} V e^{\frac{i\gamma}{2}\mathcal Q_{12}},
\qquad
\tilde S = e^{-\frac{i\gamma}{2}\mathcal Q_{12}} S e^{\frac{i\gamma}{2}\mathcal Q_{12}}.
\end{equation}
To understand the effect of the deformation on these objects, we first need to look at the action of $\mathcal Q_{12}$ on asymptotic states.
Let us take the charges $Q_1$ and $Q_2$ to act diagonally in the space of flavours, which is possible since these two charges commute and are simultaneously diagonalisable.
On one-particle states we write
\begin{equation}
Q_i\ket{p}_\mu=q_i^\mu\ket{p}_\mu,
\end{equation}
where $q_i^\mu$ is the charge of the particle with flavour $\mu$.
Introducing creation and annihilation operators satisfying canonical commutation relations $[a_\mu(p),a_\nu^\dagger (q)]=\delta_{\mu\nu}\delta(p-q)$, so that $\ket{p}_\mu=a_\mu^\dagger (p)\ket{0}$, we may represent the quantum charges as $Q_i = \int dp\ \sum_{\mu} q_i^\mu \ a^\dagger_\mu(p)a_\mu(p)$, where we sum over all flavours.

The action of $\mathcal Q_{12}$ on the multiparticle asymptotic states can now be constructed.
First, consider the spatial line along which the particles are distributed, and partition it into a collection of intervals $I_n$ with $n=1,\ldots,N$, where each interval $I_n$ contains only the wave-packet $n$.
In the definition of $\mathcal Q_{12}$ we have integrals over the spatial coordinate that we can write as the sum of integrals over the intervals $I_n$.
It is then clear that, despite the non-local nature of $\mathcal Q_{12}$, its action on asymptotic states is given by sums of products of local charges.
Explicitly, we have
\begin{equation}
\begin{aligned}
\mathcal Q_{12}\ket{p_1,\ldots,p_N}_{\mu_1,\ldots,\mu_N}
&=\int_{-\infty}^\infty d\sigma\ \int_{-\infty}^\sigma d\sigma'\ (J^\tau_1(\sigma)J^\tau_2(\sigma')-J^\tau_2(\sigma)J^\tau_1(\sigma'))\ket{p_1,\ldots,p_N}_{\mu_1,\ldots,\mu_N}\\
&=\sum_{n=1}^N\int_{I_n}d\sigma\ \sum_{m=1}^{n-1}\int_{I_m}d\sigma'\ (J^\tau_1(\sigma)J^\tau_2(\sigma')-J^\tau_2(\sigma)J^\tau_1(\sigma'))\ket{p_1,\ldots,p_N}_{\mu_1,\ldots,\mu_N}\\
&=\sum_{n=1}^N \sum_{m=1}^{n-1} (q_1^{\mu_n}q_2^{\mu_m}-q_2^{\mu_n}q_1^{\mu_m})\ket{p_1,\ldots,p_N}_{\mu_1,\ldots,\mu_N}.
\end{aligned}
\end{equation}
Note that thanks to antisymmetry we do not need to worry about the potentially problematic integration over the intervals $I_n$ and $I_m$ when $n=m$.
It follows that
\begin{equation}
e^{\frac{i\gamma}{2}\mathcal Q_{12}}\ket{p_1,p_2,\ldots,p_N}_{\mu_1,\mu_2,\ldots,\mu_N} = e^{-\frac{i\gamma}{2}\sum_{m<n}\epsilon^{ij}q_i^{\mu_m}q_j^{\mu_n}}\ket{p_1,p_2,\ldots,p_N}_{\mu_1,\mu_2,\ldots,\mu_N},
\end{equation}
where we recall $\epsilon^{12}=1$.
In the case of two-particle states, we have
\begin{equation}
e^{\frac{i\gamma}{2}\mathcal Q_{12}}\ket{p_1,p_2}_{\mu_1,\mu_2} = e^{-\frac{i\gamma}{2}\epsilon^{ij}q_i^{\mu_1}q_2^{\mu_2}}\ket{p_1,p_2}_{\mu_1,\mu_2},
\end{equation}
which we may rewrite as
\begin{equation} \label{eq:twist-charge}
e^{\frac{i\gamma}{2}Q_{12}}\ket{p_1,p_2}_{\mu_1,\mu_2} =
e^{-\frac{i\gamma}{2}(Q_1\wedge Q_2)}\ket{p_1,p_2}_{\mu_1,\mu_2} =e^{-\frac{i\gamma}{2}(Q_1\otimes Q_2-Q_2\otimes Q_1)}\ket{p_1,p_2}_{\mu_1,\mu_2} ,
\end{equation}
where it is understood that the first and second spaces of the tensor product act on the first and second particles respectively.
The generalisation to the case of $N$-particle states is
\begin{equation}
e^{\frac{i\gamma}{2}Q_{12}}\ket{p_1,p_2,\ldots,p_N}_{\mu_1,\mu_2,\ldots,\mu_N} =
e^{-\frac{i\gamma}{2}\sum_{m<n}(Q_{1,2;m,n}-Q_{2,1;m,n})}\ket{p_1,p_2,\ldots,p_N}_{\mu_1,\mu_2,\ldots,\mu_N} ,
\end{equation}
where
\begin{equation}
Q_{i,j;m,n}=\mathbf 1_1\otimes\cdots\otimes\mathbf 1_{m-1}\otimes (Q_i)_m\otimes\mathbf 1_{m+1}\otimes \cdots\otimes (Q_j)_n\otimes\cdots\otimes \mathbf 1_N.
\end{equation}

Since the action of $\mathcal Q_{12}$ on asymptotic states is diagonal, and the free Hamiltonian $H_2$ also acts diagonally on asymptotic states (e.g.~$H_2\ket{p}_\mu=\omega_p^\mu\ket{p}_\mu$), it follows that these two operators commute when acting on asymptotic states
\begin{equation}
[H_2,Q_{12}]\ket{p_1,p_2,\ldots,p_N}_{\mu_1,\mu_2,\ldots,\mu_N} =0.
\end{equation}
From this we conclude that we can effectively take the free part of the deformed and undeformed Hamiltonians to be equal, $\tilde{H}_2\simeq H_2$.
Strictly speaking, we have not proved that these operators are equal, only that they have the same action on asymptotic states, but this will be sufficient for the following arguments.

We finally turn to the deformation of the S-matrix.
Taking into account the simple action of $\mathcal Q_{12}$ on asymptotic states and eq.~\eqref{eq:Sgamma}, we can write
\begin{equation}
\tilde S_{\mu_1\mu_2\cdots \mu_N}^{\nu_N\nu_{N-1}\cdots \nu_1} = e^{\frac{i\gamma}{2}\sum_{m>n}\epsilon^{ij}q_i^{\nu_m}q_j^{\nu_n}}e^{-\frac{i\gamma}{2}\sum_{m<n}\epsilon^{ij}q_i^{\mu_m}q_j^{\mu_n}}S_{\mu_1\mu_2\cdots \mu_N}^{\nu_N\nu_{N-1}\cdots \nu_1}.
\end{equation}
The only subtle point is that, because of the action of $S$, the outgoing states labelled by the momenta $p_1,\ldots ,p_N$ have a spatial ordering that is reversed compared to that of the incoming states.
For this reason, the exponential coming from the action of $e^{-\frac{i\gamma}{2}\mathcal Q_{12}}$ has a summation with $m>n$ instead of $m<n$.
As anticipated, in the case of the two-body S-matrix we find
\begin{equation}
\tilde S_{\mu_1\mu_2}^{\nu_2\nu_1}
= e^{-\frac{i\gamma}{2}\epsilon^{ij}(q_i^{\nu_1}q_j^{\nu_2}+q_i^{\mu_1}q_j^{\mu_2})}S_{\mu_1\mu_2}^{\nu_2\nu_1}.
\end{equation}
As already discussed, this formula can be matched with the known deformations of the S-matrix under $T\bar T$ and $JT$ deformations.
For example, in the case of $T\bar T$, $Q_1$ would measure minus the worldsheet momentum and $Q_2$ the energy,
\unskip\footnote{If we take, for example,
\begin{equation*}
x^\mu= \frac{1}{\sqrt{2\pi}} \int \frac{dp}{\sqrt{2\omega_p^\mu}}\left(a^\mu(p,\tau)\, e^{ip \sigma}+a^{\mu\dagger}(p,\tau)\, e^{-ip\sigma}\right),\qquad
p_\mu=\frac{1}{\sqrt{2\pi}} \int \frac{dp}{\sqrt{2\omega_p^\mu}}(-i\omega_p^\mu)\left(a_{\mu}(p,\tau)\, e^{ip\sigma}-a_{\mu}^\dagger(p,\tau)\, e^{-ip \sigma}\right).
\end{equation*}
where we allow for different dispersion relations $\omega_p^\mu=\sqrt{m_\mu^2+p^2}$ for each flavour and $[a^\mu(p,\tau),a_{\nu}^\dagger(p',\tau)]=\delta^\mu{}_\nu\ \delta(p-p')$, then
one has
\begin{equation*}
\int d\sigma\ T^\tau{}_\sigma =\int d\sigma\ p_\mu x'{}^\mu =\int dp\ \sum_\mu (-p) a^\dagger_\mu(p)a_\mu(p),\qquad
\int d\sigma\ T^\tau{}_\tau =\int d\sigma\ \mathcal H = \int dp\ \sum_{\mu}\ \omega_{p}^\mu\ a^\dagger_\mu(p)a_\mu(p),
\end{equation*}
where we assume that the Hamiltonian is that of massive Klein-Gordon with mass $m_\mu$.}
so that the $T\bar T$ deformation of the S-matrix is
\begin{equation}
\tilde S_{\mu_1\mu_2}^{\nu_2\nu_1}
= e^{i\gamma(p_1\omega_2-\omega_1p_2)} S_{\mu_1\mu_2}^{\nu_2\nu_1}.
\end{equation}
This matches, for example, with~\cite{Frolov:2019nrr}, taking into account that the parameter $a$ of the $a$-gauge and $\gamma$ are related as $a=1/2-\gamma$.
Similarly, specifying to the case of the $JT_\sigma$ deformation, if $J$ has a conserved charge $Q$ with eigenvalues $q^\mu$, then one finds
\begin{equation}
\begin{aligned}
\tilde S_{\mu_1\mu_2}^{\nu_2\nu_1}
&= e^{\frac{i\gamma}{2}(p_1q^{\nu_2}-q^{\nu_1}p_2+p_1q^{\mu_2}-q^{\mu_1}p_2)} S_{\mu_1\mu_2}^{\nu_2\nu_1},
\end{aligned}
\end{equation}
which agrees with~\cite{Anous:2019osb}.
We have also verified these formulae with the tree-level results of \secref{s:JT1-S-matrix}.

To conclude, let us consider the case of the $\tilde JT_\tau$ deformation.
We denote the eigenvalues of the topological charge for the current $\tilde J$ by $w^\mu$, so that
\begin{equation}
\begin{aligned}
\tilde S_{\mu_1\mu_2}^{\nu_2\nu_1}
&= e^{\frac{i\gamma}{2}(\omega_1 (w^{\nu_2}+w^{\mu_2})-(w^{\nu_1}+w^{\mu_1})\omega_2)}S_{\mu_1\mu_2}^{\nu_2\nu_1}.
\end{aligned}
\end{equation}
Taking into account that the topological charge is given by
\begin{equation}
W = \int^{+\infty}_{-\infty} d\sigma\ \tilde J^\tau = -\int^{+\infty}_{-\infty} d\sigma\ \partial_\sigma c=0,
\end{equation}
hence vanish in the decompactification limit where we assume that all fields have fall off at infinity, we find that in the case of the $\tilde JT_\tau$ deformation the S-matrix is not modified,
\begin{equation}
\tilde S_{\mu_1\mu_2}^{\nu_2\nu_1}
=S_{\mu_1\mu_2}^{\nu_2\nu_1},
\end{equation}
again in agreement with the tree-level results.

\subsubsection{The \texorpdfstring{$J^\tau$}{Jτ} deformation}\label{sec:Smat-Jtau}

Let us now consider the case of the $J^\tau$ deformation, where
\begin{equation}
\tilde{ H}= H+\gamma Q.
\end{equation}
Here $Q$ is the charge for the current $J$, and it only acts on transverse fields.
As before, we need to separate the undeformed and deformed Hamiltonians into free and interacting parts.
Our aim is to show that the effect of the deformation can be completely absorbed in the free part of the Hamiltonian, so that the interacting part remains undeformed,
\begin{equation}
\tilde H_2= H_2+\gamma Q,\qquad
\tilde V=V,
\end{equation}
allowing us to conclude that the S-matrix is independent of $\gamma$,
\begin{equation}
\tilde S=S .
\end{equation}

To show this, we will make some mild assumptions.
In particular, we assume that the Lagrangian density of the reduced model before the deformation admits a perturbative expansion in powers of fields such that its quadratic part is described by $M$ Klein-Gordon fields, each with its own mass,
\begin{equation}
\mathcal L_2 = -\frac12 \sum_{\mu=1}^M\left(\partial_\alpha x_\mu\partial^\alpha x_\mu+m_\mu^2 x_\mu^2\right) .
\end{equation}
This Lagrangian density gives the free Hamiltonian $H_2$.
The charge $Q$ should then come from an internal global symmetry that is compatible with perturbation theory.
Therefore, we do not consider the possibility that any fields are massless ($m_\mu=0$ for some $\mu$), in which case $\mathcal L_2$ would be invariant under constant shifts of these fields, but we would not have a perturbative description of the scattering problem.
Instead, we consider the setup in which $m_\mu=m_\nu\neq 0$ for $\mu,\nu=1,\ldots ,d\leq M$, so that we have $d$ massive fields with $SO(d)$ invariance.
We will also assume that the interacting Hamiltonian respects this symmetry, but for the moment we will focus on the free theory.
The fields $x_\mu,\,\mu=1,\ldots,d$ transform in the vector representation of $SO(d)$.
The generators of $SO(d)$ can be realised with matrices $(T_{\mu\nu})^i{}_j\propto (\delta_\mu^i\delta_{\nu j}-\delta_\nu^i\delta_{\mu j})$, so that $T_{\mu\nu}$ rotates $x_\mu$ and $x_\nu$, leaving the other fields invariant.

Let us consider one such rotation, $T_{12}$, and focus on $x_1$ and $x_2$ since the other fields are simply spectators.
From the infinitesimal rotation $\delta x_1=\lambda x_2$, $\delta x_2=-\lambda x_1$, we find the Noether current $J^\alpha= x_2\partial^\alpha x_1- x_1\partial^\alpha x_2$.
In particular, we have $J^\tau= x_1p_2- x_2 p_1$, where $p_\mu=\dot x_\mu$.
It is convenient to introduce the complex field $\phi=\frac{1}{\sqrt{2}}(x_1+ix_2)$, $\phi^\dagger = \frac{1}{\sqrt{2}}(x_1-ix_2)$,
such that the quadratic Lagrangian becomes
\begin{equation}
\mathcal L_2 = -\left(\partial_\alpha\phi^\dagger\partial^\alpha\phi+m^2\phi^\dagger\phi\right).
\end{equation}
The conjugate momenta are $\pi=\frac{1}{\sqrt{2}} (p_1 - i p_2)$ and $\pi^\dag=\frac{1}{\sqrt{2}} (p_1 + i p_2)$.
Now the infinitesimal transformation reads $\delta \phi=-i\lambda \phi,$ and the Noether current is $J^\alpha=i(\phi^\dagger\partial^\alpha\phi-\partial^\alpha\phi^\dagger\phi)$ with $J^\tau=-i( \phi^\dagger \pi^\dag -\pi \phi)$.

Following the tree-level discussion in~\secref{s:JT1-S-matrix}, we consider the $J^\tau$ deformation
\begin{equation}
\tilde{ \mathcal H }_2= \mathcal H_2+ \gamma J^\tau = \pi^\dagger\pi+\phi^\dagger{}'\phi'+m^2\phi^\dagger\phi-i\gamma(\phi^\dagger\pi^\dagger -\pi \phi) .
\end{equation}
Computing the Hamilton equation $\dot\phi = \{\tilde{H}_2,\phi\}$ we find that in the deformed theory the identification of the conjugate momenta is modified
\begin{equation}
\pi=\dot\phi^\dagger + i\gamma\phi^\dagger,\qquad
\pi^\dagger=\dot\phi- i\gamma\phi.
\end{equation}
To quantise the theory we let
\begin{equation}
\phi=\frac{1}{\sqrt{2\pi}}\int\frac{dp}{\sqrt{2f_p}}\left(b_p\ e^{-i(\omega^b_p\tau-p\sigma)}+d^\dagger_p\ e^{i(\omega^d_p\tau-p\sigma)}\right),
\end{equation}
which implies
\begin{equation}
\pi =\frac{-i}{\sqrt{2\pi}}\int\frac{dp}{\sqrt{2f_p}}\left((\omega^d_p-\gamma)d_p\ e^{-i(\omega^d_p\tau-p\sigma)}-(\omega^b_p+\gamma)b^\dagger_p\ e^{i(\omega^b_p\tau-p\sigma)}\right),
\end{equation}
where $f_p$, $\omega^b_p$ and $\omega^d_p$ are real functions of $p$ to be determined.
Similar formulae are obtained for the complex conjugates of the fields.
If we demand that $\phi,\,\pi$, $b,\,b^\dagger$ and $d,\,d^\dagger$ all satisfy canonical commutation relations, then we obtain the relations
\begin{equation}
\omega^b_p = f_p-\gamma,\qquad \omega^d_p = f_p+\gamma.
\end{equation}
Assuming that $f_p$ is an even function of the momentum ($f_{-p}=f_p$) so that $\omega^b_{-p}=\omega^b_p$ and $\omega^d_{-p}=\omega^d_p$ as well, one finds that the Hamiltonian is
\begin{equation}
\begin{aligned}
\tilde H_2 = \int \frac{dp}{2f_p}\left[Z_p\ \left(b_pd_{-p}e^{-i\tau(\omega^b_p+\omega^d_{p})}+d^\dagger_pb^\dagger_{-p}e^{i\tau(\omega^b_p+\omega^d_{p})}\right)+W^b_pb^\dagger_pb_p+W^d_pd^\dagger_pd_p\right],
\end{aligned}
\end{equation}
where we use normal ordering and
\begin{equation}
\begin{gathered}
Z_p=-(\omega^b_p+\gamma)(\omega^d_p-\gamma)+p^2+m^2+\gamma(\omega^d_p-\gamma)-\gamma (\omega^b_p+\gamma),\\
W^b_p= (\omega^b_p+\gamma)^2+p^2+m^2-2\gamma(\omega^b_p+\gamma),\qquad
W^d_p= (\omega^d_p-\gamma)^2+p^2+m^2+2\gamma(\omega^d_p-\gamma).
\end{gathered}
\end{equation}
To have a diagonal action of $\tilde{H}_2$ we require $Z_p = 0$.
To solve this we take
\begin{equation}
f_p=\sqrt{m^2+p^2}\quad\implies\quad\omega^b_p =\sqrt{m^2+p^2}-\gamma,\qquad
\omega^d_p =\sqrt{m^2+p^2}+\gamma,
\end{equation}
such that
\begin{equation}
W^b_p = 2\sqrt{m^2+p^2}(\sqrt{m^2+p^2}-\gamma),\qquad
W^d_p = 2\sqrt{m^2+p^2}(\sqrt{m^2+p^2}+\gamma),
\end{equation}
and the Hamiltonian is
\begin{equation}
\tilde H_2 = \int dp\left(\omega^b_p\ b^\dagger_pb_p+\omega^d_p
d^\dagger_pd_p\right).
\end{equation}
In other words, particles and antiparticles receive a correction to the dispersion relation that depends on their charge.
Nevertheless, the Fourier decomposition of the fields is
\begin{equation}
\begin{aligned}
\phi&=\frac{1}{\sqrt{2\pi}}\int\frac{dp}{\sqrt{2\omega_p}}\left(b_p\ e^{-i(\omega_p^-\tau-p\sigma)}+d^\dagger_p\ e^{i(\omega_p^+\tau-p\sigma)}\right),\\
\pi &=\frac{-i}{\sqrt{2\pi}}\int\frac{dp}{\sqrt{2\omega_p}}\omega_p\left(d_p\ e^{-i(\omega_p^+\tau-p\sigma)}-b^\dagger_p\ e^{i(\omega_p^-\tau-p\sigma)}\right),
\end{aligned}
\end{equation}
where $\omega_p=\sqrt{m^2+p^2}$ is the standard relativistic dispersion relation and $\omega_p^\pm=\omega_p\pm\gamma$.
Note that the modified dispersion relation only enters in the plane-wave exponentials.
Let us also add that the charge $Q=-i\int d\sigma (\pi^\dagger\phi^\dagger -\phi\pi)$ is equal to
\begin{equation}
Q = \int dp\ (d_p^\dagger d_p-b^\dagger_pb_p),
\end{equation}
so that $d$-particles have charge $1$ and $b$-particles, charge $-1$.
This explains the modified dispersion relations $\omega^+_p=\omega^d_p$ and $\omega^-_p=\omega^b_p$, which can be interpreted as the relativistic dispersion relation shifted by $\gamma$ multiplied by the charge of the particle.
The above analysis all fully agrees with the tree-level considerations in~\secref{s:JT1-S-matrix}.

Let us now turn to scattering and discuss the claim that the S-matrix remains undeformed because $\tilde V=V$.
We will see how this works at tree-level and argue that it extends to all loops.
When computing the tree-level $2\to 2$ S-matrix in the undeformed case we evaluate expressions such as
\begin{equation}
\int dp_1dp_2dp_3dp_4\ \delta(\omega_1+\omega_2-\omega_3-\omega_4)\delta(p_1+p_2-p_3-p_4)\mathcal M(p_1,p_2,p_3,p_4),
\end{equation}
where the delta functions enforcing conservation of energy and momentum come from the integration over $\tau$ and $\sigma$ of the products of plane-wave exponentials, and $\mathcal M(p_1,p_2,p_3,p_4)$ is written in terms of creation and annihilation operators, the momenta $p_i$ and the corresponding dispersion relations.
Integrating the two delta functions over the outgoing momenta $p_3,p_4$ one finds
\begin{equation}
\int dp_1dp_2 \ \left|\frac{p_1}{\omega_1}-\frac{p_2}{\omega_2}\right|^{-1}(\mathcal M(p_1,p_2,p_1,p_2)+\mathcal M(p_1,p_2,p_2,p_1)),
\end{equation}
where we have evaluated the Jacobian using $d\omega_p/dp = p/\omega_p$.
In the deformed case the situation is similar and one evaluates expressions such as
\begin{equation}
\int dp_1dp_2dp_3dp_4\ \delta(\tilde\omega_1+\tilde\omega_2-\tilde\omega_3-\tilde\omega_4)\delta(p_1+p_2-p_3-p_4)\tilde{\mathcal M}(p_1,p_2,p_3,p_4),
\end{equation}
where now $\tilde \omega_p$ denotes the deformed dispersion relation, which, depending on the type of particle, equals $\omega_p,\,\omega_p+\gamma$ or $\omega_p-\gamma$.
Since $\tilde V=V$ we have that
\begin{equation}
\tilde{\mathcal M}(p_1,p_2,p_3,p_4)=\mathcal M(p_1,p_2,p_3,p_4).
\end{equation}
Indeed, since the modified dispersion relation only appears in the plane-wave exponentials, the deformation parameter appears in the delta function but not in $\mathcal M(p_1,p_2,p_3,p_4)$.
However, we also have
\begin{equation}
\delta(\tilde \omega_1+\tilde \omega_2-\tilde \omega_3-\tilde \omega_4) = \delta(\omega_1+\omega_2-\omega_3-\omega_4 +\gamma( q_1+q_2-q_3-q_4))= \delta(\omega_1+\omega_2-\omega_3-\omega_4 ),
\end{equation}
where we have used that charge conservation for $Q$ implies $q_1+q_2=q_3+q_4$.
The same conclusion can be reached by noticing that the modification of the dispersion relation is such that
\begin{equation}
\frac{d\tilde\omega_p}{dp}=\frac{d\omega_p}{dp}=\frac{p}{\omega_p},
\end{equation}
and the Jacobian is the same as in the undeformed case.
Therefore, all expressions reduce to those of the undeformed case with $\gamma=0$.
Let us note that the $\gamma$-deformation of the dispersion relation does not spoil the identification of momenta $p_1^{out}=p_1^{in},\,p_2^{out}=p_2^{in}$ and $p_1^{out}=p_2^{in},\,p_2^{out}=p_1^{in}$ as in the original integrable theory, thanks to the conservation of the charge $Q$.

To summarise, the tree-level S-matrix $\mathbf{T}$ in the deformed case is related to the undeformed one as $\tilde {\mathbf T}=\mathbf T$.
At higher loops the above reasoning should go through in a similar way.
External legs of scattering amplitudes correspond to asymptotic states with modified dispersion relations, but the elements of the scattering matrix are $\gamma$-independent.
When including quantum corrections one has to integrate loops in which \textit{off-shell} particles run, so the modified dispersion relation plays no role.
To conclude, in the case of the $J^\tau$ deformation, we have argued that the S-matrix is undeformed, $\tilde S=S $.

\subsection{S-matrix and symmetries}\label{sec:Smat-symm}

As discussed in \secref{s:res-symms}, fixing uniform light-cone gauge breaks symmetries of the string \sm.
Assuming classical integrability survives quantisation, the exact two-body S-matrix can be bootstrapped (up to overall dressing factors) by requiring compatibility with the symmetries of the light-cone gauge-fixed theory.
To describe the scattering of states that do not respect the level-matching condition, it is necessary to consider the off-shell symmetry algebra.
This is an extension of the subalgebra of the original string \sm{} symmetry algebra that survives gauge-fixing.
For instance, for strings propagating in an $\AdS_5 \times \Sp^5$ background, in the standard light-cone gauge the symmetry breaking pattern is
\begin{equation}
\alg{psu}(2,2|4) \rightarrow \alg{psu}(2|2)^{\oplus 2}_{c.e.},
\end{equation}
where $c.e.$ denotes a central extension of the algebra \cite{Arutyunov:2006ak}.
The same central elements are shared by the two copies of $\alg{psu}(2|2)$.
To keep the discussion in this section general we shall call $\mathcal A = \{\mathfrak J\}$ the off-shell symmetry algebra of the light-cone gauge-fixed theory, spanned by the generators $\set{\mathfrak J}$.
Assuming that these generators have a well-defined action on the asymptotic states, in operator notation the bootstrap equation then reads
\begin{equation} \label{eq:sym-S-cop}
\Delta(\mathfrak J) S = S \Delta(\mathfrak J), \qquad \forall \mathfrak J \in \mathcal A,
\end{equation}
where $\Delta(\mathfrak J)$ denotes the co-product associated with the symmetry algebra $\mathcal A$ (or rather its Hopf algebra).
It encodes how the symmetry generators $\mathfrak J$ act on two-particle states.
To make the link with the notation in the previous section, we have
\begin{equation}
S \ket{p_1,p_2}_{\mu_1, \mu_2} = S_{\mu_1 \mu_2}^{\nu_2 \nu_1} \ket{p_2,p_1}_{\nu_2,\nu_1}, \qquad \Delta(\mathfrak J) \ket{p_1,p_2}_{\mu_1,\mu_2} = \Delta(\mathfrak J)_{\mu_1 \mu_2}^{\nu_1 \nu_2} \ket{p_1, p_2}_{\nu_1 ,\nu_2}.
\end{equation}
Note in particular that in our conventions the S-matrix exchanges the order of the particles, but this is not the case for the co-product.
\unskip\footnote{Writing the momentum dependence explicitly, the bootstrap equation would read $\mathfrak J_{12}(q,p)S_{12}(p,q)=S_{12}(p,q)\mathfrak J_{12}(p,q)$, where $p,q$ are the two momenta and the indices denote the two vector spaces in $V\otimes V$ where the operators act.
Defining $R=\Pi S$ with $\Pi$ the (graded) permutation, one obtains an operator $R$ that reduces to the identity when interactions are switched off, and that satisfies the bootstrap equation in the form $\Delta^{op}(\mathfrak J) R = R \Delta(\mathfrak J)$, where $\Delta^{op}$ is the opposite co-product, or more explicitly $\mathfrak J_{21}(q,p)R_{12}(p,q)=R_{12}(p,q)\mathfrak J_{12}(p,q)$.\label{f:op-cop}}

For a different light-cone gauge-fixing that results in a current-current type deformation, we have seen in the previous section that the two-body S-matrix changes as
\begin{equation}
\tilde{S}_{\mu_1 \mu_2}^{\nu_2 \nu_1} = e^{- i \frac{\gamma}{2} \epsilon^{ij} (q_i^{\nu_1} q_j^{\nu_2} + q_i^{\mu_1} q_j^{\mu_2})} S_{\mu_1 \mu_2}^{\nu_2 \nu_1}.
\end{equation}
In operator form, we can recast this relation in the language of Drinfel'd-Reshetikhin twists~\cite{drinfeld1983constant,Reshetikhin:1990ep}, see also~\cite{Giaquinto:1994jx}.
In particular, from eq.~\eqref{eq:twist-charge}, we can write
\unskip\footnote{More explicitly, using the notation of \fref{f:op-cop}, this reads $\tilde{S}_{12}(p,q) = F_{12}(q,p)\, S_{12}(p,q) \, F^{-1}_{12}(p,q)$.
Note that the operator $R=\Pi S$ is twisted as $\tilde R=F^{op}\, R\, F^{-1}$, where $F^{op}$ is the conjugation of the twist by the (graded) permutation.
Therefore, $\tilde{R}_{12}(p,q) = F_{21}(q,p)\, R_{12}(p,q) \, F^{-1}_{12}(p,q)$.}
\begin{equation} \label{eq:SF}
\tilde{S} = F\, S \, F^{-1},
\end{equation}
where we have defined the twist
\begin{equation} \label{eq:twist}
F=e^{i \frac{\gamma}{2} Q_1 \wedge Q_2}.
\end{equation}
Therefore, assuming that the S-matrix $S$ satisfies the bootstrap equation \eqref{eq:sym-S-cop} with co-product $\Delta$, then the S-matrix $\tilde{S}$ associated to a different gauge-fixing satisfies the bootstrap equation with the twisted co-product
\begin{equation} \label{eq:twisted-cop}
\tilde{\Delta}(\mathfrak J) =F \Delta(\mathfrak J)F^{-1}.
\end{equation}
The discussion above is mainly relevant for the $JT_\sigma$ deformation.
The $T\bar T$ deformation produces a twist that is proportional to the identity, hence only affects the dressing factor not the symmetries, while for the $J^\tau$ deformation, the S-matrix is left invariant, $\tilde{S} = S$, hence the co-product also remains the same, $\tilde{\Delta}(\mathfrak J) = \Delta(\mathfrak J)$.

\medskip

For concreteness, let us focus on the $AdS_5\times S^5$ string.
In this case we know that in the standard light-cone gauge the off-shell symmetry algebra is $ \alg{psu}(2|2)^{\oplus 2}_{c.e.}$.
It follows from the result above that, even in a non-standard light-cone gauge, the S-matrix is still invariant under a $ \alg{psu}(2|2)^{\oplus 2}_{c.e.}$ algebra, albeit in a twisted form.
In particular, for the $JT_\sigma$ deformation, the action of generators on two-particle states will depend on momentum-dependent factors.
Such factors already appear in the co-product of the supercharges in the usual realisation of the $ \alg{psu}(2|2)^{\oplus 2}_{c.e.}$ algebra~\cite{Arutyunov:2006ak,Arutyunov:2009ga}.
Here, after fixing a non-standard light-cone gauge, the co-product of the bosonic generators may also contain momentum-dependent factors.

It is interesting to ask how this result is compatible with the discussion of symmetries in~\secref{s:res-symms}.
There, the on-shell symmetry algebra of the light-cone gauge-fixed theory was argued to be given by $\alg{c}_+$, the centraliser in $\mathfrak{psu}(2,2|4)$ of $\Lambda_+$.
\unskip\footnote{In a general gauge we identify $\Lambda_+$ and $\Lambda_-$, see also below, through the relation $t \Lambda_\mathfrak{a}+\varphi \Lambda_\mathfrak{s} = x^+ \Lambda_+ + x^- \Lambda_-$.}
This identifies the charges in the gauged-fixed model that have no explicit dependence on $x^+=\tau$~\cite{Arutyunov:2006ak}, hence Poisson-commute with the Hamiltonian.
\unskip\footnote{The charges may be divided into ``kinematical'' (if they do not depend on $x^-$) or ``dynamical'' (if they depend on $x^-$).}
In the standard light-cone gauge, $\alg{c}_+= \alg{psu}(2|2)^{\oplus 2}\oplus \mathfrak u(1)^{\oplus 2}$, which after relaxing the level-matching condition is centrally-extended to $ \alg{psu}(2|2)^{\oplus 2}_{c.e.}$.
Since these symmetries have a well-defined action on the asymptotic states (up to exponentials of $x^-$, which are reinterpreted as exponentials of the worldsheet momentum~\cite{Arutyunov:2006ak}) this centrally-extended algebra can be identified with the symmetry algebra $\mathcal A$ of the S-matrix.

For a general light-cone gauge, the relation between $\mathfrak c_+$ and $\mathcal A$ may not be as straightforward.
First, for a generic choice of light-cone gauge, the action of $\mathfrak c_+$ will not necessarily have a well-defined action on asymptotic states.
A priori, it is not obvious how such a symmetry would constrain the two-body S-matrix.
A second important point is that in the full \sm{}, as well as light-cone gauge-fixing the bosonic fields, the fermionic $\kappa$-symmetry should also be fixed.
It is then necessary to understand how the $\kappa$-gauge affects the identification of $\alg c_+$~\cite{Arutyunov:2006ak}.
Furthermore, as happens for the supercharges in the standard light-cone gauge, one should keep in mind that generators that do not commute with $\Lambda_-$ will give rise to charges with an explicit dependence on $x^-$ and their action on one-particle states can be non-trivial.

Nevertheless, knowing that different light-cone gauge-fixings lead to different algebras $\alg{c}_+$, let us assume that they also lead to different S-matrix symmetry algebras $\mathcal A$.
For example, consider a light-cone gauge-fixing ``$A$'' with S-matrix $S_A$ invariant under the symmetry algebra $\mathcal A_A$, and a light-cone gauge-fixing ``$B$'' with $S_B$ invariant under $\mathcal A_B$.
The S-matrix $S_A$ should then actually be invariant under a larger symmetry algebra that includes $\mathcal A_A$ and a twisted version of $\mathcal A_B$.
It would be interesting to verify explicitly whether this scenario is correct and if, patching together all possible light-cone gauge-fixings, the full symmetry algebra of the theory before gauge-fixing, i.e.~$\alg{psu}(2,2|4)$ for the $\AdS_5 \times \Sp^5$ superstring, can be recovered.

\section{Gauge-invariance of the spectrum}\label{sec:spectrum}

Despite the fact that the Hamiltonian and the S-matrix of the gauge-fixed model are (almost by definition) gauge-dependent objects, the spectrum of the string \sm{} should be independent of the gauge.
In this section we check this explicitly, assuming that the asymptotic spectrum (i.e.~up to wrapping corrections due to the finite string length $L$) is encoded in a set of Bethe equations constructed from the worldsheet S-matrix.
This is the case for integrable models of interest such as strings on $AdS_5\times S^5$ and $AdS_3\times S^3\times T^4$.
Without going into the details of these specific models, we consider a toy-example with a nested Bethe ansatz that has the necessary level of complication to demonstrate the gauge invariance of the spectrum.

\subsection{A toy-example of nested Bethe ansatz}\label{sec:toy}

For this discussion we will use the coordinate Bethe ansatz (see, e.g., the reviews~\cite{Levkovich-Maslyuk:2016kfv,Retore:2021wwh}).
We assume that we have two particle flavours denoted by $\phi$ and $\chi$.
For simplicity, we take $\phi$ to be a boson, although this is not necessary for the following discussion.
Let us suppose that in the two-particle basis $\ket{\phi\phi},\ket{\phi\chi},\ket{\chi\phi},\ket{\chi\chi}$ the S-matrix is
\begin{equation}
S = \left(\begin{array}{cccc}
A & 0 & 0 & 0 \\
0 & B & C & 0 \\
0 & D & E & 0 \\
0 & 0 & 0 & F
\end{array}\right).
\end{equation}
Braiding unitarity, that is the condition $S_{21}S_{12}=1$, implies various relations including $A_{12}A_{21}=1$.
To adapt to standard conventions, we adopt a different notation here to that used in \secref{sec:Smat}.
For example, the two-particle states are related as $\ket{\phi_1 \chi_2} := \ket{p_1,p_2}_{\phi \chi}$ and the action of the S-matrix is such that $S \ket{\phi_1 \chi_2} = B_{12} \ket{\phi_2 \chi_1} + D_{12} \ket{\chi_2 \phi_1} $.
In particular, the subscripts denote the momenta of the scattered particles.
We then construct $N$-particle states as
\begin{equation}
\ket{\phi_1\phi_2\cdots\phi_N} = \sum_{\sigma_1\ll\sigma_2\ll\cdots\sigma_N}
e^{i\sum_{j=1}^Np_i\sigma_j}\ket{\phi_{\sigma_1}\phi_{\sigma_2}\cdots\phi_{\sigma_N}},
\end{equation}
where the states on the left-hand-side have well-defined momenta ordered as $p_1>p_2>\cdots >p_N$, and on the right-hand-side we create wave-packets centred around the positions $\sigma_i$.
Here the formula is written for the case when all of the particles have flavour $\phi$, but the generalisation is straightforward.

The Bethe equations are obtained by requiring periodicity of the wave function for eigenstates of the S-matrix.
Let us start with the case of two particles of flavour $\phi$.
Because they simply scatter as $S\ket{\phi_1\phi_2} = A_{12} \ket{\phi_2\phi_1}$, it is sufficient to consider the state
\begin{equation}
\ket{\Psi} = \ket{\phi_1\phi_2} + A_{12} \ket{\phi_2\phi_1},
\end{equation}
and it follows from braiding unitarity that $S\ket{\Psi}=\ket{\Psi}$.
If we write
\begin{equation}
\ket{\Psi} = \sum_{\sigma_1\ll\sigma_2}\psi(\sigma_1,\sigma_2)\ket{\phi_{\sigma_1}\phi_{\sigma_2}},
\end{equation}
we then identify the wave function as
\begin{equation}
\psi(\sigma_1,\sigma_2) = e^{i(p_1\sigma_1+p_2\sigma_2)}+A_{12} e^{i(p_2\sigma_1+p_1\sigma_2)},
\end{equation}
and the periodicity condition $\psi(\sigma_2,\sigma_1+L)=\psi(\sigma_1,\sigma_2) $ implies the two Bethe equations
\begin{equation}
e^{ip_1L} = A_{21},\qquad
e^{ip_2L} = A_{12}.
\end{equation}
The generalisation to the case of $N^I$ particles of flavour $\phi$ is
\begin{equation}
e^{ip_kL} = \prod_{\substack{j = 1\\j \neq k}}^{N^I} A_{jk}.
\end{equation}
To include particles of flavour $\chi$ we need to introduce two ``levels.''
We interpret the states constructed with $\phi$ as belonging to level I only.
On top of level I, we construct level II excitations to account for $\chi$.
The difficulty now comes from the non-diagonal scattering of $\phi$ and $\chi$, for example,
\begin{equation}
S\ket{\phi_1\chi_2} =B_{12}\ket{\phi_2\chi_1}+D_{12}\ket{\chi_2\phi_1}.
\end{equation}
To construct eigenstates of the S-matrix, e.g.~in the case of two particles, we first take
\begin{equation}
\ket{\mathcal Y_y} = f(y,p_1)\ket{\chi_1\phi_2}+f(y,p_2)S^{II,I}(y,p_1)\ket{\phi_1\chi_2},
\end{equation}
where $f(y,p)$ and $S^{II,I}(y,p)$ are functions of an auxiliary root $y$ and the momentum $p$.
The function $S^{II,I}(y,p)$ can be interpreted as the scattering element between level I and II excitations.
Both functions $f(y,p)$ and $S^{II,I}(y,p)$ are determined by demanding that
\begin{equation}\label{eq:eig-levII}
S\ket{\mathcal Y_y}=A_{12}\ket{\mathcal Y_y}_\pi,
\end{equation}
where $\ket{\mathcal Y_y}_\pi$ is obtained from $\ket{\mathcal Y_y}$ by exchanging $p_1$ and $p_2$.
Let us explicitly write down the constraints imposed by this equation, since they will be useful later
\begin{equation}\label{eq:SII-I}
\begin{aligned}
&f(y,p_1)C_{12}+f(y,p_2)S^{II,I}(y,p_1)B_{12} = A_{12}f(y,p_1)S^{II,I}(y,p_2),\\
&f(y,p_1)E_{12}+f(y,p_2)S^{II,I}(y,p_1)D_{12} = A_{12}f(y,p_2).
\end{aligned}
\end{equation}
These are functional equations whose solutions will depend on the coefficients $A_{12}$, $B_{12}$, $C_{12}$, $D_{12}$, $E_{12}$ and $F_{12}$, hence will be model-dependent.
If~\eqref{eq:eig-levII} is satisfied, then $\ket{\Psi} = \ket{\mathcal Y_y}+A_{12}\ket{\mathcal Y_y}_\pi$ is an eigenstate of the S-matrix.
In this case periodicity of the wave function leads to the new Bethe equations
\begin{equation}
e^{ip_1L} = A_{21}S^{II,I}(y,p_1),\qquad
e^{ip_2L} = A_{12}S^{II,I}(y,p_2),\qquad
1=S^{II,I}(y,p_1)S^{II,I}(y,p_2).
\end{equation}
In principle, there may also be non-trivial scattering among level II excitations, which can be found by constructing states such as
\begin{equation}
\ket{\mathcal Y_{y_1}\mathcal Y_{y_2}} = f(y_1,p_1)f(y_2,p_2)S^{II,I}(y_2,p_1)\ket{\chi_1\chi_2}+f(y_2,p_1)f(y_1,p_2)S^{II,I}(y_1,p_1)S^{II,II}(y_1,y_2)\ket{\chi_1\chi_2}.
\end{equation}
Demanding $S\ket{\mathcal Y_{y_1}\mathcal Y_{y_2}}=A_{12} \ket{\mathcal Y_{y_1}\mathcal Y_{y_2}}_\pi$, where $\ket{\mathcal Y_{y_1}\mathcal Y_{y_2}}_\pi $ is obtained by exchanging $p_1$ and $p_2$, one finds the functional equation
\begin{equation}\label{eq:SII-II}
\begin{aligned}
&\left[f(y_1,p_1)f(y_2,p_2)S^{II,I}(y_2,p_1)+f(y_2,p_1)f(y_1,p_2)S^{II,I}(y_1,p_1)S^{II,II}(y_1,y_2)\right]F_{12}\\
&=
\left[f(y_1,p_2)f(y_2,p_1)S^{II,I}(y_2,p_2)+f(y_2,p_2)f(y_1,p_1)S^{II,I}(y_1,p_2)S^{II,II}(y_1,y_2)\right]A_{12}.
\end{aligned}
\end{equation}
As before, we will not need the explicit model-dependent solution to this equation.
In the general case the Bethe equations are given by
\begin{equation}
\begin{aligned}
e^{ip_kL} &= \prod_{\substack{j = 1\\j \neq k}}^{N} A_{jk}\prod_{j=1}^{N^{II}}S^{II,I}(y_j,p_k),\qquad &&k=1,\ldots,N,\\
1 &= \prod_{\substack{j = 1\\j \neq k}}^{N^{II}}S^{II,II}(y_k,y_j)\prod_{j=1}^{N}S^{II,I}(y_k,p_j),\qquad &&k=1,\ldots,N^{II},
\end{aligned}
\end{equation}
where $N^I$ is the number of excitations of flavour $\phi$, $N^{II}$ the number of excitations of flavour $\chi$, and $N=N^I+N^{II}$.
Note that multiplying all the Bethe equations gives $e^{ip^{tot}L}=1$, where $p^{tot} = \sum_{k=1}^Np_k$.
We will take the level-matching condition $p^{tot}=0$.

Once a state is fixed and the corresponding solution to the Bethe equations is found, that is a list of values $p_1,\ldots,p_N$, the conserved charge $E-J$ is given by the sum of the magnon energies
\begin{equation}
E-J = \sum_{i=1}^N \mathcal E_i,\qquad
\mathcal E_i = \sqrt{m^2_i+4h^2\sin^2\frac{p_i}{2}},
\end{equation}
where $m_i$ is the mass of the excitation, $p_i$ its momentum, and $h$ is a function of the string tension.

\subsection{Invariance for current-current gauge transformations}

When the gauge transformation is a current-current deformation, the invariance of the spectrum comes from the fact that, in addition to the Hamiltonian and S-matrix of the reduced model, the length $L$ of the string is also gauge-dependent.
Taking this into account ensures that the Bethe equations, and therefore the spectrum, are gauge-independent.
For the $T\bar T$ gauge deformation this has been discussed in the literature, in particular see~\cite{Baggio:2018gct,Frolov:2019nrr} and~\cite{Arutyunov:2009ga}.
To the best of our knowledge the case of the $JT_\sigma$ gauge deformation has not been discussed before.
We first briefly review the case of the $T\bar T$ gauge deformation below, before discussing the more involved $JT_\sigma$ gauge deformation.

Let us note that the invariance of the spectrum is a consequence of interpreting the deformations as gauge transformations.
For genuine $T\bar T$ or $JT_\sigma$ deformations, the length $L$ is fixed to be $\gamma$-independent, and the spectrum would be $\gamma$-dependent.

\subsubsection{\texorpdfstring{$T\bar T$}{TTbar}}

In the case of the $T\bar T$ gauge transformation we know that the S-matrix changes by an overall factor,
\begin{equation}
\tilde S_{12} = e^{i\gamma(p_1\omega_2-p_2\omega_1)}S_{12}.
\end{equation}
Working with the toy-example of~\secref{sec:toy}, this means that $\tilde A_{12} = e^{i\gamma(p_1\omega_2-p_2\omega_1)}A_{12}$, and similarly for all the other entries.
It is easy to see that, given the $\gamma$-dependent factor is common to all entries of the S-matrix, it drops out of eqs.~\eqref{eq:SII-I} and~\eqref{eq:SII-II}, so that the functions $f(y,p)$, $S^{II,I}(y,p)$, $S^{II,II}(y_k,y_j)$ can be taken to be the same as in the undeformed case.
At the same time, we should take into account the $\gamma$-dependence of the length of the string.
In particular, integrating the relation $\tilde p_- = p_-+\gamma p_+$ from eq.~\eqref{eq:prom-for-TT}, it follows that $\tilde L = L-\gamma \mathcal E^{tot}$, where $\mathcal E^{tot}=\sum_{k=1}^N\mathcal E_k$ is the total energy.
Therefore,
\begin{equation}
\begin{aligned}
e^{ip_k\tilde L} = e^{ip_k L} e^{-i\gamma p_k \mathcal E^{tot}} &= \prod_{\substack{j = 1\\j \neq k}}^{N} e^{i\gamma(p_j\omega_k-p_k\omega_j)} A_{jk}\prod_{j=1}^{N^{II}} S^{II,I}(y_j,p_k)\\
&= \cancel{e^{i\gamma \omega_k p^{tot}}}e^{-i\gamma p_k \mathcal E^{tot}}\prod_{\substack{j = 1\\j \neq k}}^{N} A_{jk}\prod_{j=1}^{N^{II}} S^{II,I}(y_j,p_k),
\end{aligned}
\end{equation}
where we have used $p^{tot}=0$.
The factor $e^{-i\gamma p_k \mathcal E^{tot}}$ appears on both sides of the equation, hence cancels and the Bethe equations for $p_k$ are $\gamma$-independent.
It is immediate to see that the equations for the auxiliary roots are also independent of the deformation.

\subsubsection{\texorpdfstring{$JT_\sigma$}{JTσ}}

In the case of a $JT_\sigma$ gauge transformation, verifying that the spectrum is invariant is more involved.
First, we notice that
\begin{equation}
\begin{aligned}
\tilde A_{12} = e^{i\gamma q_\phi(p_1-p_2)}A_{12},\qquad
\tilde B_{12} = e^{\frac{i}{2}\gamma (q_\phi+q_\chi)(p_1-p_2)}B_{12},\qquad
\tilde C_{12} = e^{i\gamma (p_1q_\phi-p_2q_\chi)}C_{12},\\
\tilde F_{12} = e^{i\gamma q_\chi(p_1-p_2)}F_{12},\qquad
\tilde E_{12} = e^{\frac{i}{2}\gamma (q_\phi+q_\chi)(p_1-p_2)}E_{12},\qquad
\tilde D_{12} = e^{i\gamma (p_1q_\chi-p_2q_\phi)}D_{12},
\end{aligned}
\end{equation}
where $q_\phi$ and $q_\chi$ denote the charges of $\phi$ and $\chi$ under the symmetry corresponding to the current $J$.
Note that we assume that $\phi$ and $\chi$ are eigenstates of the charge.
Consider now the equations in~\eqref{eq:SII-I}.
We have similar equations in the deformed case, but with tildes.
The equations without tildes imply those with tildes if we take
\begin{equation}
\tilde f(y,p)= f(y,p)e^{\frac{i}{2}\gamma p(q_\phi-q_\chi)},\qquad
\tilde S^{II,I}(y,p)= S^{II,I}(y,p)e^{i\gamma p(q_\phi-q_\chi)}.
\end{equation}
We also note that, with this identification and with $\tilde S^{II,II}(y_k,y_j)=S^{II,II}(y_k,y_j)$, equation~\eqref{eq:SII-II} is automatically solved in the presence of the deformation.

Now let us look at the Bethe equations, starting with those for the momenta $p_k$.
Knowing that $\tilde p_-=p_--\gamma J^\tau$ from eq.~\eqref{eq:CT-JT-2}, we conclude that $\tilde L = L-\gamma q^{tot}$, where $q^{tot}$ is the total charge for all the excitations.
Therefore, the Bethe equations become
\begin{equation}
\begin{aligned}
e^{ip_k\tilde L} = e^{ip_k L} e^{-i\gamma p_k q^{tot}} &= \prod_{\substack{j = 1\\j \neq k}}^{N} e^{i\gamma q_\phi(p_j-p_k)} A_{jk}\prod_{j=1}^{N^{II}}e^{i\gamma p_k(q_\phi-q_\chi)} S^{II,I}(y_j,p_k)\\
&= \cancel{e^{i\gamma q_\phi p^{tot}}}e^{-i\gamma p_k[Nq_\phi-N^{II}(q_\phi-q_\chi)]} \prod_{\substack{j = 1\\j \neq k}}^{N} A_{jk}\prod_{j=1}^{N^{II}} S^{II,I}(y_j,p_k).
\end{aligned}
\end{equation}
The $\gamma$-independence of the equation is a consequence of $p^{tot}=0$ and $q^{tot} = N^Iq_\phi+N^{II}q_{\chi} = Nq_\phi-N^{II}(q_\phi-q_\chi)$, where we recall $N=N^I+N^{II}$.
The Bethe equations for the auxiliary roots \begin{equation}
\begin{aligned}
1 = \prod_{\substack{j = 1\\j \neq k}}^{N^{II}}S^{II,II}(y_k,y_j)\prod_{j=1}^{N}e^{i\gamma p_j(q_\phi-q_\chi)}S^{II,I}(y_k,p_j),
\end{aligned}
\end{equation}
are $\gamma$-independent thanks to $p^{tot}=0$.

\subsection{Invariance for \texorpdfstring{$J^\tau$}{Jτ} deformations}

The invariance of the spectrum under a $J^\tau$ gauge deformation is even simpler to see.
Before the gauge transformation, we compute the eigenvalues of the Hamiltonian $H$, which are identified by the solutions to the Bethe equations constructed from the S-matrix $S$.
As already mentioned, given the solution $p_1,\ldots,p_N$ for a certain state, the eigenvalue of the Hamiltonian is then $\mathcal E=\sum_{k=1}^N\mathcal E_k$, where $\mathcal E_k = \sqrt{m^2_k+4h^2\sin^2\frac{p_k}{2}}$.

After the $J^\tau$ gauge deformation, we compute the eigenvalues of the Hamiltonian $\tilde H=H+\gamma Q$.
These are found by identifying the solutions to the Bethe equations constructed from the S-matrix $\tilde S$, which, in this case, is equal to the undeformed S-matrix, $\tilde S=S$.
Hence, both the Bethe equations and their solutions are trivially $\gamma$-independent.

From the point of view of the scattering problem, the dispersion relations of the asymptotic states are modified by shifts proportional to their charges, as we saw explicitly in~\secref{s:JT1-S-matrix} and~\secref{sec:Smat-Jtau}.
Therefore, the eigenvalue of the Hamiltonian $\tilde H$ will now be obtained by computing $\tilde{\mathcal E}=\sum_{k=1}^N\tilde{\mathcal E}_k$ where $\tilde{\mathcal E}_k =\mathcal E_k +\gamma q_k$, with $q_k$ the charge of the excitation.
It is clear that the $\gamma$-dependence of the spectrum of $\tilde H$ is spurious: it is a consequence of the fact that the definition of $\tilde H$ itself depends on $\gamma$.
Even in this gauge, if we were computing the eigenvalues of $H=\tilde H-\gamma Q$, we would find a $\gamma$-independent spectrum.

\section{Conclusions and outlook}\label{sec:concl}

In this paper we analysed inequivalent uniform light-cone gauges for string \sms{} with at least two commuting isometries, one timelike and one spacelike.
By implementing target-space coordinate transformations before light-cone gauge-fixing, we found four classes of inequivalent gauge-fixings, which can be understood as $T\bar{T}$, $JT_\sigma$, $J^\tau$ and $\tilde{J}T_\tau$ deformations.
We further demonstrated that of these, only the $T\bar T$ and $JT_\sigma$ deformations modify the worldsheet S-matrix.
In the context of string \sms, they are understood simply as different gauge choices, so that the spectrum remains invariant, see~\secref{sec:spectrum}.

In~\secref{sec:in-sym-cosets} we investigated the moduli space of inequivalent light-cone gauge-fixings for spacetimes given by the Cartesian product of two rank-1 symmetric spaces $\Man_{\alg{a}} \times \Man_{\alg{s}}$, of which $AdS_n\times S^n$ is an important example.
In particular, we explicitly constructed part of this moduli space for the unique (up to global symmetries) point-like string solution with momentum in both $\Man_{\alg{a}}$ and $\Man_{\alg{s}}$, confirming the expected freedom related to $T\bar{T}$, $JT_\sigma$, $J^\tau$ and $\tilde{J}T^\tau$ deformations.
There is also the option of starting from massless geodesics on $\Man_{\alg{a}}$.
In the case of $AdS$ space this leads to the $AdS$ light-cone gauge~\cite{Giombi:2009gd}, which we have not discussed.
It would also be interesting to study more general spacetimes, including higher-rank cosets.
Since $S^3 \times S^3$ is a rank-2 coset, this would be important for the $AdS_3\times S^3\times S^3\times S^1$ background.
In this case there will no longer be a unique point-like string solution with momentum in both $\Man_{\alg{a}}$ and $\Man_{\alg{s}}$ up to global symmetries.
For example, in the case of $S^3 \times S^3$ we have a one-parameter family of solutions, distinguished by the ratio of momenta on the two spheres.
Nevertheless, once a choice of point-like string has been made, the classification of inequivalent gauges should follow the pattern explained in this paper.

\medskip

We have focused on fixing uniform light-cone gauge for bosonic $AdS_n\times S^n$ backgrounds, i.e.~realised in terms of symmetric spaces.
It would be interesting to extend our systematic analysis to semisymmetric spaces and the Green-Schwarz superstring, where in addition to fixing worldsheet diffeomorphisms, one should also fix the gauge of the local fermionic $\kappa$-symmetry transformations (see, e.g.,~\cite{Arutyunov:2009ga,Demulder:2023bux} for reviews).
Since the $\kappa$-symmetry commutes with the superisometries, its gauge-fixing will not affect the identification of the residual symmetries in the light-cone gauge-fixed theory.
Nevertheless, $\kappa$-symmetry is important for understanding how the residual superalgebra acts on the transverse theory, hence it would be interesting to incorporate this analysis.

In general, after gauge-fixing the original supersymmetry algebra is reduced to a residual superalgebra.
In the standard setup this is a centrally-extended $\mathfrak{psu}(2|2)^{\oplus 2}$ for $AdS_5\times S^5$ and a central extension of $[\mathfrak{u}(1) \ltimes \mathfrak{psu}(1|1)^{\oplus 2}]^{\oplus 2}$ for $AdS_3\times S^3\times T^4$ (ignoring the torus directions and their superpartners).
As shown in \secref{s:res-symms} (c.f.~\tabref{tab:ca}, \tabref{tab:cs}, and eq.~\eqref{eq:largest-lcsymms}), the residual symmetry algebra may change depending on the choice of gauge.
It would be interesting to understand if in general the worldsheet S-matrix is uniquely fixed by the residual symmetries up to an overall factor, as in~\cite{Beisert:2005tm,Arutyunov:2006iu} for the standard choice.
For this, it would be necessary to understand how the action of the residual generators is realised on the transverse theory, as well as the effect of $\kappa$-symmetry, which we expect to be non-trivial.

As discussed in~\secref{sec:Smat-symm}, if we consider, for example, $AdS_5\times S^5$, the centrally-extended $\mathfrak{psu}(2|2)^{\oplus 2}$ symmetry is not actually broken under the light-cone gauge transformation; instead, it undergoes a twist.
Since different gauges have residual symmetries that are different subalgebras of $\mathfrak{psu}(2,2|4)$, it may be possible to identify a larger invariance of the worldsheet $AdS_5\times S^5$ S-matrix going beyond the usual centrally-extended $\mathfrak{psu}(2|2)^{\oplus 2}$, possibly corresponding to a non-standard action of the inherent $\mathfrak{psu}(2,2|4)$ symmetry on the transverse fields and their S-matrix.

\medskip

Our motivation for the analysis in this paper came from the study of integrable deformations of $AdS_n\times S^n$ \sms, their worldsheet S-matrices and quantum integrability descriptions.
Thinking of an undeformed string \sm{} as a point in a space of theories, its continuous deformations can be pictured as lines departing from this point.
As we have seen, the undeformed model may have a moduli space of inequivalent light-cone gauge-fixings, each describing the same \sm, with an unchanged string spectrum.
However, an integrable deformation may break some symmetries, resulting in a smaller moduli space of light-cone gauge-fixings.
In other words, in order to be able to deform the gauge-fixed model, we would need to restrict to a subspace of light-cone gauge-fixings.
Correspondingly, to be able to deform the worldsheet S-matrix we may first need to apply a $JT_\sigma$ transformation.
We refer to~\cite{Hoare:2023zti,inpreparation-RiccardoSibylle} for realisations of this scenario.

Knowing that inequivalent light-cone gauges play an important role in the integrability formulation of integrable deformations of the string \sms, it would be interesting to understand how this is paralleled in the spin-chain description of the dual gauge theories~\cite{Minahan:2002ve,Beisert:2003jj}.
This would be the starting point to construct deformations of the spin chain corresponding to deformations of the string theory background.
The case of the homogeneous Yang-Baxter deformations, which are expected to be implemented by Drinfel'd twists, should be particularly tractable.
Starting with~\cite{vanTongeren:2015uha,vanTongeren:2016eeb}, there has been substantial progress in the identification of the deformations of the gauge theory that are dual to homogeneous Yang-Baxter deformations of the string, see in particular the recent~\cite{Meier:2023kzt,Meier:2023lku}.
Given that the construction is under control when the deformation is based on twists of the Poincar\'e algebra, it would be interesting to understand if spin-chain constructions could help with the identification of the gauge theory duals beyond those cases.

\paragraph{Acknowledgements } S.D., A.L.R., and F.S. thank the participants of the Workshop “Integrability in Low Supersymmetry Theories”
in Filicudi, Italy, for stimulating discussions that initiated this project.
We thank Sergey Frolov for discussions on the connection of light-cone gauge-fixing and $JT_\sigma$-deformations, and J.~Luis Miramontes for discussions on \cite{Miramontes:2008wt}.
R.B. is supported by the grant RYC2021-032371-I (funded by MCIN/AEI/10.13039/501100011033 and by the European Union ``NextGenerationEU''/PRTR) and by the grant 2023-PG083 with reference code ED431F 2023/19 funded by Xunta de Galicia. R.B. also acknowledges AEI-Spain (under project PID2020-114157GB-I00 and Unidad de Excelencia Mar\'\i a de Maetzu MDM-2016-0692), Xunta de Galicia (Centro singular de investigaci\'on de Galicia accreditation 2019-2022, and project ED431C-2021/22), and the European Union FEDER.
S.D. is supported by the Swiss National Science Foundation through the NCCR SwissMAP, and thanks the group at IGFAE, U.~Santiago de Compostela, for hospitality during a visit where part of this work was carried out.
The work of B.H. and A.L.R. was supported by a UKRI Future Leaders Fellowship (grant number MR/T018909/1).
F.S. is supported by the European Union’s Horizon 2020 research and innovation programme under the Marie Sk\l{}odowska-Curie grant agreement number 101027251.

\appendix

\section{Conventions and review of light-cone gauge-fixing}\label{sec:appendix}
We consider a string \sm{} on a $D$-dimensional background parametrised by the coordinates $x^M$ with $M=0,\ldots,D-1$
\begin{equation}\label{eq:S-Poly}
\Act =-\frac{T}{2}\int_{-\tfrac{L}{2}}^{\tfrac{L}{2}}d\tau d\sigma\ \left(\gamma^{\alpha\beta}G_{MN}-\epsilon^{\alpha\beta}B_{MN}\right)\partial_\alpha x^M\partial_\beta x^N ,
\end{equation}
where $T$ denotes the string tension and $L$ the length of the string.
Moreover, $\gamma^{\alpha\beta}$ is the Weyl-invariant combination of the worldsheet metric, and we use the convention $\epsilon^{\tau\sigma} = -1$.
The \sm{} couplings are the target-space metric $G_{MN}$ and the $B$-field $B_{MN}$.
We assume that the background possesses at least two abelian isometries realised by shifts of two coordinates $x^0=t$ and $x^1=\varphi$.
Here $t$ is a timelike and $\varphi$ a spacelike coordinate.
The remaining coordinates will be called transverse and are denoted by $x^\mu$ with $\mu=2,\ldots,D-1$.

Under the above assumptions, a solution to the equations of motion of the \sm{} is
\begin{equation}
\bar t = \kappa \tau,\qquad
\bar\varphi = \tau,\qquad
\bar x^\mu=0.
\end{equation}
Here the bar denotes a field evaluated on the classical solution.
In this solution the velocity of $\bar \varphi$ is fixed to 1 (e.g.,~by redefining $\tau$).
In principle, $\bar x^\mu$ can be a collection of non-vanishing constants, but these can be set to zero by redefining $x^\mu$.

The Virasoro constraints fix the value of the parameter $\kappa$.
To see this, let us construct the stress-energy tensor of the \sm{}~\eqref{eq:S-Poly}
\begin{equation}
\mathcal T_{\alpha\beta} = \partial_\alpha x^M G_{MN} \partial_\beta x^N - \frac{1}{2} \gamma_{\alpha\beta} \gamma^{\gamma\delta} \partial_\gamma x^M G_{MN} \partial_\delta x^N .
\end{equation}
If we rewrite our classical solution as $\bar x^M=a^M\tau$ with $a^0=\kappa$, $a^1=1$ and $a^\mu=0$, then the components of the stress-energy tensor on the classical solution read
\begin{equation}
\bar{\mathcal T}_{\tau\tau}=\mathscr C\left(1-\frac{1}{2} \gamma_{\tau\tau} \gamma^{\tau\tau}\right),\qquad
\bar{\mathcal T}_{\tau\sigma}=-\frac{1}{2}\mathscr C \gamma_{\tau\sigma} \gamma^{\tau\tau},\qquad
\bar{\mathcal T}_{\sigma\sigma}=-\frac{1}{2}\mathscr C \gamma_{\sigma\sigma} \gamma^{\tau\tau},
\end{equation}
where
\begin{equation}
\mathscr C= \bar G_{MN}a^Ma^N = \bar G_{00}\kappa^2+\bar G_{11}.
\end{equation}
Here we assumed $\bar G_{01}=0$, which can be achieved by redefining $t$ and $\varphi$.
On the classical solution the Virasoro constraints $	\bar{\mathcal T}_{\alpha\beta}=0$ are satisfied if $\mathscr C=0$.
We solve this by taking
\begin{equation}
\kappa=\sqrt{-\frac{\bar G_{11}}{\bar G_{00}}},
\end{equation}
where we are making a choice for the sign of $\kappa$.
Finally, rescaling the field $t$ by $\kappa$, we can work with a classical solution of the form $\bar t = \tau$, $\bar\varphi = \tau$ and $\bar x^\mu=0$, so that we effectively set $\kappa=1$.

Let us now review how to fix uniform light-cone gauge in the Hamiltonian formalism following the review~\cite{Arutyunov:2009ga}, see also~\cite{Arutyunov:2004yx,Arutyunov:2005hd,Arutyunov:2006gs}.
Starting from the classical \sm{} action~\eqref{eq:S-Poly},
we define the conjugate momenta as
\begin{equation}
p_M=\frac{\delta \Act}{\delta \dot x^M}=-T\gamma^{\tau\beta}\partial_\beta x^NG_{MN}-Tx'{}^NB_{MN}.
\end{equation}
Here, and in the rest of the paper, a dot denotes the time derivative $\dot x^M = \partial_\tau x^M$ and a prime, the spatial derivative $x'{}^M = \partial_\sigma x^M$.
On the classical solution the momenta read
\begin{equation}
\bar p_0=-T\bar \gamma^{\tau\tau} \bar G_{00},\qquad
\bar p_1=-T\bar \gamma^{\tau\tau} \bar G_{11}, \qquad
\bar p_\mu=-T\bar \gamma^{\tau\tau} ( \bar G_{\mu 0}+\bar G_{11}).
\end{equation}
In principle, $\bar p_\mu$ can be a non-vanishing constant vector, but from now on we assume that $\bar p_\mu=0$.
In~\secref{s:a-lcgf} we show that we can always redefine our fields to achieve this, and that when doing so we end up with an equivalent gauge-fixing.
We also fix $\bar\gamma^{\tau\tau}=-(T\bar G_{11})^{-1}$ so that $\bar p_1=1$.
To summarise, thus far we have
\begin{equation}
\begin{aligned}
&\bar t = \tau,\qquad &&\bar\varphi=\tau,\qquad &&&\bar x^\mu=0,\\
&\bar p_0 = -1,\qquad &&\bar p_1=1,\qquad &&&\bar p_\mu=0.
\end{aligned}
\end{equation}
We now introduce light-cone coordinates as
\unskip\footnote{We could use
\begin{equation}
\begin{alignedat}{3}
x^+&=(1-a)t+a\varphi,\qquad
&&x^-&&=\varphi- t, \\
p_+ &= p_0+p_1,\qquad
&&p_- &&=-a p_0+(a-1)p_1,
\end{alignedat}
\end{equation}
and the classical solution would still remain the same.
However, here we set $a=1/2$ and the parameter $a$ will instead be recovered from the discussion in~\secref{s:a-lcgf}.\label{f:xpxm-with-a}}
\begin{equation}
x^+=\frac12\left(t+\varphi\right),\qquad
x^-=\varphi- t,
\end{equation}
so that
\begin{equation}
p_+ = p_0+p_1,\qquad
p_- = \frac12(- p_0+p_1).
\end{equation}
On the classical solution we have
\begin{equation} \label{eq:can-momenta-class}
\begin{aligned}
&\bar x^+ = \tau,\qquad &&\bar x^-=0,\qquad &&&\bar x^\mu=0,\\
&\bar p_+ = 0,\qquad &&\bar p_-=1,\qquad &&&\bar p_\mu=0.
\end{aligned}
\end{equation}
After introducing the momenta $p_M$, the action can be rewritten as
\begin{equation}
\Act = \int d\tau d\sigma \left( p_M \dot{x}^M + \frac{\gamma^{\tau\sigma}}{\gamma^{\tau \tau}} C_1 + \frac{1}{2T \gamma^{\tau \tau }} C_2 \right) ,
\end{equation}
where
\begin{equation}
\begin{aligned}
C_1&=p_Mx'{}^M,\\
C_2&=G^{MN}p_Mp_N+T^2 G_{MN}x'{}^Mx'{}^N-2T p_MG^{MN}B_{NQ}x'{}^Q+T^2G^{MN}B_{MP}B_{NQ}x'{}^Px'{}^Q,
\end{aligned}
\end{equation}
and $\gamma^{\tau\sigma}$ and $\gamma^{\tau \tau}$ are Lagrange multipliers imposing the Virasoro constraints are $C_1=C_2=0$.
On these constraints the action is simply
\begin{equation}
\Act = \int d\tau d\sigma \ p_M \dot{x}^M = \int d\tau d\sigma \left( p_+ \dot{x}^++p_-\dot{x}^-+p_\mu \dot{x}^\mu \right).
\end{equation}

We now expand the fields around their classical values as $x^M=\bar x^M+\hat x^M$ and $p_M=\bar p_M+\hat p_M$, where the hats denote fluctuations.
We expand around a classical solution to ensure the Lagrangian and Hamiltonian start at quadratic order in the fluctuating fields.
Because of the reparametrisation invariance on the worldsheet, we can choose a gauge where two fluctuations are set to zero, and we take
\begin{equation}
\hat x^+=0=\hat p_-.
\end{equation}
All other fields are allowed to fluctuate.
Taking into account the classical solution and the gauge choice, we have
\begin{equation}\label{eq:recap-lcg}
\begin{aligned}
& x^+ = \tau,\qquad && x^-=\hat x^-,\qquad &&&x^\mu=\hat x^\mu,\\
& p_+ = \hat p_+,\qquad &&p_-=1,\qquad &&&p_\mu=\hat p_\mu.
\end{aligned}
\end{equation}
Since each field either coincides with its classical value or with its fluctuation, the notation is unambiguous if we omit the bars and hats, and we will do so from now on.
The expansion of the action around the classical solution is therefore
\begin{equation}
\Act_{g.f.} = \int d\tau d\sigma \left( p_+ +\dot{x}^-+p_\mu \dot{x}^\mu \right) = \int d\tau d\sigma \left( p_+ + p_\mu \dot{x}^\mu \right),
\end{equation}
where in the second step we dropped a total derivative.
We recognise the action for the transverse fields $x^\mu$, $p_\mu$ with Hamiltonian density $\mathcal H=-p_+$.
Indeed, $p_+$ is expressed in terms of transverse fields once we solve the Virasoro constraints $C_1=C_2=0$ for the fluctuations $x^-$ and $p_+$.
The first equation is solved by
\begin{equation}\label{eq:vir1}
x^-{}'=-p_\mu x^\mu{}'.
\end{equation}
The second equation is quadratic in $p_+$.
If we introduce indices $m,n=-,\mu$ (i.e.~all except $+$) then writing the equation as $C_2=Ap_+^2+Bp_++C=0$, where
\begin{equation}
\begin{aligned}
A&=G^{++},\\
B&= 2G^{+m}p_m-2T G^{+M}B_{Mn}x'{}^n,\\
C&= G^{mn}p_mp_n+T^2 G_{mn}x'{}^mx'{}^n-2T p_mG^{mN}B_{Nq}x'{}^q+T^2G^{MN}B_{Mp}B_{Nq}x'{}^px'{}^q,
\end{aligned}
\end{equation}
we take the solution to be
\begin{equation}
p_+=\frac{-B+\sqrt{B^2-4AC}}{2A},
\end{equation}
where the sign is chosen to give the correct Hamiltonian.
In this expression $p_-$ is replaced by its classical value $\bar p_-=1$ and $x^-{}'$ using~\eqref{eq:vir1}.
The solution for the Hamiltonian density is therefore
\begin{equation}
\mathcal H=\frac{B-\sqrt{B^2-4AC}}{2A}.
\end{equation}
If we define charges $Q_M=\int_{-\frac{L}{2}}^{\frac{L}{2}}d\sigma\, p_M$, then we have the relations $Q_+=-\int_{-\frac{L}{2}}^{\frac{L}{2}}d\sigma\, \mathcal H=-H$ where $H$ is the Hamiltonian, and $Q_-=L$.

\addtocontents{toc}{\protect\setcounter{tocdepth}{-1}}
\begin{bibtex}[\jobname]

@article{Hoare:2023zti,
author = "Hoare, Ben and Retore, Ana L. and Seibold, Fiona K.",
title = "{Elliptic deformations of the $AdS_{3} \times S^{3} \times T^{4}$ string}",
eprint = "2312.14031",
archivePrefix = "arXiv",
primaryClass = "hep-th",
reportNumber = "Imperial-TP-FS-2023-02",
doi = "10.1007/JHEP04(2024)042",
journal = "JHEP",
volume = "04",
pages = "042",
year = "2024"
}

@article{Arutyunov:2009ga,
author = "Arutyunov, Gleb and Frolov, Sergey",
title = "{Foundations of the $AdS_5 \times S^5$ Superstring. Part I}",
eprint = "0901.4937",
archivePrefix = "arXiv",
primaryClass = "hep-th",
reportNumber = "ITP-UU-09-05, SPIN-09-05, TCD-MATH-09-06, HMI-09-03",
doi = "10.1088/1751-8113/42/25/254003",
journal = "J. Phys. A",
volume = "42",
pages = "254003",
year = "2009"
}

@article{Arutyunov:2004yx,
author = "Arutyunov, Gleb and Frolov, Sergey",
title = "{Integrable Hamiltonian for classical strings on $AdS_5 \times S^5$}",
eprint = "hep-th/0411089",
archivePrefix = "arXiv",
reportNumber = "AEI-2004-105",
doi = "10.1088/1126-6708/2005/02/059",
journal = "JHEP",
volume = "02",
pages = "059",
year = "2005"
}

@article{Arutyunov:2005hd,
author = "Arutyunov, Gleb and Frolov, Sergey",
title = "{Uniform light-cone gauge for strings in $AdS_5 \times S^5$: Solving SU(1|1) sector}",
eprint = "hep-th/0510208",
archivePrefix = "arXiv",
reportNumber = "ITP-UU-05-47, SPIN-05-32, AEI-2005-160",
doi = "10.1088/1126-6708/2006/01/055",
journal = "JHEP",
volume = "01",
pages = "055",
year = "2006"
}

@article{Arutyunov:2006gs,
author = "Arutyunov, Gleb and Frolov, Sergey and Zamaklar, Marija",
title = "{Finite-size Effects from Giant Magnons}",
eprint = "hep-th/0606126",
archivePrefix = "arXiv",
reportNumber = "AEI-2006-049, ITP-UU-06-28, SPIN-06-24",
doi = "10.1016/j.nuclphysb.2006.12.026",
journal = "Nucl. Phys. B",
volume = "778",
pages = "1--35",
year = "2007"
}

@article{Dubovsky:2023lza,
author = "Dubovsky, Sergei and Negro, Stefano and Porrati, Massimo",
title = "{Topological gauging and double current deformations}",
eprint = "2302.01654",
archivePrefix = "arXiv",
primaryClass = "hep-th",
doi = "10.1007/JHEP05(2023)240",
journal = "JHEP",
volume = "05",
pages = "240",
year = "2023"
}

@article{Doyon:2021tzy,
author = "Doyon, Benjamin and Durnin, Joseph and Yoshimura, Takato",
title = "{The Space of Integrable Systems from Generalised $T\bar{T}$-Deformations}",
eprint = "2105.03326",
archivePrefix = "arXiv",
primaryClass = "hep-th",
doi = "10.21468/SciPostPhys.13.3.072",
journal = "SciPost Phys.",
volume = "13",
number = "3",
pages = "072",
year = "2022"
}

@article{Miramontes:2008wt,
author = "Miramontes, J. Luis",
title = "{Pohlmeyer reduction revisited}",
eprint = "0808.3365",
archivePrefix = "arXiv",
primaryClass = "hep-th",
doi = "10.1088/1126-6708/2008/10/087",
journal = "JHEP",
volume = "10",
pages = "087",
year = "2008"
}

@article{Frolov:2019xzi,
author = "Frolov, Sergey",
title = "{$T{\overline T}$, $\widetilde JJ$, $JT$ and $\widetilde JT$ deformations}",
eprint = "1907.12117",
archivePrefix = "arXiv",
primaryClass = "hep-th",
doi = "10.1088/1751-8121/ab581b",
journal = "J. Phys. A",
volume = "53",
number = "2",
pages = "025401",
year = "2020"
}

@article{Frolov:2019nrr,
author = "Frolov, Sergey",
title = "{$T\overline{T}$ Deformation and the Light-Cone Gauge}",
eprint = "1905.07946",
archivePrefix = "arXiv",
primaryClass = "hep-th",
reportNumber = "TCD-MATH-19-06",
doi = "10.1134/S0081543820030098",
journal = "Proc. Steklov Inst. Math.",
volume = "309",
pages = "107--126",
year = "2020"
}

@article{Baggio:2018gct,
author = "Baggio, Marco and Sfondrini, Alessandro",
title = "{Strings on NS-NS Backgrounds as Integrable Deformations}",
eprint = "1804.01998",
archivePrefix = "arXiv",
primaryClass = "hep-th",
doi = "10.1103/PhysRevD.98.021902",
journal = "Phys. Rev. D",
volume = "98",
number = "2",
pages = "021902",
year = "2018"
}

@article{vanTongeren:2021jhh,
author = "van Tongeren, Stijn J. and Zimmermann, Yannik",
title = "{Do Drinfeld twists of $AdS_5 \times S^5$ survive light-cone quantization?}",
eprint = "2112.10279",
archivePrefix = "arXiv",
primaryClass = "hep-th",
doi = "10.21468/SciPostPhysCore.5.2.028",
journal = "SciPost Phys. Core",
volume = "5",
pages = "028",
year = "2022"
}

@article{Anous:2019osb,
author = "Anous, Tarek and Guica, Monica",
title = "{A general definition of $JT_a$-deformed QFTs}",
eprint = "1911.02031",
archivePrefix = "arXiv",
primaryClass = "hep-th",
doi = "10.21468/SciPostPhys.10.4.096",
journal = "SciPost Phys.",
volume = "10",
number = "4",
pages = "096",
year = "2021"
}

@article{Arutyunov:2014jfa,
author = "Arutyunov, Gleb and van Tongeren, Stijn J.",
title = "{Double Wick rotating Green-Schwarz strings}",
eprint = "1412.5137",
archivePrefix = "arXiv",
primaryClass = "hep-th",
doi = "10.1007/JHEP05(2015)027",
journal = "JHEP",
volume = "05",
pages = "027",
year = "2015"
}

@article{Bena:2003wd,
author = "Bena, Iosif and Polchinski, Joseph and Roiban, Radu",
title = "{Hidden symmetries of the $AdS_5 \times S^5$ superstring}",
eprint = "hep-th/0305116",
archivePrefix = "arXiv",
reportNumber = "NSF-KITP-03-34, UCLA-03-TEP-14",
doi = "10.1103/PhysRevD.69.046002",
journal = "Phys. Rev. D",
volume = "69",
pages = "046002",
year = "2004"
}

@article{Maldacena:1997re,
author = "Maldacena, Juan Martin",
title = "{The Large N limit of superconformal field theories and supergravity}",
eprint = "hep-th/9711200",
archivePrefix = "arXiv",
reportNumber = "HUTP-97-A097, HUTP-98-A097",
doi = "10.4310/ATMP.1998.v2.n2.a1",
journal = "Adv. Theor. Math. Phys.",
volume = "2",
pages = "231--252",
year = "1998"
}

@article{Babichenko:2009dk,
author = "Babichenko, A. and Stefanski, Jr., B. and Zarembo, K.",
title = "{Integrability and the AdS(3)/CFT(2) correspondence}",
eprint = "0912.1723",
archivePrefix = "arXiv",
primaryClass = "hep-th",
reportNumber = "ITEP-TH-59-09, LPTENS-09-36, UUITP-25-09",
doi = "10.1007/JHEP03(2010)058",
journal = "JHEP",
volume = "03",
pages = "058",
year = "2010"
}

@article{Arutyunov:2006ak,
author = "Arutyunov, Gleb and Frolov, Sergey and Plefka, Jan and Zamaklar, Marija",
title = "{The Off-shell Symmetry Algebra of the Light-cone $AdS_5 \times S^5$ Superstring}",
eprint = "hep-th/0609157",
archivePrefix = "arXiv",
reportNumber = "AEI-2006-071, HU-EP-06-31, ITP-UU-06-39, SPIN-06-33, TCDMATH-06-13",
doi = "10.1088/1751-8113/40/13/018",
journal = "J. Phys. A",
volume = "40",
pages = "3583--3606",
year = "2007"
}

Cite Article
@article{Borsato:2014hja,
author = "Borsato, Riccardo and Ohlsson Sax, Olof and Sfondrini, Alessandro and Stefanski, Bogdan",
title = "{The complete $AdS_{3} \times S^3 \times T^4$ worldsheet S matrix}",
eprint = "1406.0453",
archivePrefix = "arXiv",
primaryClass = "hep-th",
reportNumber = "IMPERIAL-TP-OOS-2014-03, HU-MATHEMATIK-2014-11, HU-EP-14-19, SPIN-14-15, ITP-UU-14-17",
doi = "10.1007/JHEP10(2014)066",
journal = "JHEP",
volume = "10",
pages = "066",
year = "2014"
}

@article{Lloyd:2014bsa,
author = "Lloyd, Thomas and Ohlsson Sax, Olof and Sfondrini, Alessandro and Stefa\'nski, Jr., Bogdan",
title = "{The complete worldsheet S matrix of superstrings on $AdS_3 \times S^3 \times T^4$ with mixed three-form flux}",
eprint = "1410.0866",
archivePrefix = "arXiv",
primaryClass = "hep-th",
reportNumber = "IMPERIAL-TP-OOS-2014-04, HU-MATHEMATIK-2014-21, HU-EP-14-34",
doi = "10.1016/j.nuclphysb.2014.12.019",
journal = "Nucl. Phys. B",
volume = "891",
pages = "570--612",
year = "2015"
}

@article{Giombi:2009gd,
author = "Giombi, S. and Ricci, R. and Roiban, R. and Tseytlin, A. A. and Vergu, C.",
title = "{Quantum $AdS_5 \times S^5$ superstring in the AdS light-cone gauge}",
eprint = "0912.5105",
archivePrefix = "arXiv",
primaryClass = "hep-th",
reportNumber = "IMPERIAL-TP-RR-02-2009, BROWN-HET-1597",
doi = "10.1007/JHEP03(2010)003",
journal = "JHEP",
volume = "03",
pages = "003",
year = "2010"
}

@article{vanTongeren:2015soa,
author = "van Tongeren, Stijn J.",
title = "{On classical Yang-Baxter based deformations of the $AdS_5 \times S^5$ superstring}",
eprint = "1504.05516",
archivePrefix = "arXiv",
primaryClass = "hep-th",
reportNumber = "HU-EP-15-18, HU-MATH-15-05",
doi = "10.1007/JHEP06(2015)048",
journal = "JHEP",
volume = "06",
pages = "048",
year = "2015"
}

@article{Kawaguchi:2014qwa,
author = "Kawaguchi, Io and Matsumoto, Takuya and Yoshida, Kentaroh",
title = "{Jordanian deformations of the $AdS_5 \times S^5$ superstring}",
eprint = "1401.4855",
archivePrefix = "arXiv",
primaryClass = "hep-th",
reportNumber = "KUNS-2477, ITP-UU-14-05, SPIN-14-05",
doi = "10.1007/JHEP04(2014)153",
journal = "JHEP",
volume = "04",
pages = "153",
year = "2014"
}

@article{Lacroix:2023qlz,
author = "Lacroix, Sylvain and Wallberg, Anders",
title = "{An elliptic integrable deformation of the Principal Chiral Model}",
eprint = "2311.09301",
archivePrefix = "arXiv",
primaryClass = "hep-th",
reportNumber = "CERN-TH-2023-205",
month = "11",
year = "2023"
}

@article{Beisert:2005tm,
author = "Beisert, Niklas",
title = "{The SU(2|2) dynamic S-matrix}",
eprint = "hep-th/0511082",
archivePrefix = "arXiv",
reportNumber = "PUTP-2181, NSF-KITP-05-92",
doi = "10.4310/ATMP.2008.v12.n5.a1",
journal = "Adv. Theor. Math. Phys.",
volume = "12",
pages = "945--979",
year = "2008"
}

@article{Arutyunov:2006iu,
author = "Arutyunov, G. and Frolov, S.",
title = "{On $AdS_5 \times S^5$ String S-matrix}",
eprint = "hep-th/0604043",
archivePrefix = "arXiv",
reportNumber = "ITP-UU-06-15, SPIN-06-13",
doi = "10.1016/j.physletb.2006.06.064",
journal = "Phys. Lett. B",
volume = "639",
pages = "378--382",
year = "2006"
}

@article{Minahan:2002ve,
author = "Minahan, J. A. and Zarembo, K.",
title = "{The Bethe ansatz for N=4 superYang-Mills}",
eprint = "hep-th/0212208",
archivePrefix = "arXiv",
reportNumber = "UUITP-17-02, ITEP-TH-73-02",
doi = "10.1088/1126-6708/2003/03/013",
journal = "JHEP",
volume = "03",
pages = "013",
year = "2003"
}

@article{Beisert:2003jj,
author = "Beisert, Niklas",
title = "{The complete one loop dilatation operator of N=4 superYang-Mills theory}",
eprint = "hep-th/0307015",
archivePrefix = "arXiv",
reportNumber = "AEI-2003-056",
doi = "10.1016/j.nuclphysb.2003.10.019",
journal = "Nucl. Phys. B",
volume = "676",
pages = "3--42",
year = "2004"
}

@article{Borsato:2022ubq,
author = "Borsato, Riccardo and Driezen, Sibylle",
title = "{All Jordanian deformations of the $AdS_5 \times S^5$ superstring}",
eprint = "2212.11269",
archivePrefix = "arXiv",
primaryClass = "hep-th",
doi = "10.21468/SciPostPhys.14.6.160",
journal = "SciPost Phys.",
volume = "14",
number = "6",
pages = "160",
year = "2023"
}

@article{inpreparation-RiccardoSibylle,
author = "Borsato, Riccardo and Driezen, Sibylle",
note = "{in preparation}",
}

@article{Minahan:2010js,
author = "Minahan, Joseph A.",
title = "{Review of AdS/CFT Integrability, Chapter I.1: Spin Chains in N=4 Super Yang-Mills}",
eprint = "1012.3983",
archivePrefix = "arXiv",
primaryClass = "hep-th",
reportNumber = "UUITP-38-10",
doi = "10.1007/s11005-011-0522-9",
journal = "Lett. Math. Phys.",
volume = "99",
pages = "33--58",
year = "2012"
}

@article{Kruczenski:2004cn,
author = "Kruczenski, Martin and Tseytlin, Arkady A.",
title = "{Semiclassical relativistic strings in $S^5$ and long coherent operators in N=4 SYM theory}",
eprint = "hep-th/0406189",
archivePrefix = "arXiv",
reportNumber = "BRX-TH-543",
doi = "10.1088/1126-6708/2004/09/038",
journal = "JHEP",
volume = "09",
pages = "038",
year = "2004"
}

@article{Zarembo:2009au,
author = "Zarembo, K.",
title = "{Worldsheet spectrum in AdS(4)/CFT(3) correspondence}",
eprint = "0903.1747",
archivePrefix = "arXiv",
primaryClass = "hep-th",
reportNumber = "ITEP-TH-11-09, LPTENS-09-05, UUITP-08-09",
doi = "10.1088/1126-6708/2009/04/135",
journal = "JHEP",
volume = "04",
pages = "135",
year = "2009"
}

@article{vanTongeren:2015uha,
author = "van Tongeren, Stijn J.",
title = "{Yang-Baxter deformations, AdS/CFT, and twist-noncommutative gauge theory}",
eprint = "1506.01023",
archivePrefix = "arXiv",
primaryClass = "hep-th",
reportNumber = "HU-EP-15-27, HU-MATH-15-08",
doi = "10.1016/j.nuclphysb.2016.01.012",
journal = "Nucl. Phys. B",
volume = "904",
pages = "148--175",
year = "2016"
}

@article{Kruczenski:2004kw,
author = "Kruczenski, M. and Ryzhov, A. V. and Tseytlin, Arkady A.",
title = "{Large spin limit of $AdS_5 \times S^5$ string theory and low-energy expansion of ferromagnetic spin chains}",
eprint = "hep-th/0403120",
archivePrefix = "arXiv",
reportNumber = "BRX-TH-537",
doi = "10.1016/j.nuclphysb.2004.05.028",
journal = "Nucl. Phys. B",
volume = "692",
pages = "3--49",
year = "2004"
}

@article{Hoare:2021dix,
author = "Hoare, Ben",
title = "{Integrable deformations of sigma models}",
eprint = "2109.14284",
archivePrefix = "arXiv",
primaryClass = "hep-th",
doi = "10.1088/1751-8121/ac4a1e",
journal = "J. Phys. A",
volume = "55",
number = "9",
pages = "093001",
year = "2022"
}

@article{Delduc:2013qra,
author = "Delduc, Francois and Magro, Marc and Vicedo, Benoit",
title = "{An integrable deformation of the $AdS_5 \times S^5$ superstring action}",
eprint = "1309.5850",
archivePrefix = "arXiv",
primaryClass = "hep-th",
doi = "10.1103/PhysRevLett.112.051601",
journal = "Phys. Rev. Lett.",
volume = "112",
number = "5",
pages = "051601",
year = "2014"
}

@article{Klimcik:2002zj,
author = "Klimcik, Ctirad",
title = "{Yang-Baxter sigma models and dS/AdS T duality}",
eprint = "hep-th/0210095",
archivePrefix = "arXiv",
reportNumber = "IML-02-XY",
doi = "10.1088/1126-6708/2002/12/051",
journal = "JHEP",
volume = "12",
pages = "051",
year = "2002"
}

@article{Cherednik:1981df,
author = "Cherednik, I. V.",
title = "{Relativistically Invariant Quasiclassical Limits of Integrable Two-dimensional Quantum Models}",
doi = "10.1007/BF01086395",
journal = "Theor. Math. Phys.",
volume = "47",
pages = "422--425",
year = "1981"
}

@article{Dubovsky:2017cnj,
author = "Dubovsky, Sergei and Gorbenko, Victor and Mirbabayi, Mehrdad",
title = "{Asymptotic fragility, near AdS$_{2}$ holography and $ T\overline{T} $}",
eprint = "1706.06604",
archivePrefix = "arXiv",
primaryClass = "hep-th",
doi = "10.1007/JHEP09(2017)136",
journal = "JHEP",
volume = "09",
pages = "136",
year = "2017"
}

@article{Hassan:1992gi,
author = "Hassan, S. F. and Sen, Ashoke",
title = "{Marginal deformations of WZNW and coset models from O(d,d) transformation}",
eprint = "hep-th/9210121",
archivePrefix = "arXiv",
reportNumber = "TIFR-TH-92-61",
doi = "10.1016/0550-3213(93)90429-S",
journal = "Nucl. Phys. B",
volume = "405",
pages = "143--165",
year = "1993"
}

@article{Borsato:2023dis,
author = "Borsato, Riccardo",
title = "{Lecture notes on current-current deformations}",
eprint = "2312.13847",
archivePrefix = "arXiv",
primaryClass = "hep-th",
month = "12",
year = "2023"
}

@article{Henningson:1992rn,
author = "Henningson, Mans and Nappi, Chiara R.",
title = "{Duality, marginal perturbations and gauging}",
eprint = "hep-th/9301005",
archivePrefix = "arXiv",
reportNumber = "IASSNS-HEP-92-88",
doi = "10.1103/PhysRevD.48.861",
journal = "Phys. Rev. D",
volume = "48",
pages = "861--868",
year = "1993"
}

@article{Levkovich-Maslyuk:2016kfv,
author = "Levkovich-Maslyuk, Fedor",
title = "{The Bethe ansatz}",
eprint = "1606.02950",
archivePrefix = "arXiv",
primaryClass = "hep-th",
doi = "10.1088/1751-8113/49/32/323004",
journal = "J. Phys. A",
volume = "49",
number = "32",
pages = "323004",
year = "2016"
}

@article{Retore:2021wwh,
author = "Retore, Ana L.",
title = "{Introduction to classical and quantum integrability}",
eprint = "2109.14280",
archivePrefix = "arXiv",
primaryClass = "hep-th",
doi = "10.1088/1751-8121/ac5a8e",
journal = "J. Phys. A",
volume = "55",
number = "17",
pages = "173001",
year = "2022"
}

@article{Meier:2023kzt,
author = "Meier, Tim and van Tongeren, Stijn J.",
title = "{Quadratic Twist-Noncommutative Gauge Theory}",
eprint = "2301.08757",
archivePrefix = "arXiv",
primaryClass = "hep-th",
reportNumber = "HU-EP-23/03-RTG",
doi = "10.1103/PhysRevLett.131.121603",
journal = "Phys. Rev. Lett.",
volume = "131",
number = "12",
pages = "121603",
year = "2023"
}

@article{Meier:2023lku,
author = "Meier, Tim and van Tongeren, Stijn J.",
title = "{Gauge theory on twist-noncommutative spaces}",
eprint = "2305.15470",
archivePrefix = "arXiv",
primaryClass = "hep-th",
reportNumber = "HU-EP-23/11-RTG",
doi = "10.1007/JHEP12(2023)045",
journal = "JHEP",
volume = "12",
pages = "045",
year = "2023"
}

@article{Demulder:2023bux,
author = "Demulder, Saskia and Driezen, Sibylle and Knighton, Bob and Oling, Gerben and Retore, Ana L. and Seibold, Fiona K. and Sfondrini, Alessandro and Yan, Ziqi",
title = "{Exact approaches on the string worldsheet}",
eprint = "2312.12930",
archivePrefix = "arXiv",
primaryClass = "hep-th",
reportNumber = "NORDITA 2023-083",
month = "12",
year = "2023"
}

@article{vanTongeren:2016eeb,
author = "van Tongeren, Stijn J.",
title = "{Almost abelian twists and AdS/CFT}",
eprint = "1610.05677",
archivePrefix = "arXiv",
primaryClass = "hep-th",
reportNumber = "HU-EP-16-35, HU-MATH-16-19",
doi = "10.1016/j.physletb.2016.12.002",
journal = "Phys. Lett. B",
volume = "765",
pages = "344--351",
year = "2017"
}

@article{drinfeld1983constant,
title={Constant quasiclassical solutions of the Yang--Baxter quantum equation},
author={Drinfeld, Vladimir Gershonovich},
journal={Doklady Akademii Nauk},
volume={273},
number={3},
pages={531--535},
year={1983},
organization={Russian Academy of Sciences}
}

@article{Reshetikhin:1990ep,
author = "Reshetikhin, N.",
title = "{Multiparameter quantum groups and twisted quasitriangular Hopf algebras}",
doi = "10.1007/BF00626530",
journal = "Lett. Math. Phys.",
volume = "20",
pages = "331--335",
year = "1990"
}

@article{Giaquinto:1994jx,
author = "Giaquinto, Anthony and Zhang, James J.",
title = "{Bialgebra actions, twists, and universal deformation formulas}",
eprint = "hep-th/9411140",
archivePrefix = "arXiv",
doi = "10.1016/S0022-4049(97)00041-8",
journal = "J. Pure Appl. Algebra",
volume = "128",
pages = "133--151",
year = "1998"
}

@article{Kamefuchi:1961sb,
author = "Kamefuchi, S. and O'Raifeartaigh, L. and Salam, Abdus",
title = "{Change of variables and equivalence theorems in quantum field theories}",
doi = "10.1016/0029-5582(61)90056-6",
journal = "Nucl. Phys.",
volume = "28",
pages = "529--549",
year = "1961"
}

@article{Sfondrini:2019smd,
author = "Sfondrini, Alessandro and van Tongeren, Stijn J.",
title = "{$T\bar{T}$ deformations as $TsT$ transformations}",
eprint = "1908.09299",
archivePrefix = "arXiv",
primaryClass = "hep-th",
doi = "10.1103/PhysRevD.101.066022",
journal = "Phys. Rev. D",
volume = "101",
number = "6",
pages = "066022",
year = "2020"
}

@article{Harmark:2014mpa,
author = "Harmark, Troels and Orselli, Marta",
title = "{Spin Matrix Theory: A quantum mechanical model of the AdS/CFT correspondence}",
eprint = "1409.4417",
archivePrefix = "arXiv",
primaryClass = "hep-th",
doi = "10.1007/JHEP11(2014)134",
journal = "JHEP",
volume = "11",
pages = "134",
year = "2014"
}

\end{bibtex}

\bibliographystyle{nb}
\bibliography{\jobname}

\end{document}